\documentclass[aps,pra,amssymb,twocolumn,amsmath,superscriptaddress,showpacs,10pt]{revtex4}
\usepackage{graphicx,color}
\usepackage{dcolumn} 
\usepackage{bm} 

\newcommand{\Tr}{{\rm Tr}}
\newcommand{\beq}{\begin{equation}}
\newcommand{\eeq}{\end{equation}}

\newcommand{\bra}[1]{\big<#1|}
\newcommand{\ket}[1]{|#1\big>}
\newcommand{\barr}{\begin{eqnarray}}
\newcommand{\earr}{\end{eqnarray}}
\newcommand{\Ord}[1]{{\cal O}\left( #1\right)}

\newcommand{\REV}[1]{\textrm{\color{red}[[#1]]}}

\def\cD{{\cal D}}
\def\cZ{{\cal Z}}

\begin{document}
\author{A. De Pasquale}
\affiliation{Dipartimento di Fisica, Universit\`a di Bari, I-70126
Bari, Italy} \affiliation{INFN, Sezione di Bari, I-70126 Bari,
Italy} \affiliation{MECENAS, Universit\`a Federico II di Napoli, Via
Mezzocannone 8, I-80134 Napoli, Italy}

\author{P. Facchi}

\affiliation{INFN, Sezione di Bari, I-70126 Bari, Italy}
\affiliation{Dipartimento di Matematica, Universit\`a di Bari,
        I-70125  Bari, Italy}

\author{G. Parisi}
\affiliation{Dipartimento di Fisica, Universit\`{a} di Roma ``La
Sapienza", Piazzale Aldo Moro 2, 00185 Roma, Italy}
\affiliation{Centre for Statistical Mechanics and Complexity (SMC),
CNR-INFM, 00185 Roma,
Italy\\
INFN, Sezione di Roma,  00185 Roma, Italy}
\author{S. Pascazio} \affiliation{Dipartimento di Fisica,
Universit\`a di Bari,
        I-70126  Bari, Italy}
\affiliation{INFN, Sezione di Bari, I-70126 Bari, Italy}
\author{A. Scardicchio} \affiliation{Abdus Salam International Center for Theoretical Physics, Strada Costiera 11, I-34014 Trieste, Italy}\affiliation{INFN, Sezione di Trieste, I-34014 Trieste, Italy}

\title{Phase transitions and metastability in the distribution of the bipartite entanglement of a 
large quantum system 
}

\date{\today}

\begin{abstract}
We study the distribution of the Schmidt coefficients of the reduced density matrix of a quantum system in a pure state.
By applying general methods of statistical mechanics, we introduce a fictitious temperature and a partition function and
translate the problem in terms of the distribution of the eigenvalues of random matrices. We investigate the appearance of two phase transitions, one at a positive temperature, associated to very entangled states, and one at a negative temperature, signalling the appearance of a significant factorization in the many-body wave function. We also focus on the presence of metastable states (related to 2-D quantum gravity) and study the finite size corrections to the saddle point solution.
\end{abstract}


\maketitle
\section{Introduction}
Entanglement is an important resource in quantum information
processing and quantum enabled technologies \cite{nielsen,benenticasati}.
Besides its important applications in relatively simple systems, that can be described in terms of a few effective quantum variables, it is also
widely investigated in many-body systems \cite{Amico08,h4}, where the bipartite entanglement can be given a satisfactory quantitative
definition in terms of entropy and its linearized versions, such as purity
\cite{woot,entanglement2}. The characterization of the global
features of entanglement is more involved, unveiling in general different features of the many-body wave function \cite{multipart,mmes}, and it is
becoming clear that the multipartite entanglement of a large system cannot be fully characterized in terms of a single (or a few) measure(s).

Entanglement measures the nonclassical correlations between the different components of a quantum system and unearths
different characteristics of its many-body wave-function. When the quantum system is large,
it becomes therefore interesting to scrutinize the features of the
distribution of some bipartite entanglement measure,
such as the purity or the Von Neumann entropy. Besides being of interest for applications, this is
an interesting problem in statistical mechanics. In \cite{paper1} we
tackled this problem by studying a random matrix model that
describes the statistical properties of the eigenvalues of the
reduced density matrix of a subsystem $A$ of 
dimension $N$
(the complementary subsystem $B$ having the same 
dimension as $A$).
 In the limit of large system dimension $N$,
we introduced a partition function for the canonical ensemble as a
function of a fictitious temperature. The role of energy is
played by purity: different temperatures correspond to
different entanglement.  The most important result of our
analysis was the proof that the different regions of entanglement,
corresponding to different ranges of the fictitious temperature, are
separated by phase transitions.

One puzzle was left open in that paper: in the region of negative
temperatures our solution suddenly ceased to exist at a critical
$\beta_g$ where the average purity of subsytem $A$ was $\pi_{AB}=9/4N$ (a phenomenon
quite common in random matrix theory, as this critical point
corresponds to tesselations of random surfaces, or 2-D quantum
gravity). As the partition function exists for all $\beta \in
\mathbb{R}$, the region of factorizable states, where
$\pi_{AB}=\Ord{1}$, was not covered.

We will show in this paper that the solution in \cite{paper1}
becomes metastable for any $\beta<0$ (in the scaling of
\cite{paper1}) and a new stable solution appears which interpolates
smoothly from $\pi_{AB}=2/N$ to $\pi_{AB}=1$ as $\beta$ goes from
$0$ to $-\infty$. Moreover, we will also study the metastable solution that is born at $\beta=0$ and follow it through the region
$0 > \beta > -\infty$.

This paper is organized as follows. In Sec.\
\ref{sec:introductionstatisticalapproach} we introduce the notation and set the bases of the statistical mechanical approach to the problem.
In Sec.\ \ref{sec:positivetemp} we study positive temperatures, where at very low temperatures we find very entangled states.
Negative temperatures are investigated in Sec.\ \ref{sec:negativetemp}, where it is shown that two branches exist, a stable one associated to a partial factorization of the many-body wave function, and a metastable one which contains the 2D-quantum gravity point.
The finite size corrections are investigated in Sec.\ \ref{sec:finitesize}. We discuss the implications of our results for quantum information in Sec.\ \ref{sec:QI} and we conclude in Sec.\ \ref{sec:concl} by summarizing our findings and discussing them in terms of future perspectives.

\section{A statistical approach to the study of bipartite
entanglement} \label{sec:introductionstatisticalapproach} We start
off by describing a statistical approach to the study of bipartite
entanglement for large quantum systems. We will tackle this problem
by introducing a partition function \cite{paper1}.

Consider a bipartite system, composed of two subsystems $A$ and $B$.
The total system lives in the tensor product Hilbert space
$\mathcal{H}=\mathcal{H}_A\otimes\mathcal{H}_B$, with
$\dim\mathcal{H}_A=N \leq \dim\mathcal{H}_B=M$. We assume that the
system is in a pure state $\ket{\psi}\in\mathcal{H}$. The reduced
density matrix of subsystem $A$ reads
\begin{equation}
\rho_A=\Tr_B\ket{\psi}\bra{\psi}
\end{equation}
and is a Hermitian, positive, unit-trace $N\times N$ matrix. A good
measure of the entanglement between the two subsystems is given by
the purity
\begin{equation}
\pi_{AB}=\Tr_A\rho_A^2=\Tr_B\rho_B^2 
=\sum_{j=1}^N \lambda_j^2
\in [1/N,1] ,
\label{eq:purityN}
\end{equation}
whose minimum is attained when all the eigenvalues $\lambda_j$ are equal to
$1/N$ (completely mixed state and maximal entanglement between the
two bipartitions), while its maximum 
(attained when one eigenvalue is 1 and all others are 0)
detects a factorized state (no
entanglement).

In order to study the statistics of bipartite entanglement for pure
quantum states, we consider typical vector states $\ket{\psi}$
\cite{aaa,page}, sampled according to the unique, unitarily
invariant 
measure. The significance of this measure can be
understood in the following way: consider a reference state vector
$\ket{\psi_0}$ and a unitary transformation $\ket{\psi}=U
\ket{\psi_0}$. In the least set of assumptions on $U$, the measure
is chosen in a unique way, being the only left- and right-invariant
Haar (probability) measure of the unitary group $U(N^2)$. The final state $\ket{\psi}$ will hence be distributed
according to the measure mentioned above (independently of
$\ket{\psi_0}$). Notice the analogy with the maximum entropy
argument in classical statistical mechanics. By tracing over
subsystem $B$, this measure induces the following
measure over the
space of Hermitian, positive matrices of unit trace \cite{aaa,page}
\begin{eqnarray}
d\mu(\rho_A) &=& \cD\rho_A (\det \rho_A)^{M-N} \delta(1-\Tr \rho_A), \nonumber \\
&=& d^N\lambda \prod_{i<j}(\lambda_i-\lambda_j)^2
 \prod_\ell \lambda_\ell^{\eta N} \delta \left(1-\sum_k \lambda_k \right),
\label{eq:measure}
\end{eqnarray}
where $\lambda_k$ are the (positive) eigenvalues of $\rho_A$ (Schmidt
coefficients), we dropped the volume of the $U(N)$ group  (which is
irrelevant for our purposes) and $\eta N\equiv M-N$ is the difference
between the dimensions of the Hilbert spaces $\mathcal{H}_A$ and
$\mathcal{H}_B$. In order to study the statistical behavior of a large
bipartite quantum system we introduce a partition function from
which all the thermodynamic quantities, for example the entropy or
the free energy, can be computed:
\begin{equation}
\label{eq:partitionfunction}
\cZ_{AB}=\int d\mu(\rho_A)  \exp\left(-\beta N^{\alpha}
\pi_{AB}\right) ,
\end{equation}
where $\alpha$ is a positive integer (either 2 or 3, as we shall
see) and $\beta$ a fictitious temperature ``selecting" different
regions of entanglement. The value of $\alpha$ needs to be chosen in
order to yield the correct thermodynamic limit as
\begin{equation}
\label{eq:piscale}
N^\alpha\langle\pi_{AB}\rangle=\Ord{N^2},
\end{equation}
since $N^2$ is the number of degrees of freedom of the matrix $\rho_A$.
Around the maximally entangled states (for $\beta>0$ \cite{paper1})
we have $\langle\pi_{AB}\rangle=\Ord{1/N}$ so $\alpha=3$, while
around separable states (for $\beta<0$) we have
$\langle\pi_{AB}\rangle=\Ord{1}$ and hence $\alpha=2$. In the
following we will assume $\eta=0$, since this does not change the
qualitative picture (the extension to $\eta\neq0$ being
straightforward but computationally cumbersome).

Since the purity depends only on the eigenvalues of  $\rho_A$ the
partition function reads
\begin{equation}
\cZ_{AB}=\int_{\lambda_i\geq 0}
d^N\lambda\prod_{i<j}(\lambda_i-\lambda_j)^2\delta\left(1-\sum_{i=1}^N\lambda_i\right)e^{-\beta
N^\alpha \sum_i\lambda_i^2},
\end{equation}
which by introducing a Lagrange multiplier for the delta function
yields
\begin{eqnarray}\label{eq: partition function alpha}
\cZ_{AB}&=&N^2
\int_{-\infty}^\infty\frac{d\xi}{2\pi}\int_{\lambda_i\geq 0}
d^N\lambda
\nonumber\\
& & \times e^{i N^2 \xi(1-\sum_i\lambda_i)-\beta N^\alpha
\sum_i\lambda_i^2+2\sum_{i<j}\ln|\lambda_i-\lambda_j|}. \;
\end{eqnarray}
A physical interpretation of the exponent in the integrand of the
partition function can be given as follows \cite{mehta}: the
eigenvalues of $\rho_A$ can be interpreted as a gas of interacting
point charges (Coulomb gas) at positions $\lambda_i$'s, on the positive half-line
and with a quadratic potential. The solution of these integrals is known
(as Selberg's integral) for the case in which the integration limits
are $-\infty<\lambda_i<+\infty$ \cite{mehta}.

The constraint of the positivity of the eigenvalues makes the
computation of this integral far more complicated. Although a
exact solution for finite $N$ is unlikely to be found
\footnote{An exact solution can always be found by means of the
orthogonal polynomials method, but the expressions for $\cZ_{AB}$ grow
enormously in complexity with increasing $N$.} (but see \cite{Giraud,paper1} for the first few moments), the problem arising
from the constraint on the positivity of the eigenvalues can be
overcome in the large $N$ limit, as we will look for the stationary
point of the exponent with respect to both the $\lambda$'s and
$\xi$. In particular, the contour of integration for $\xi$ lies on
the real axis, but we will soon see that the saddle point for $\xi$
lies on the imaginary $\xi$ axis. It is then understood that the
contour needs to be deformed to pass by this point parallel to the
line of steepest descent. For the saddle point we need to find the
minimum of the free energy:
\begin{equation}
\beta F= \beta N^\alpha
\sum_i\lambda_i^2 - 2 \sum_{i<j}\ln|\lambda_i-\lambda_j| - i N^2 \xi \left( 1- \sum_i \lambda_i\right).
 \label{eq:freeF}
 \end{equation}
By varying $F$ we find the saddle point equations
\begin{eqnarray}
\label{eq:stat1}-2\beta N^\alpha \lambda_i+2\sum_{j\neq i}\frac{1}{\lambda_i-
\lambda_j}-iN^2 \xi&=&0,
\label{eq:normal0}
\\
\sum_i\lambda_i&=&1.
\label{eq:normal}
\end{eqnarray}
In the following sections we will separately analyze the range of
positive and negative temperatures, and unveil the presence of two
phase transitions for the system, a second order one at a positive
critical $\beta$ and a first order one at a negative critical
$\beta$.

\subsection{The global picture}

Before proceeding to a formal analysis of the phase transitions, it is convenient to give a qualitative picture of the behavior of the Schmidt coefficients as the temperature is changed. As the inverse temperature is decreased, the density matrix of subsystem $A$ changes as follows. As $\beta\to+\infty$ all eigenvalues are $=1/N$ (maximally mixed state).
As $\beta$ decreases, we encounter two phase transitions: one at a positive critical value $\beta_+$ and one at a negative critical value $\beta_-$, both critical values being to be determined.
For $\beta>\beta_+$ all eigenvalues remain $\Ord{1/N}$, their
distribution being characterized, as we shall see, by the Wigner semicircle law.
After the first phase transition, for $\beta_+>\beta>\beta_-$, the eigenvalues, all always  $\Ord{1/N}$, follow the Wishart distribution, divergent at the origin.
Finally, after the second phase transition, for $\beta<\beta_-$, one eigenvalue becomes $\Ord{1}$, ``evaporating" from the ``sea" of eigenvalues $\Ord{1/N}$: this is a signature of the emergence of factorization in the many-body wave function, the eigenvalue $\Ord{1}$ being associated with a significant separability between subsystems $A$ and $B$. For $\beta\to-\infty$ the many body wavefunction is fully factorized.
Pictorially, the typical eigenvalues vector evolves starting from $\beta=+\infty$ to $\beta=-\infty$ as
\begin{eqnarray}
\underbrace{
\left(\frac{1}{N},\frac{1}{N},...,\frac{1}{N}\right)
}_{\beta\to+\infty}
&\longrightarrow&
\underbrace{
\left(\Ord{\frac{1}{N}},...,\Ord{\frac{1}{N}}\right)
}_{+\infty>\beta>\beta_+} \nonumber \\
&\longrightarrow&
\underbrace{
\left(\Ord{\frac{1}{N}},\Ord{\frac{1}{N}},...,0,...,0\right)
}_{\beta_+>\beta>\beta_-} \nonumber \\
&\longrightarrow&
\underbrace{
\left(\Ord{1},\Ord{\frac{1}{N}},...,0,...,0\right)
}_{\beta_->\beta>-\infty} \nonumber \\
&\longrightarrow&
\underbrace{
\left(1,0,...,0\right)
}_{\beta\to-\infty} ,
\end{eqnarray}
where the zeroes in the second and third lines mean an accumulation of points around the origin, and we will find that [in the scaling of $\beta$ given by $\alpha=3$ in Eq.\ (\ref{eq:piscale})], $\beta_+=2$ and $\beta_-=-2.455/N$.

\section{Positive temperatures: towards maximally entangled states}
\label{sec:positivetemp}

In this section we will consider the range of positive temperatures:
$0 < \beta< +\infty$; in particular we will study the occurrence of a
second order phase transition at $\beta=2$ \cite{paper1}.
We will use a novel more general method, that will enable us to find all solutions and can be easily extended to negative temperatures.

From the expression of the partition function one can easily infer
that when $\beta \rightarrow + \infty$ the typical states belonging to this distribution are maximally entangled states and correspond to the case $\lambda_i=1/N$, $\forall i \in \{1, \ldots N\}$. It then follows that for this range of temperatures the right scaling exponent
in (\ref{eq:partitionfunction})-(\ref{eq:piscale}) is $\alpha=3$.

In order to estimate the thermodynamic quantities of the system we
solve the saddle point equations (\ref{eq:stat1})-(\ref{eq:normal})
in the continuous limit, by introducing the natural scaling
\begin{equation}
\lambda_i=\frac{1}{N} \lambda (t_i), \qquad 0< t_i=\frac{i}{N}\leq 1.
\label{eq:scalinglambda}
\end{equation}
In the limit $N\to\infty$, Eq.~(\ref{eq:stat1}) becomes
\begin{equation}
-\beta\lambda + P\int_0^\infty
d\lambda'\frac{\rho(\lambda')}{\lambda-\lambda'}- i\frac{\xi}{2}=0 ,
\label{eq:sadpt2}
\end{equation}
which is a singular Fredholm equation of the first kind, known as Tricomi
equation \cite{tricomi}. The function
\begin{equation}
\rho(\lambda)=\int_0^1 dt \; \delta(\lambda-\lambda(t))
\label{eq:rhodef}
\end{equation}
is the density of eigenvalues we want to determine.
A similar equation, restricted at $\beta=0$, was studied by Page
\cite{page}.

\begin{figure}
\centering
\includegraphics[width=0.8\columnwidth]{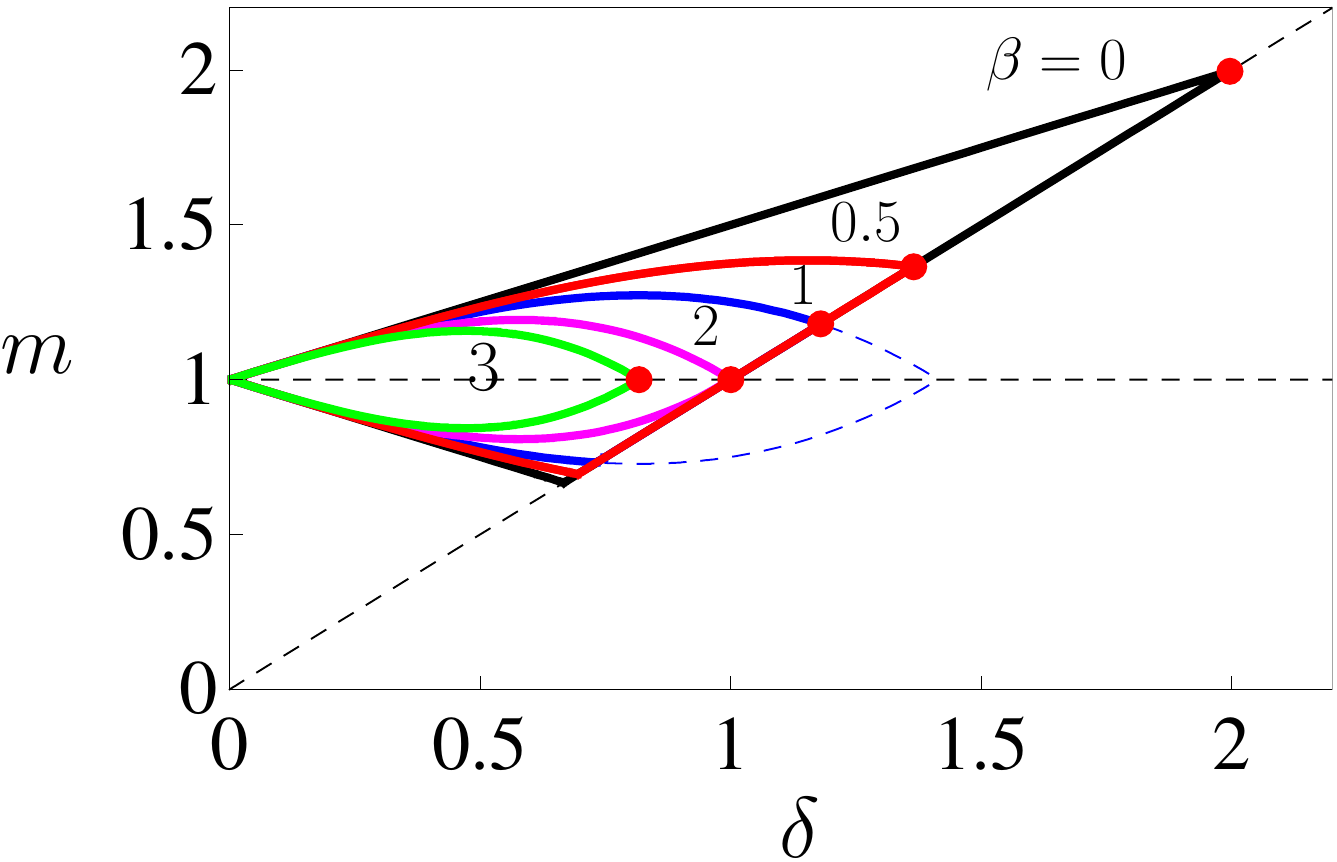}
\caption{Solution domain for different values of
temperature: for each value of $\beta$ (indicated) the relative full line encloses the region of the parameter space such that the
eigenvalue density is positive. The line $m=\delta$ corresponds to the positive eigenvalues condition.}
\label{fig:positive_eye}
\end{figure}

According to the Tricomi theorem \cite{tricomi,page} the
solution of the integral equation (\ref{eq:sadpt2})
 lies in a compact interval
$[a,b]$, ($0\leq a \leq b$).
Let us set
\begin{equation}
\label{eq:mdeltadef}
m=\frac{a+b}{2}, \qquad \delta=\frac{b-a}{2}, \qquad 0\leq\delta\leq m.
\end{equation}
We map the interval $[a,b]$ into the interval $[-1, 1]$ by
introducing the following change of variables:
\begin{equation}
\label{eq:new variable}
\lambda= m+ x \delta,\qquad
\phi(x)= \rho(\lambda ) \delta.
\end{equation}
We get
\begin{equation} \frac{1}{\pi} P\int_{-1}^1 \frac{\phi(y)}{y-x}d y =
g(x),
\label{eq:Tricomi}
\end{equation}
with
\begin{equation}
g(x)=-\frac{1}{\pi}\left(i\xi \frac{\delta}{2} +\beta \delta m +\beta\delta^2 x\right),
\end{equation}
whose normalized solution ($\int \phi\, dx=1$) is
\begin{equation}
\phi(x)=-\frac{1}{\pi}P\int_{-1}^1
\sqrt{\frac{1-y^2}{1-x^2}}\frac{g(y)}{y-x}d y+\frac{1}
{\pi\sqrt{1-x^2}},
\end{equation}
\begin{figure}
\includegraphics[width=0.7\columnwidth]{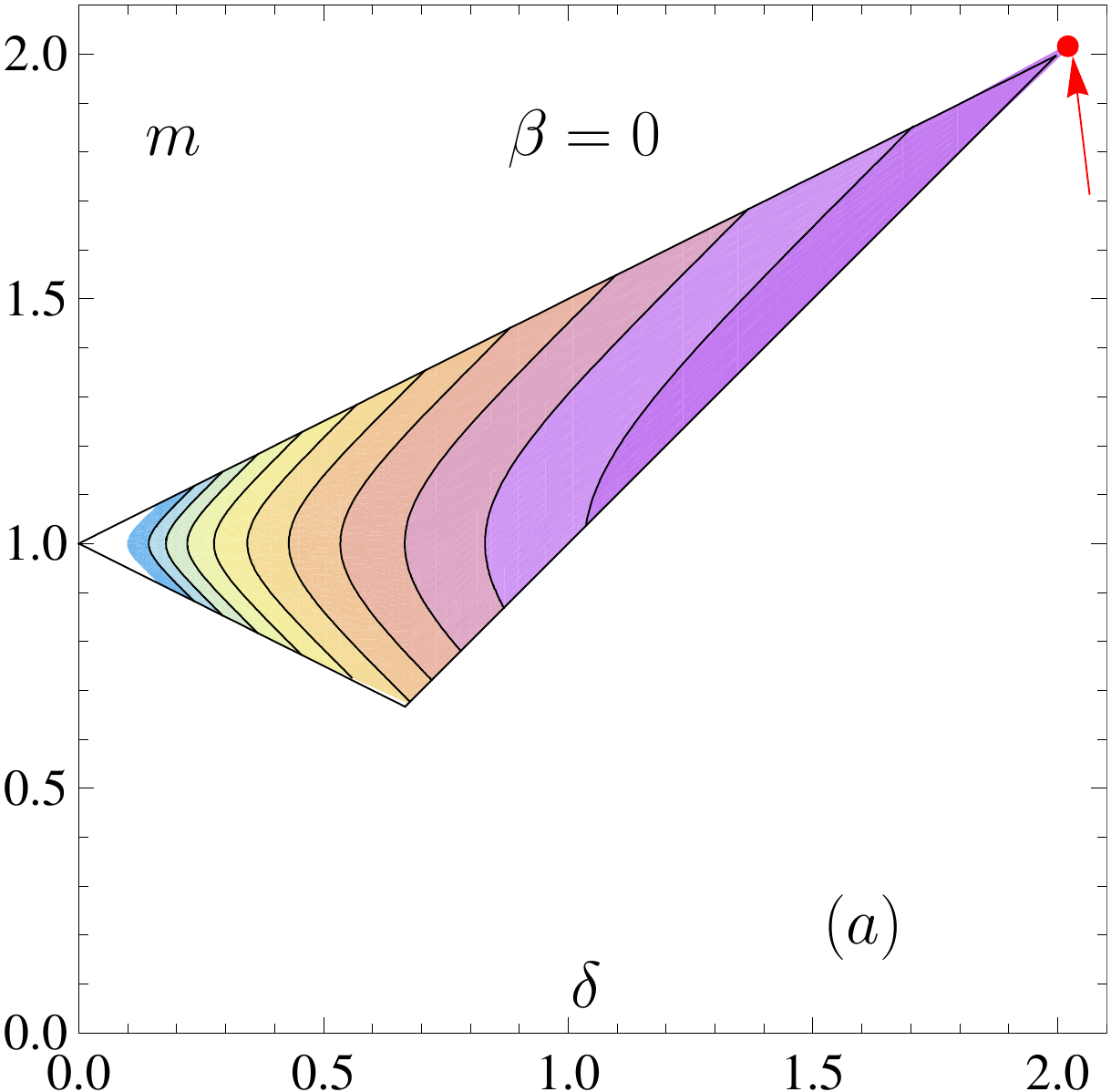}\\  \vspace{0.2cm}
\includegraphics[width=0.7\columnwidth]{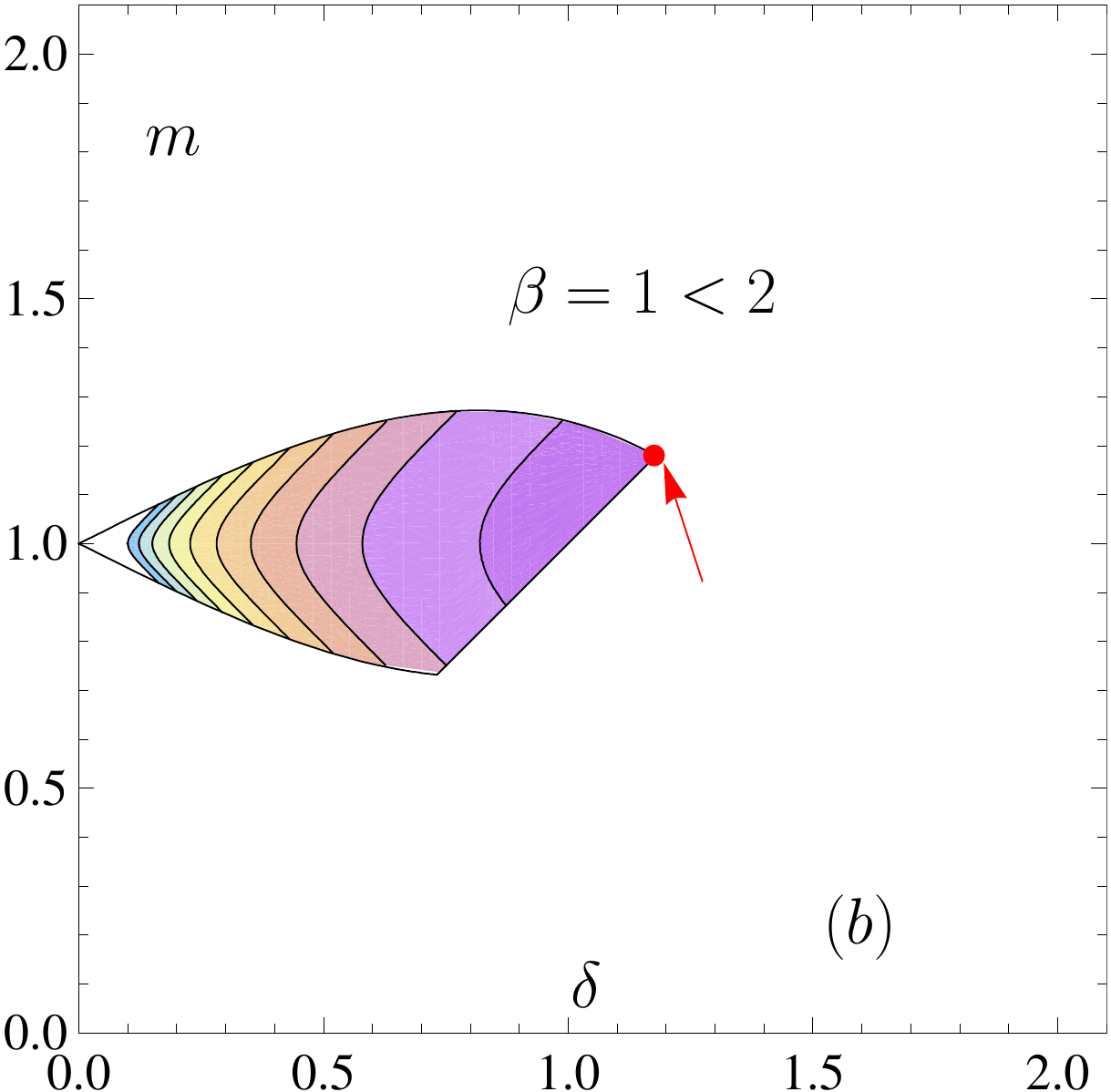}\\ \vspace{0.2cm}
\includegraphics[width=0.7\columnwidth]{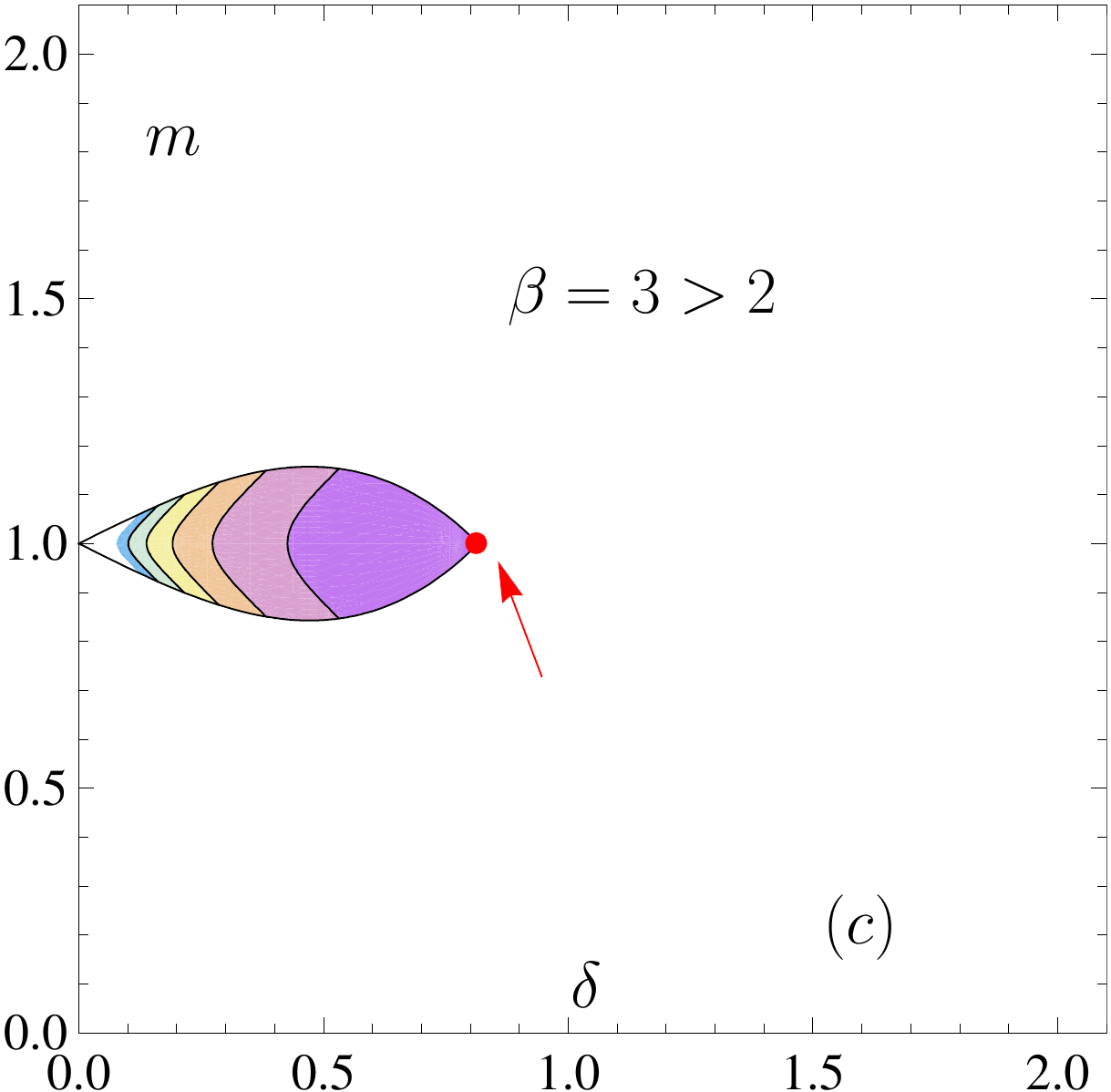}
\caption{Contour plots of the free energy in regions of the parameter
space such that $\phi(x)\geq0$, for (a) $\beta=0$, (b) $\beta=1$ and (c) $\beta=3$. Darker regions have lower free energy.
The arrows point at the  minima.}
\label{fig:contourspos}
\end{figure}
By using the constraint (\ref{eq:normal}), that is
\begin{equation}
\int_{-1}^{1} \lambda\, \phi(x)\, dx=1,
\end{equation}
we can fix the Lagrange multiplier to obtain
\begin{equation}\label{eq:phi(x)}
\phi(x)=
\frac{1}{\pi \sqrt{1-x^2}}\left[ 1+\frac{\beta \delta^2}{2}
+\frac{2(1-m)}{\delta}x -\beta \delta^2 x^2\right].
\end{equation}

The physical solutions must have a density $\phi(x)$ that is nonnegative for all $x\in(-1,1)$.
Let us look at the points where the density vanishes, $\phi(x)=0$. From (\ref{eq:phi(x)}) one gets
\begin{equation}
\label{eq:roots}
x_{\pm}=\frac{1}{\beta\delta^2} \left(\frac{1-m}{\delta}\pm\sqrt{\Delta}\right),
\end{equation}
where
\begin{equation}
\label{eq:discriminant}
\Delta=\left(\frac{1-m}{\delta}\right)^2+\beta\delta^2 \left(1+\frac{\beta \delta^2}{2}\right).
\end{equation}
For $\beta\geq 0$ one gets that $\Delta \geq 0$ for every $m$ and $\delta$,
and $\phi(x)\geq0$ for $x\in [x_-,x_+]$.
The level curves $x_{\pm}=\pm 1$ are given by $m=\Gamma_1^{\pm} (\delta,\beta) $, where
\begin{eqnarray}
\label{Gamma1}
\Gamma_1^{\pm} (\delta,\beta) = 1\pm\frac{\delta}{2}\left(1-\frac{\beta\delta^2}{2}\right).
\end{eqnarray}
They are symmetric with respect to the line $m=1$ and intersect at $\delta=0$ and at $\delta=\sqrt{2/\beta}$.
Therefore, the condition $(-1,1)\subset [x_1,x_2]$ implies
that the points $(\delta, m)$ should be restricted to a (possibly cut) ``eye-shaped'' domain
given by
\begin{equation}\label{eq:positive_temp_eyesShapeDomain}
\max\left\{ \delta,\Gamma_1^{-} (\delta,\beta)\right\} \leq m \leq \Gamma_1^{+} (\delta,\beta),
\end{equation}
(recall the constraint $m\geq\delta$ in (\ref{eq:mdeltadef}) that expresses the positivity of eigenvalues). The right corner of the eye is at
\begin{equation}
(\delta,m)=\left(\sqrt{\frac{2}{\beta}},1\right),
\label{eq:beta>2}
\end{equation}
and belongs to the boundary as long as $\beta\geq2$. For $\beta<2$ the eye is cut by the line $m=\delta$. See Fig.\ \ref{fig:positive_eye}.

Let  us remark that all points inside the region correspond to solutions of the saddle point equations. In other
words we have a \emph{two parameter  continuous family} of solutions.
We will look at the eigenvalue density that
minimizes the free energy density of the system.
From Eqs.\ (\ref{eq:freeF}) and (\ref{eq:normal}) with $\alpha=3$ by applying the scaling (\ref{eq:scalinglambda}) we get
\begin{eqnarray}
f_N &=& \frac{F}{N^2} =
 \frac{1}{N}
\sum_i\lambda(t_i)^2 - \frac{2}{N^2\beta} \sum_{i<j}\ln|\lambda(t_i)-\lambda(t_j)| 
\nonumber \\ 
& & + \frac{2}{N^2 \beta} \sum_{i<j}\ln N
\nonumber\\
&=&  u - \frac{1}{\beta} s + \ln N + \Ord{\frac{\ln N }{N}}
\nonumber\\
&=& f + \ln N + \Ord{\frac{\ln N }{N}}.
\end{eqnarray}
Here,
\begin{equation}
f= \lim_{N \to \infty}\left(f_N-\frac{1}{\beta}\ln N\right)
\end{equation}
is the free energy density in the thermodynamic limit, which reads
\begin{equation}\label{eq:formal free energy from energy density and entropy density}
 \beta f= \beta u - s,
\end{equation}
in terms of  the internal energy density $u$ and the entropy density $s$,
\begin{eqnarray}
u&=&\int_{-1}^{1}\lambda^2 \phi(x)d x,
\nonumber \\
s&=& \int_{-1}^{1} d x \int_{-1}^{1} d y \  \phi(x)\phi(y)\ln(\delta | x - y|) .
\end{eqnarray}

\begin{figure}
\includegraphics[width=0.7\columnwidth]{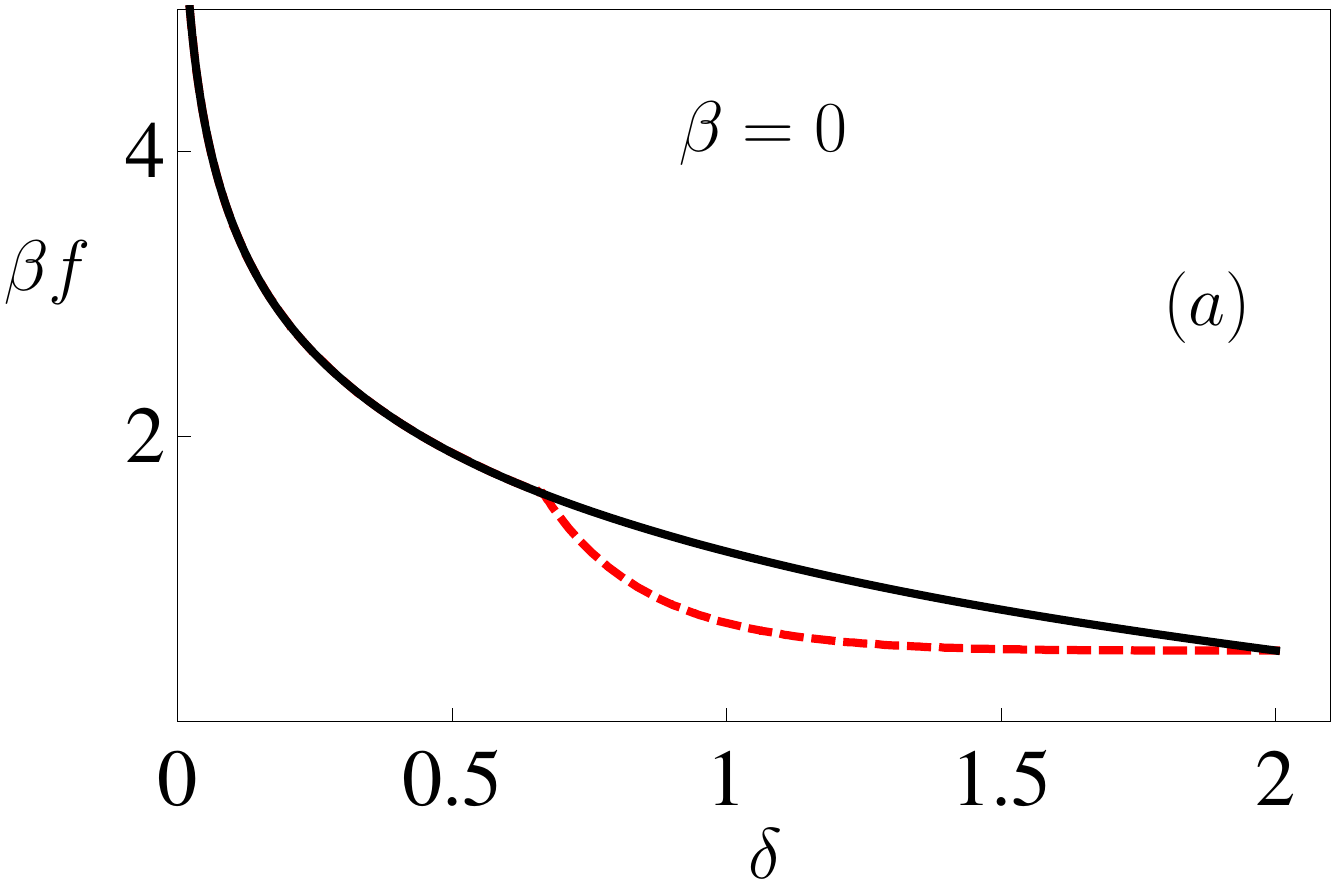}\\ \vspace{0.2cm}
\hspace{-0.3cm}\includegraphics[width=0.73\columnwidth]{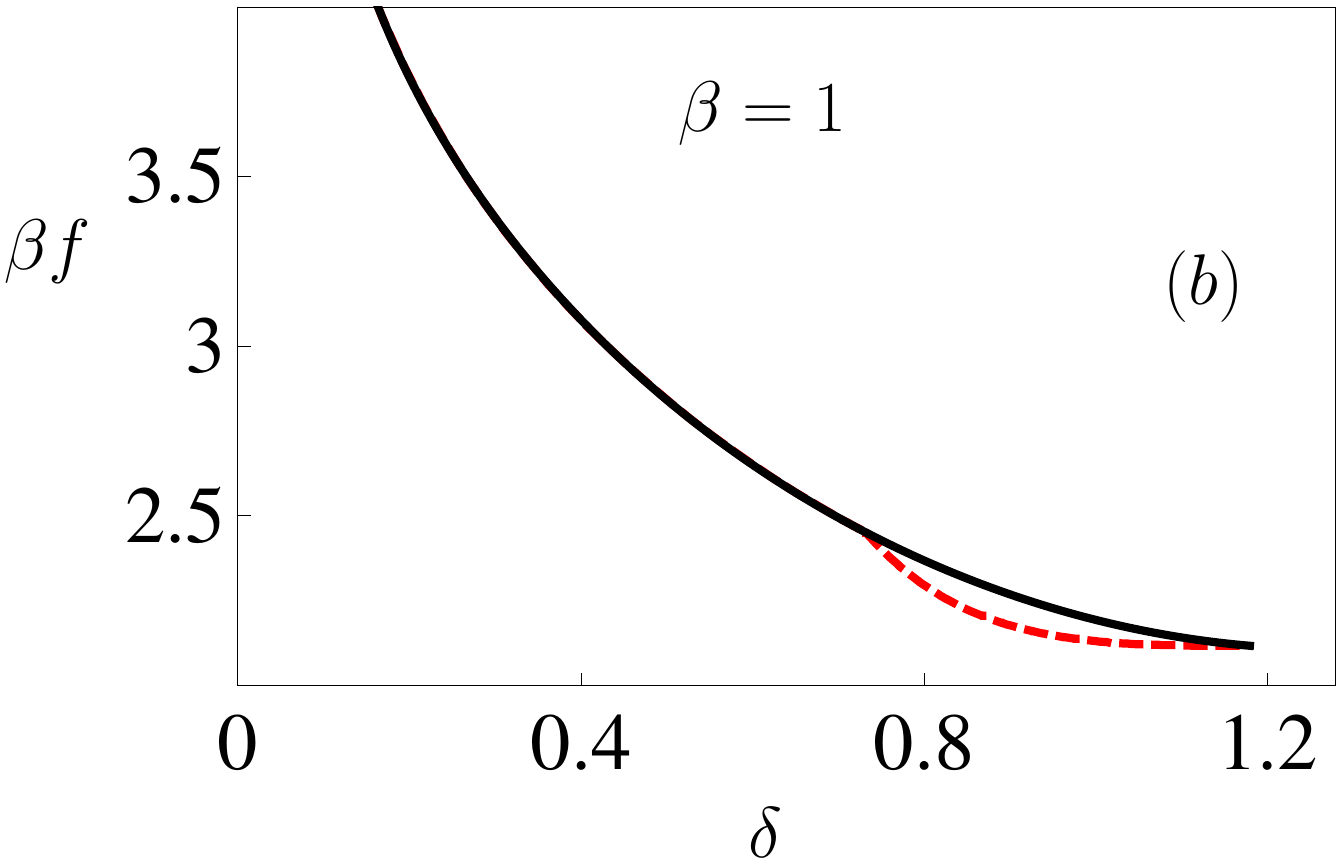}\\ \vspace{0.2cm}
\includegraphics[width=0.7\columnwidth]{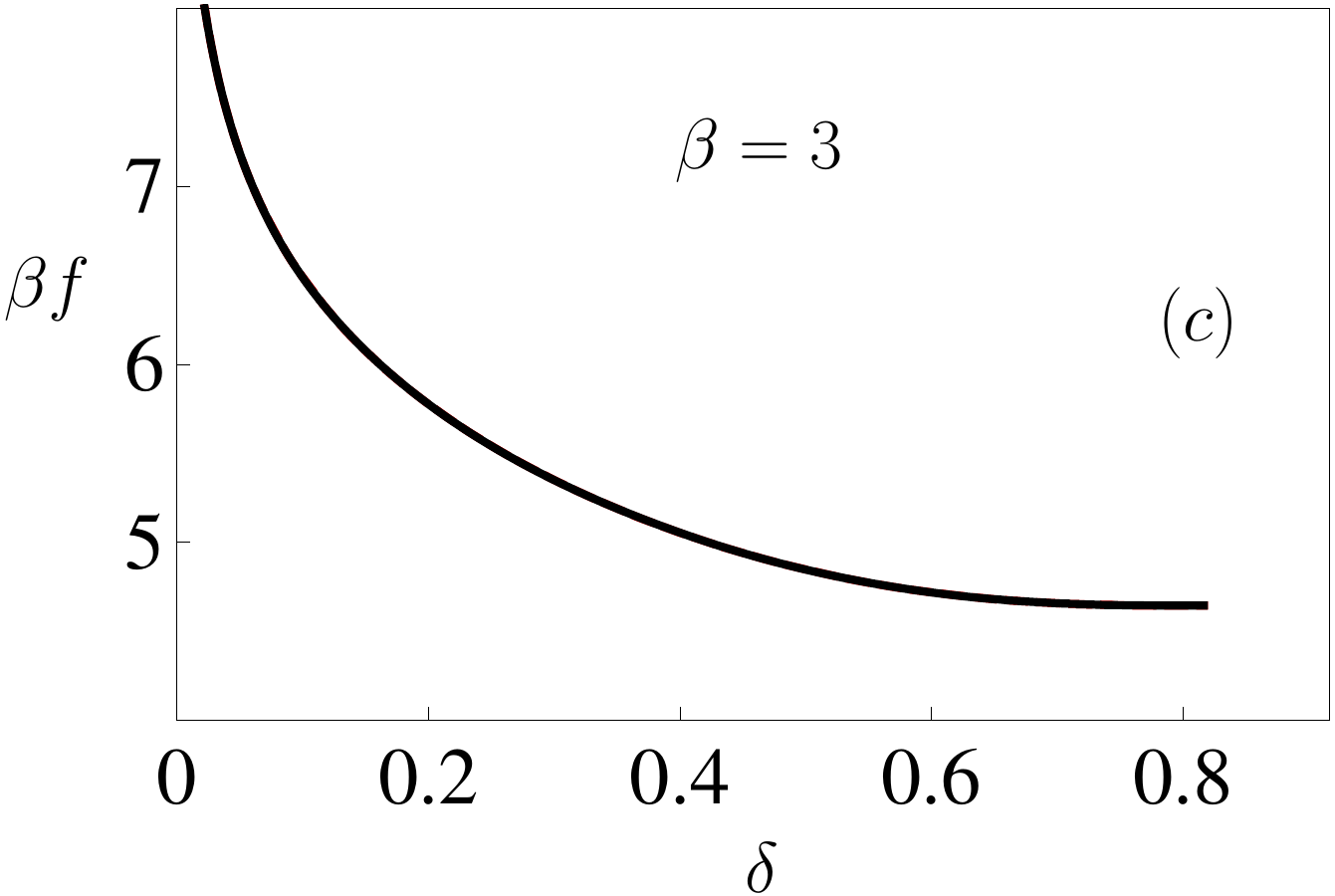}
\caption{Free energy on the boundary of the region of the domain
where $\phi(x)\geq0$, for different temperatures (indicated).
Dashed line: free energy $f$ on the lower boundary of the eye-shaped domain;
full line: free energy on the upper boundary. The sought minima of the free energy can be inferred from the graph and coincide with the dots in Figs.\ \ref{fig:positive_eye} and \ref{fig:contourspos}.}
\label{fig:3free}
\end{figure}

In order to compute the entropy density one
integrates the Tricomi equation (\ref{eq:Tricomi}) and obtains
\begin{eqnarray}
& & \int_{-1}^{1} \phi(x) d x  \int_{-1}^{1} \phi(y) d y \ln |y-x|
\nonumber \\
& & =
\int_{-1}^{1} d x \phi(x)\ln (x+1)
\nonumber\\ 
& &\quad
- \pi \int_{-1}^{1} d x
\phi(x)\int_{-1}^{x} g(y) dy.
\end{eqnarray}
We get
\begin{eqnarray}\label{eq:energy density1}
u (\delta,m,\beta) &=& 1-(1-m)^2+ \frac{\delta^2}{2}-\frac{\beta \delta^4}{8},\\
\label{eq:entrophy density1}
s (\delta,m,\beta) &=&-\frac{2(1-m)^2}{\delta^2}-\frac{\beta^2 \delta^4 }{16} +\ln
{\frac{\delta}{2}},
\end{eqnarray}
and thus
\begin{eqnarray}
\beta f(\delta,m,\beta)&=&\beta -\beta (1-m)^2 +  \frac{2(1-m)^2}{\delta^2}+
\frac{\beta\delta^2}{2}\nonumber\\
& & - \frac{\beta^2 \delta^4}{16}-\ln {\frac{\delta}{2}}.
\label{eq:freeenergy}
\end{eqnarray}
The contour plots of the free energy are shown in Fig.\ \ref{fig:contourspos}.
Note that $f$, as well as $u$ and $s$, is symmetric with respect to the line $m=1$. This $\mathbb{Z}_2$ symmetry will play a major role in the following.
The only stationary point (a saddle point) of the free energy density
$f$ is at  the right corner of the eye  (\ref{eq:beta>2}), see Figs.\ \ref{fig:positive_eye}
and \ref{fig:contourspos}.
Thus, the absolute minimum is on the boundary.

For $\beta\geq\beta_+$, where
\begin{equation}
\label{ eq:beta+def}
\beta_+=2,
\end{equation}
the point (\ref{eq:beta>2}) is also the absolute minimum,
whereas for $0<\beta<\beta_+$ the absolute minimum is at the right upper corner of the allowed region,
$\delta=\Gamma_1^+(\delta,\beta)$,
namely at
\begin{equation}
m=\delta, \quad\mbox{with} \quad \beta \frac{\delta^3}{4} +\frac{\delta}{2} -1 =0.
\label{eq:beta<2}
\end{equation}
See the dots in Figs.\ \ref{fig:positive_eye} and \ref{fig:contourspos}. The behavior of the free energy at the boundaries of the allowed domain is shown in Fig.\ \ref{fig:3free} for different temperatures.

\begin{figure}
\hspace{-0.8cm}\includegraphics[width=0.7\columnwidth]{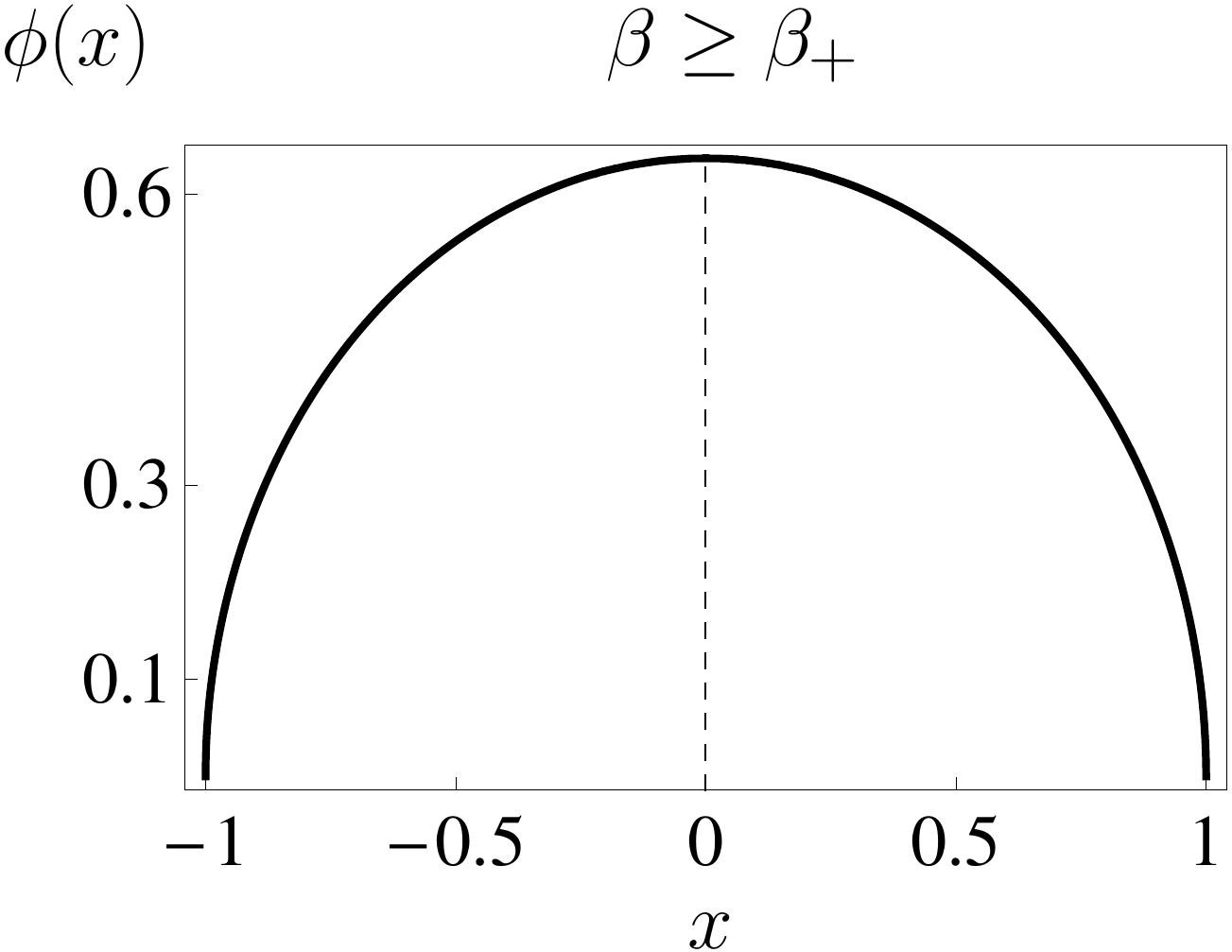}\\ 
\includegraphics[width=0.7\columnwidth]{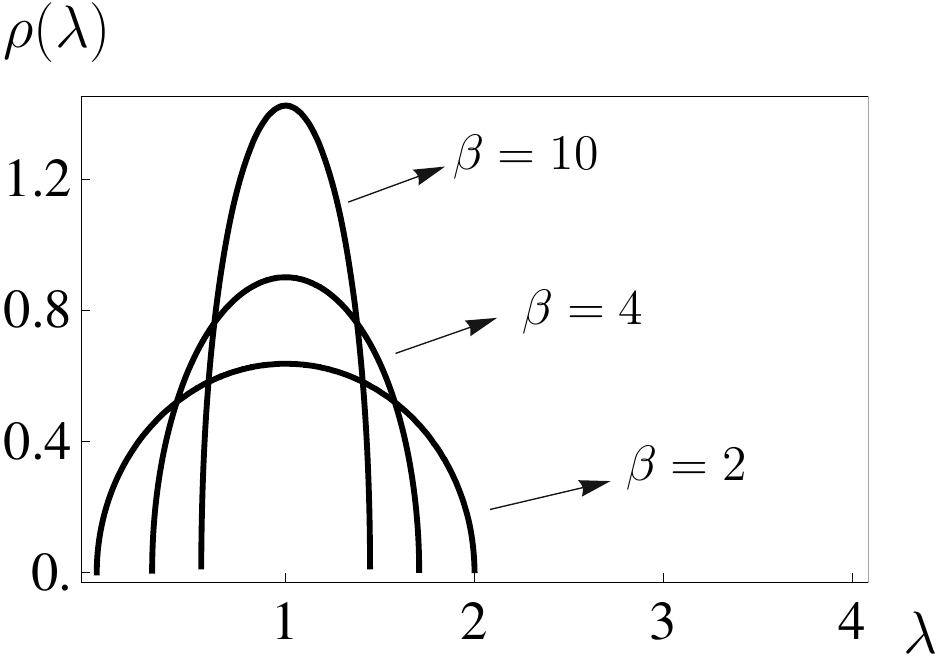}
\caption{(Up) Density of the eigenvalues for $\beta\geq\beta_+=2$. (Down) Density of the eigenvalues for $\beta=2, 4$ and
$10$. In the temperatures range $\beta \in [2,\infty]$ the solution is given by the semicircle
law.}\label{fig:semicircle}
\end{figure}

We will study the behavior of our system starting from high values of $\beta$, that is low values of internal energy $u$ (purity). 
The analysis of lower values of $\beta$, down to $\beta=0$ and even below, will be done in the next section.
For $\beta>\beta_{+}$, by setting from
(\ref{eq:beta>2}) $m=1$ and
\begin{equation}
\label{eq:betasss2}
\beta=\frac{2}{\delta^2}
\end{equation}
and recalling
(\ref{eq:phi(x)}), one gets the semicircle law (see Fig.~\ref{fig:semicircle})
\begin{equation}
\phi(x)= \frac{2}{\pi}\sqrt{1-x^2},
\label{eq:semicircle0}
\end{equation}
whence, by (\ref{eq:new variable}),
\begin{equation}
\label{eq:semicircle}
\rho(\lambda)=\frac{\beta}{\pi}\sqrt{\lambda-a}\sqrt{b-\lambda},
\end{equation}
where
\begin{equation}
a=1-\delta = 1 -\sqrt{\frac{\beta_+}{\beta}}, \qquad b=1+\delta=
1+\sqrt{\frac{\beta_+}{\beta}}.
\end{equation}
This distribution is displayed in Fig.\ \ref{fig:semicircle}. Observe that as $\beta$ becomes larger the distribution becomes increasingly peaked around 1. This simply means that all eigenvalues tend to $1/N$ in the natural scaling (\ref{eq:scalinglambda}): for temperatures $T =1/\beta$ close to zero the quantum state becomes maximally entangled.
 
By plugging (\ref{eq:beta>2}) into (\ref{eq:energy
density1}) and  (\ref{eq:entrophy density1}) we get for $\beta>\beta_+$
\begin{eqnarray}
u&=&1+ \frac{\delta^2}{4} = 1+\frac{1}{2\beta} ,
\label{eq:puri2}\\
\label{eq:entropy density for semicircle}
s&=&-\frac{1}{4} + \ln \frac{\delta}{2}= -\frac{1}{4} -\frac{1}{2} \ln (2\beta),
\end{eqnarray}
and thus
\begin{equation}
\beta f= \frac{2}{\delta^2}+\frac{3}{4}- \ln \frac{\delta}{2}=
\beta+\frac{3}{4} +\frac{1}{2} \ln(2\beta).
\end{equation}

\begin{figure}
\hspace{-0.7cm}\includegraphics[width=0.7\columnwidth]{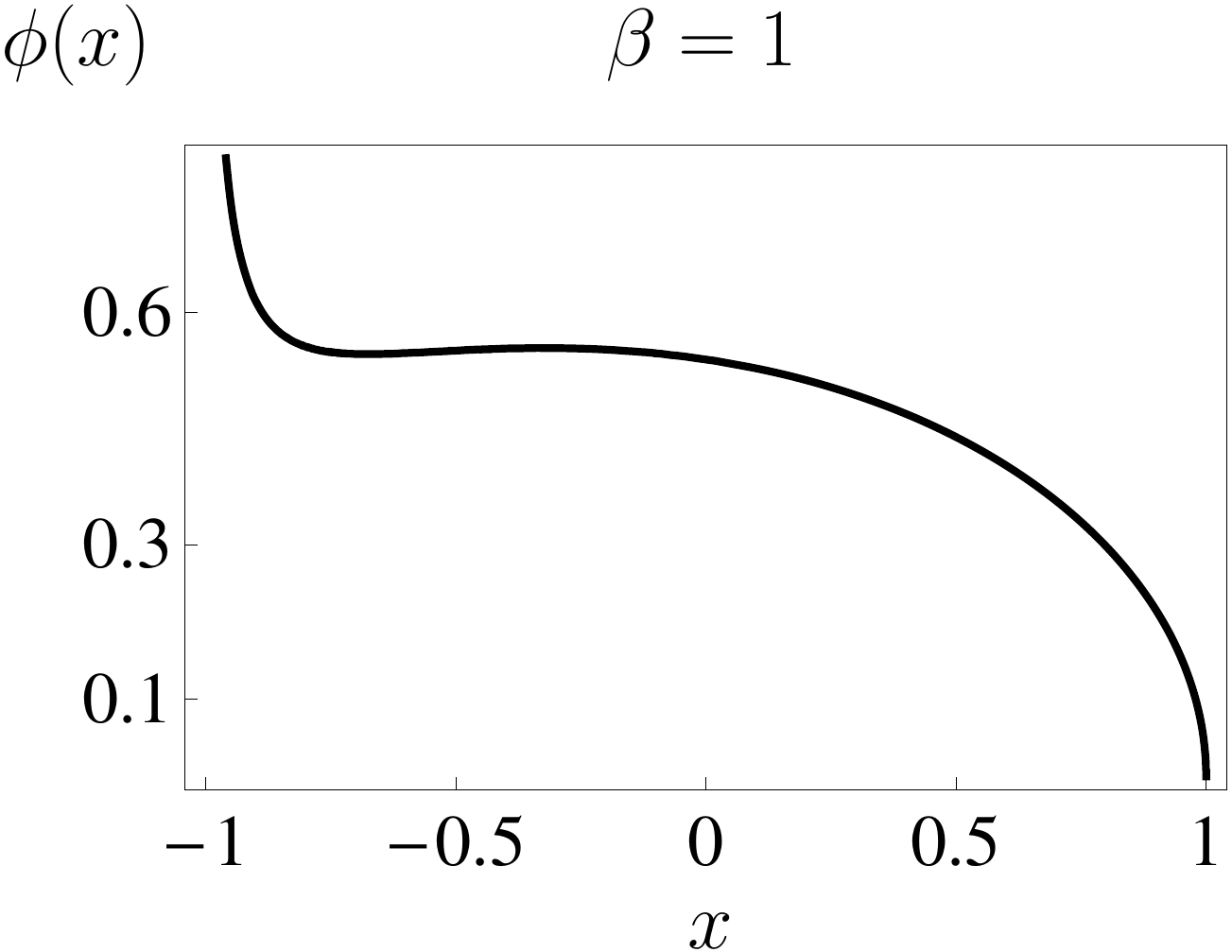}\\ 
\includegraphics[width=0.7\columnwidth]{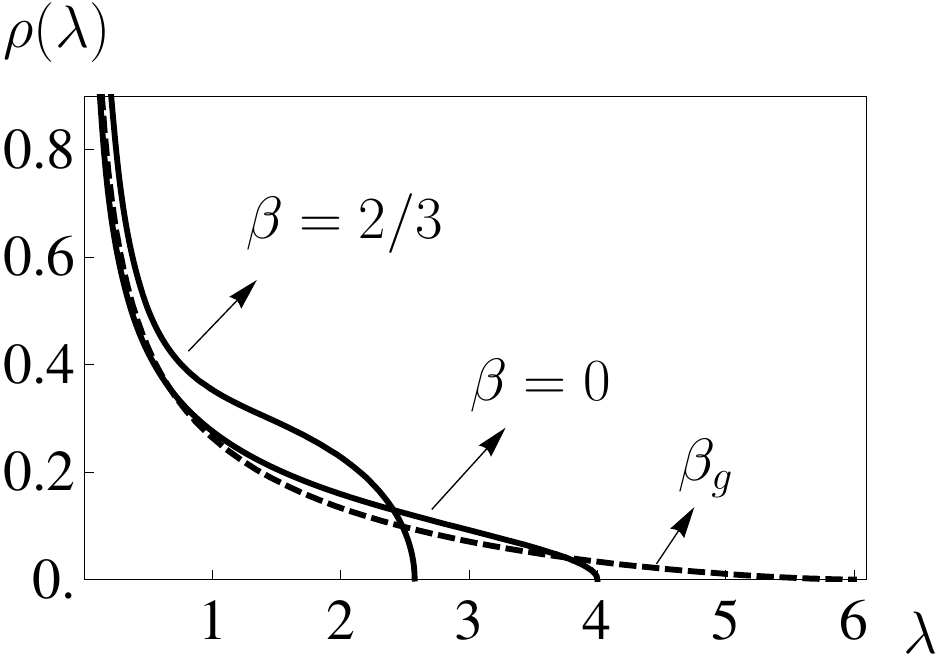}
\caption{(Up) Density of the eigenvalues for $\beta=1$. (Down) Density of eigenvalues for $\beta=0$, $\beta=2/3$, 
and $\beta=\beta_g=-2/27$ (dashed). In the range of temperatures $\beta \in (\beta_g,\beta_+)$, with $\beta_+=2$, the solution is given by the Wishart distribution.}
\label{fig:Wisharts}
\end{figure}

\begin{figure}
\includegraphics[width=0.8\columnwidth]{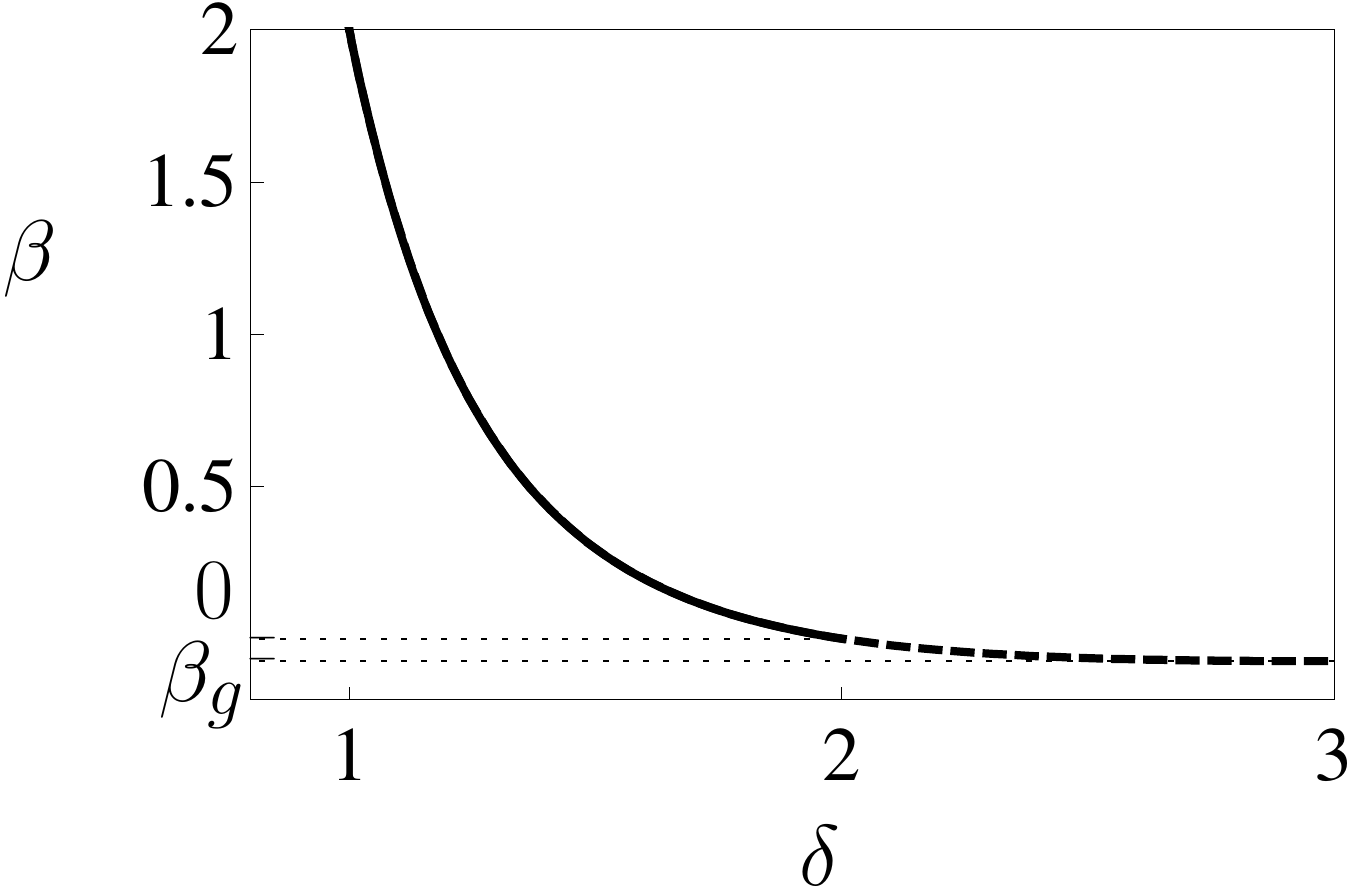}
\caption{Plot of Eq.\ (\ref{eq:betasss}) for positive  (solid line) and negative (dashed line) temperatures. The minimum $\beta_g=-2/27$ is attained at $\delta=3$.}
\label{fig:betadeltapos}
\end{figure}

At higher temperatures $0\leq \beta \leq \beta_+$ the solution 
acquires a different physiognomy. By plugging (\ref{eq:beta<2}) into (\ref{eq:phi(x)})
\begin{equation}
\phi(x)=\frac{2}{\pi\delta}\sqrt{\frac{1-x}{1+x}}\big(1+(2-\delta)x\big) ,
\label{eq:ansatz1phi}
\end{equation}
see Fig.\ \ref{fig:Wisharts}, yielding, by (\ref{eq:new variable}),
\begin{equation}
\label{eq:ansatz1}
\rho(\lambda)= \frac{4}{\pi b^2} \sqrt{\frac{b-\lambda}{\lambda}} \left(b-2+\frac{2(4-b)}{b}\lambda\right),
\end{equation}
with $b=2\delta$. This is a Wishart distribution.
See Fig.\ \ref{fig:Wisharts}.
The change from semicircle to Wishart is accompanied by a phase transition (the first of a series!) as we shall presently see. 

The half width $\delta=b/2$ is related to $\beta$
by (\ref{eq:beta<2})
\begin{equation}
\label{eq:betasss}
\beta=\frac{4}{\delta^3}-\frac{2}{\delta^2} ,
\end{equation}
which runs monotonically from $\beta =2$ when $\delta=1$ to $\beta=0$ when $\delta=2$. Moreover, it reaches a minimum equal to 
\begin{equation}
\label{eq:betagdef}
\beta_g=-\frac{2}{27} 
\end{equation}
at $\delta=3$. Therefore, the above solution can
be smoothly extended down to $\beta_g$, which is slightly negative, but not
below. See Fig.\ \ref{fig:betadeltapos}.
We will study the solution for negative
temperatures in the next section.
Note, incidentally, that the inverse function of (\ref{eq:betasss}) can be explicitly written
\begin{eqnarray}
\label{eq:adibeta}
\delta(\beta)= \frac{1}{\beta} \sqrt{\frac{2\beta}{3}}\left(\tilde\Delta-\frac{1}{\tilde\Delta}\right),
\end{eqnarray}
with $\tilde\Delta=(\sqrt{-\beta/\beta_g}+\sqrt{1-\beta/\beta_g})^{1/3}$.

For $\beta\leq\beta_+$ the internal energy (average purity) $u$ is obtained by plugging
(\ref{eq:betasss}) into (\ref{eq:energy
density1})
\begin{equation}
\label{eq:puri1}
u=\frac{3}{2}\delta -\frac{\delta^2}{4} .
\end{equation}
Therefore, at $\beta=0$ ($\delta=2$) one gets $u=2$,
at $\beta_+=2$ ($\delta=1$) one gets  $u=5/4$, and at $\beta_g=-2/27$ ($\delta=3$) one gets $u=9/4$ (see the next section for the
significance of these values).
From (\ref{eq:betasss}) and (\ref{eq:entrophy
density1}) one can also compute the entropy and the free energy for $\beta\leq\beta_+$
\begin{eqnarray}
\label{eq:entropydensityWishart}
s&=& -\frac{9}{4} + \frac{5}{\delta} - \frac{3}{\delta^2}
+\ln\frac{\delta}{2},\\
\beta f&=& \frac{9}{\delta^2} - \frac{9}{\delta} +\frac{11}{4} -\ln\frac{\delta}{2} ,
\label{eq:Fbeta0}
\end{eqnarray}
in terms of the function $\delta(\beta) \in (1,3]$ introduced in Eq.\ (\ref{eq:adibeta}).

Notice that $\beta f$ is the generating function for the connected
correlations of $\pi_{AB}$. The radius of convergence in the expansion
around $\beta=0$, namely $2/27$, defines the behavior of the late terms in the cumulants series. Another interesting observation is that the function $r(x)=u(\beta=-x/2)$ is the generating function of the number of rooted non-separable planar maps with $n$ edges on the sphere (Sloane's A000139 also in \cite{Tutte,BM}), namely
\begin{eqnarray}
r(x)&=&2+x+2x^2+6x^3+22x^4+91x^5+408x^6 \nonumber \\
& &+1,938x^7+9,614x^8+49,335x^9+...\ .
\end{eqnarray}
The counting of rooted planar maps on higher genus surfaces is an unsolved problem in combinatorics and we conjecture it to be related to $1/N$ corrections of our formulas.

We are now ready to unveil the presence of the first critical point
at $\beta_{+}=2$. Let consider the density of eigenvalues (\ref{eq:semicircle0}) and
(\ref{eq:ansatz1phi}) (or their counterpart (\ref{eq:semicircle}) and (\ref{eq:ansatz1})).
The phase transition at $\beta_+$ is due to
the restoration of a ${\mathbb Z}_2$ symmetry $P$ (``parity") present
in Eqs.\  (\ref{eq:energy density1}), (\ref{eq:entrophy density1}) and (\ref{eq:freeenergy}), namely the reflection of the distribution
$\rho(\lambda)$ around the center of its support ($m=\delta=b/2$ for
$\beta\leq\beta_+$ and $m=1$ for $\beta>\beta_+$). For $\beta\leq
\beta_+$ there are two solutions linked by this symmetry, and we
picked the one with the lowest $f$; at $\beta_+$ this two solutions
coincide with the semicircle (\ref{eq:semicircle}), which is
invariant under $P$ and becomes the valid and stable solution for
higher $\beta$. In order to explicitly show the presence of a second
order phase transition in the system for $\beta=\beta_{+}$ we look
at the expression of the entropy density $s=\beta(u-f)$, which
counts the number of states with a given value of the purity. The
expression for $\beta<\beta_+$ is given in Eq.\ (\ref {eq:entropydensityWishart}), 
while for $\beta\geq \beta_+$ it is given in Eq.\
(\ref{eq:entropy density for semicircle}).

\begin{figure}
\includegraphics[width=0.7\columnwidth]{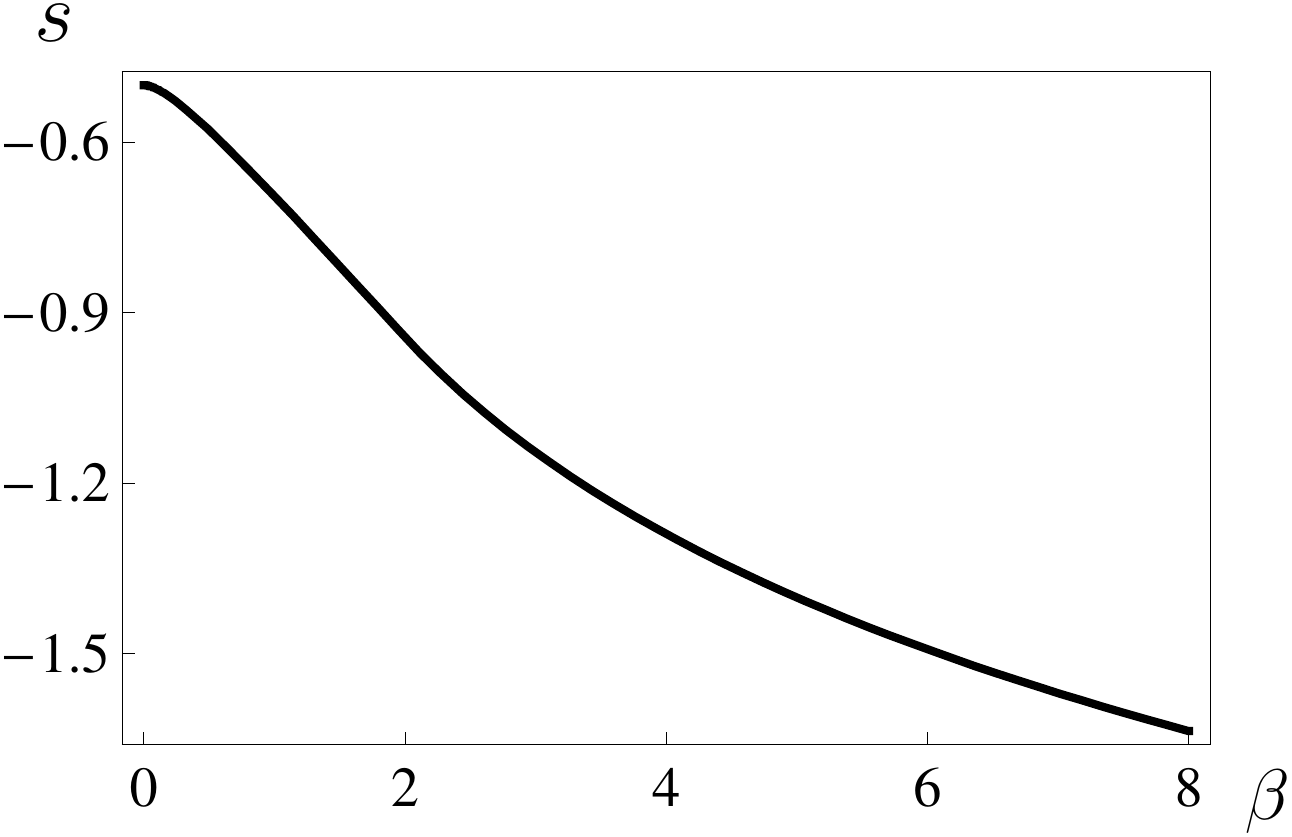}\\
\includegraphics[width=0.7\columnwidth]{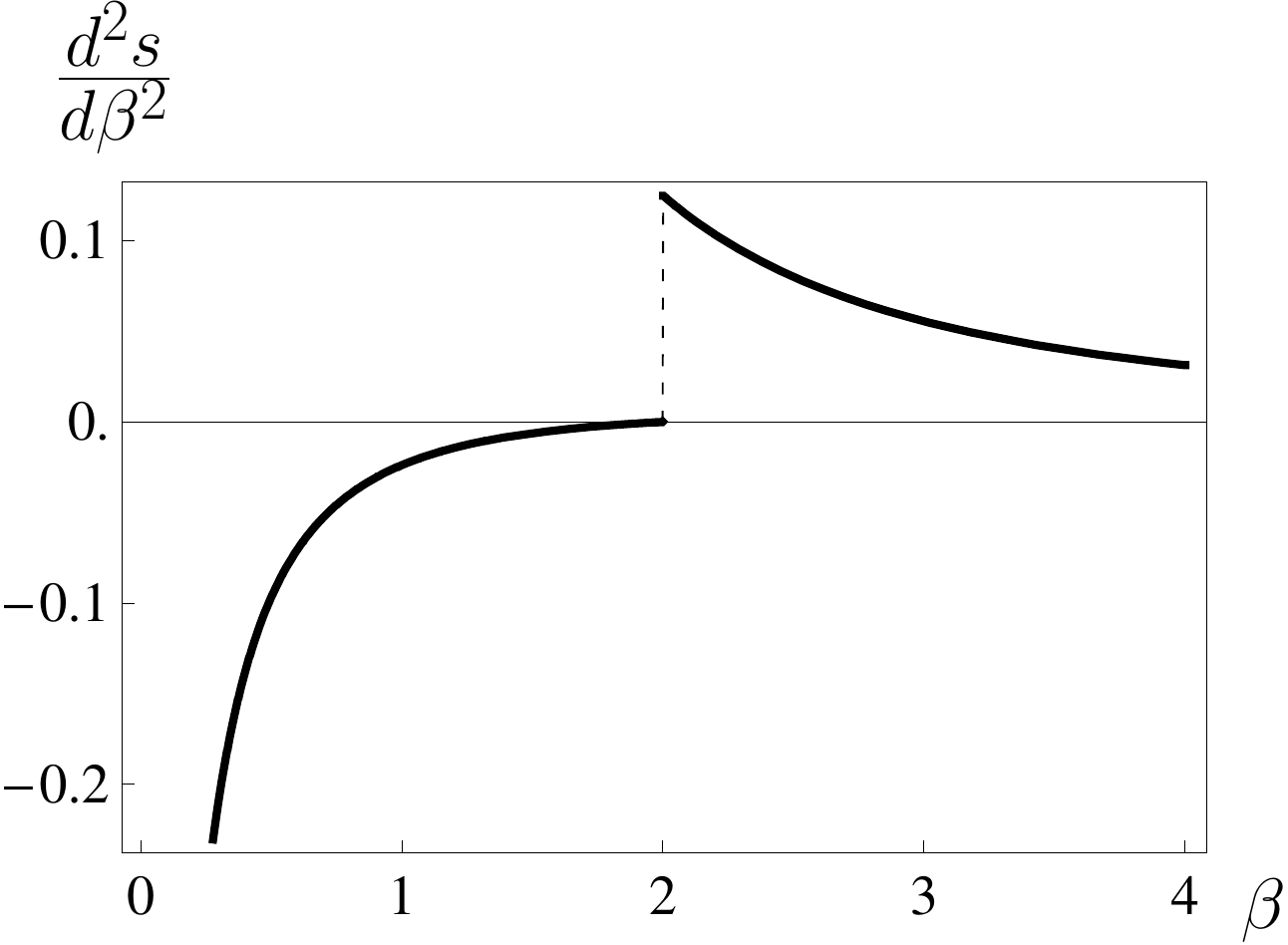}
\caption{Entropy and its second derivative with respect to $\beta$.
The entropy is continuous in $\beta_{+}$ while its second derivative
presents a finite discontinuity. }
\label{fig:entrophy at beta+}
\end{figure}

At $\beta=\beta_+$ we get $\delta=1$ and  $s=-1/4-\ln 2$. On the other hand
the first derivative of $s$ with respect to $\delta$ is discontinuous at $\delta=1$. However, also $\beta$ as a function of $\delta$, as given by (\ref{eq:betasss}) and (\ref{eq:betasss2}), has a discontinuous first derivative at $\delta=1$. By recalling that
\begin{eqnarray}
\frac{ds}{d\beta} &=& \frac{ds}{d\delta}\Big/\frac{d\beta}{d\delta},\nonumber \\
\frac{d^2s}{d\beta^2}&=& 
\frac{d^2s}{d\delta^2} \Big/\left(\frac{d\beta}{d\delta}\right)^2  - \frac{ds}{d\delta}\, \frac{d^2\beta}{d\delta^2} \Big/ \left(\frac{d\beta}{d\delta}\right)^3 ,
\end{eqnarray}
one easily obtains that the discontinuities compensate and  in the critical region, $\beta \to \beta_+$, we have
\begin{eqnarray}
s \sim -\frac{1}{4}-\ln 2 -\frac{\beta-\beta_+}{4} +
\theta(\beta-\beta_+) \frac{(\beta-\beta_+)^2}{16},\;
\end{eqnarray}
where $\theta$ is the step function. The entropy $s$ is
continuous at the phase transition, together with its first
derivative, although the second derivative is discontinuous, as shown
in Fig.\ \ref{fig:entrophy at beta+}.
Notice that the entropy is unbounded from below when $\beta \to
+\infty$. The interpretation of this result is quite
straightforward: the minimum value of $\pi_{AB}$ is reached on a
submanifold (isomorphic to $SU(N)/Z_{N}$ \cite{Kus01}) of dimension
$N^2-1$, as opposed to the typical vectors which form a
manifold of dimension $2N^2-N-1$ in the Hilbert space $\mathcal{H}$.
Since this manifold has zero volume in the original Hilbert space,
the entropy, being the logarithm of this volume, diverges.

\begin{figure}
\includegraphics[width=0.8\columnwidth]{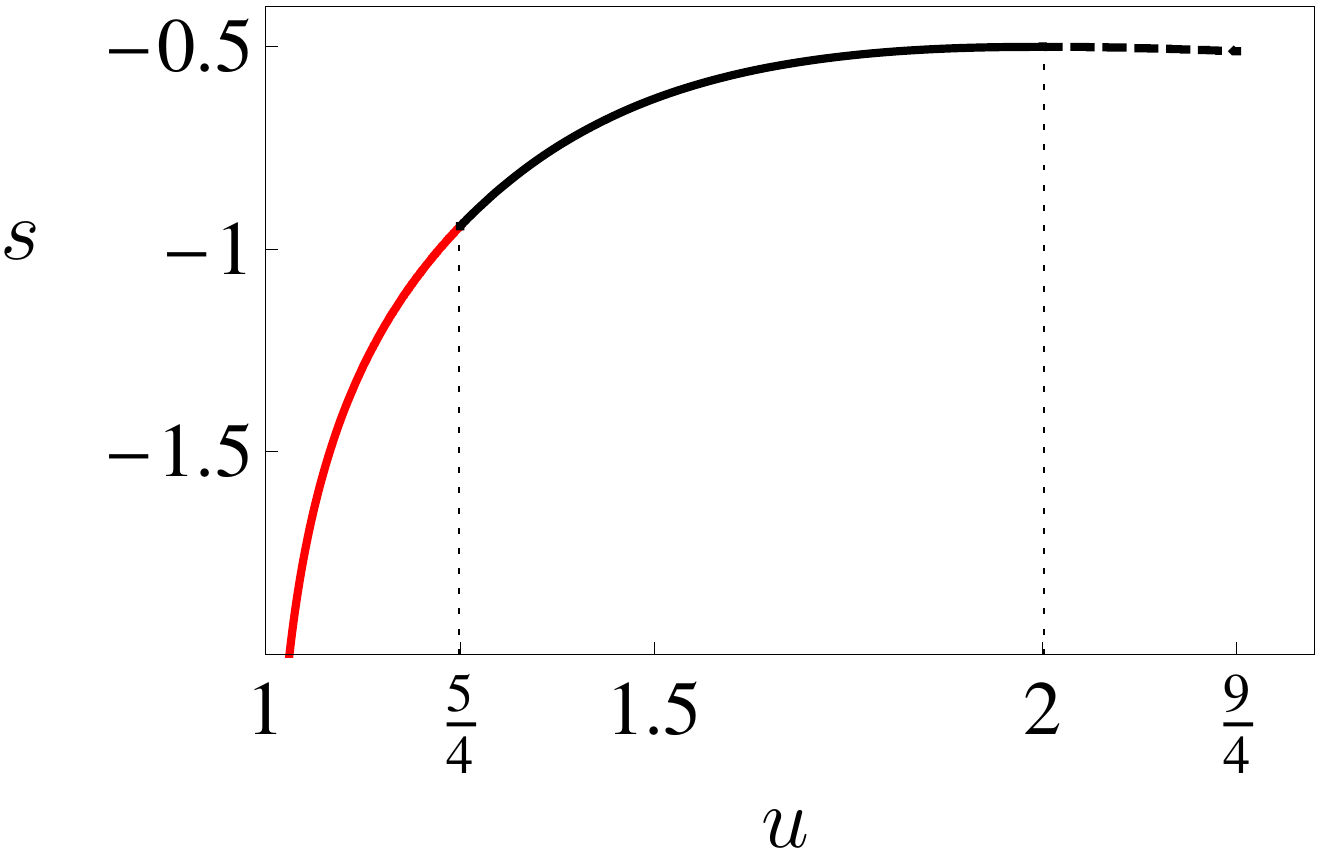}
\caption{Entropy density $s$ versus internal energy density $u=N\langle\pi_{AB}\rangle$. See Eq.\ (\ref{eq:112}).}
\label{fig:svsuN}
\end{figure}

Now we want to express the entropy density $s$ as a function of the internal energy density $u$.
From (\ref{eq:puri1}) and (\ref{eq:puri2}) we get
\begin{equation}
u  = \begin{cases}
1+\frac{\delta^2}{4},  & 0 < \delta \leq  1 ,  \\
\\
\frac{3}{2}\delta -\frac{\delta^2}{4},  & 1 < \delta \leq 2,
\end{cases}
\end{equation}
that can  be easily inverted
\begin{equation}
\delta  = \begin{cases}
2\sqrt{u-1},  & 1 < u \leq  \frac{5}{4} ,  \\
\\
3-\sqrt{9-4u},  & \frac{5}{4} < u \leq 2.
\end{cases}
\label{eq:deltavsu}
\end{equation}
From (\ref{eq:entropydensityWishart}) and (\ref{eq:entropy density for semicircle}) one gets the entropy density as a function of $\delta$
\begin{equation}
s  = \begin{cases}
-\frac{1}{4} + \ln \frac{\delta}{2},  & 0 < \delta \leq  1 ,  \\
\\
 -\frac{9}{4} + \frac{5}{\delta} - \frac{3}{\delta^2}
+\ln\frac{\delta}{2},  & 1 < \delta \leq 2.
\end{cases}
\label{eq:svsdeltatot}
\end{equation}
Finally, by plugging (\ref{eq:deltavsu}) into (\ref{eq:svsdeltatot}), we obtain the entropy of the submanifold of fixed purity, $s = (\ln V)/N^2$ as a function of its internal energy $u= N \pi_{AB}$:
\begin{widetext}
\begin{eqnarray}
s(u)  = \begin{cases}
\frac{1}{2} \ln (u-1) -\frac{1}{4},  & 1 \leq u \leq  \frac{5}{4} ,  \\
\\
 \ln \left( \frac{3}{2}  - \sqrt{\frac{9}{4} -u}
\right) - \frac{9}{4} + \frac{5}{2\left( \frac{3}{2}  - \sqrt{\frac{9}{4} -u}
\right)} -
\frac{3}{4\left( \frac{3}{2}  - \sqrt{\frac{9}{4} -u}
\right)^2},
& \frac{5}{4} \leq u \leq 2.
\end{cases}
\label{eq:112}
\end{eqnarray}
\end{widetext}
This function is plotted in Fig.\ \ref{fig:svsuN}.

Let us discuss the significance of these results. The present section was devoted to the study of positive temperatures $T=1/\beta >0$. In this range of temperatures, the eigenvalues of the reduced density matrix of our 
$N^2$-dimensional system
are always of $\Ord{1/N}$. 
As a consequence, the value of energy (purity) in Eq.\ (\ref{eq:purityN}) 
\begin{equation}
\pi_{AB}= \sum_{j=1}^N \lambda_j^2 \simeq \frac{1}{N}\int \lambda^2 \rho(\lambda) d\lambda = \Ord {\frac{1}{N}}
\label{eq:puritysmall}
\end{equation}
is always small: there is therefore a lot of entanglement in our system.
There are however, important differences as purity changes (it is important to keep in mind that in the statistical mechanical approach pursued here, the Lagrange multiplier $\beta$ fixes the value of energy/purity). When $1/N < \pi_{AB} < 5/4N$ the eigenvalues are distributed according to the semicircle law (Fig.\ \ref{fig:semicircle}), while for $5/4N < \pi_{AB} < 2/N$ they follow the Wishart distribution (Fig.\ \ref{fig:Wisharts}), the two regimes being separated by a second-order phase transition. The value $\pi_{AB} = 2/N$ corresponds to infinite temperatures $\beta=0$ and therefore to typical vectors in the Hilbert space (according to the Haar measure). One is therefore tempted to extend these results to negative temperatures \cite{paper1} and one can indeed do so up to $\pi_{AB} = 9/4N$, corresponding to the slightly negative temperature $\beta_g = -2/27$. However, as we have seen, a mathematical difficulty emerges, as this value represents the radius of convergence of an expansion around $\beta=0$ and no smooth continuation of this solution seems possible beyond $\beta_g$. In the next section we will see that two branches exist for negative $\beta$: one containing the point $\beta=\beta_g$ and in which purity is always of $\Ord{1/N}$ and one in which purity is of $\Ord{1}$. The latter becomes stable for sufficiently large $-\beta$'s through a first order phase transition.

Before continuing, we remind that larger values of purity, towards the regime $\pi_{AB} = \Ord{1}$ yield separable (factorized) states. We are therefore going to look at the behavior of our quantum
system towards separability (regime of small entanglement).

\section{Negative temperatures}
\label{sec:negativetemp}

\subsection{Metastable branch (quantum gravity)}
\label{sec:negativetemp1}

By analytic continuation, the solution at positive $\beta$ of the previous subsection can be turned into a solution for negative $\beta$, satisfying the constraints of positivity and normalization. In this section we will study this analytic continuation and its phase transitions, but we anticipate that this is metastable for sufficiently large $-\beta$'s (namely for $\beta<-2.455/N$) and that it will play a secondary role in the thermodynamics of our model. However, our interest in it is spurred by one of its critical points, at $\beta=-2/27\equiv \beta_g$ which corresponds to the so-called 2-D quantum gravity free energy (see \cite{Di Francesco:1993nw}), provided an appropriate double-scaling limit (jointly $\beta\to\beta_g$ and $N\to\infty$) is performed.

\begin{figure}
\includegraphics[width=0.7\columnwidth]{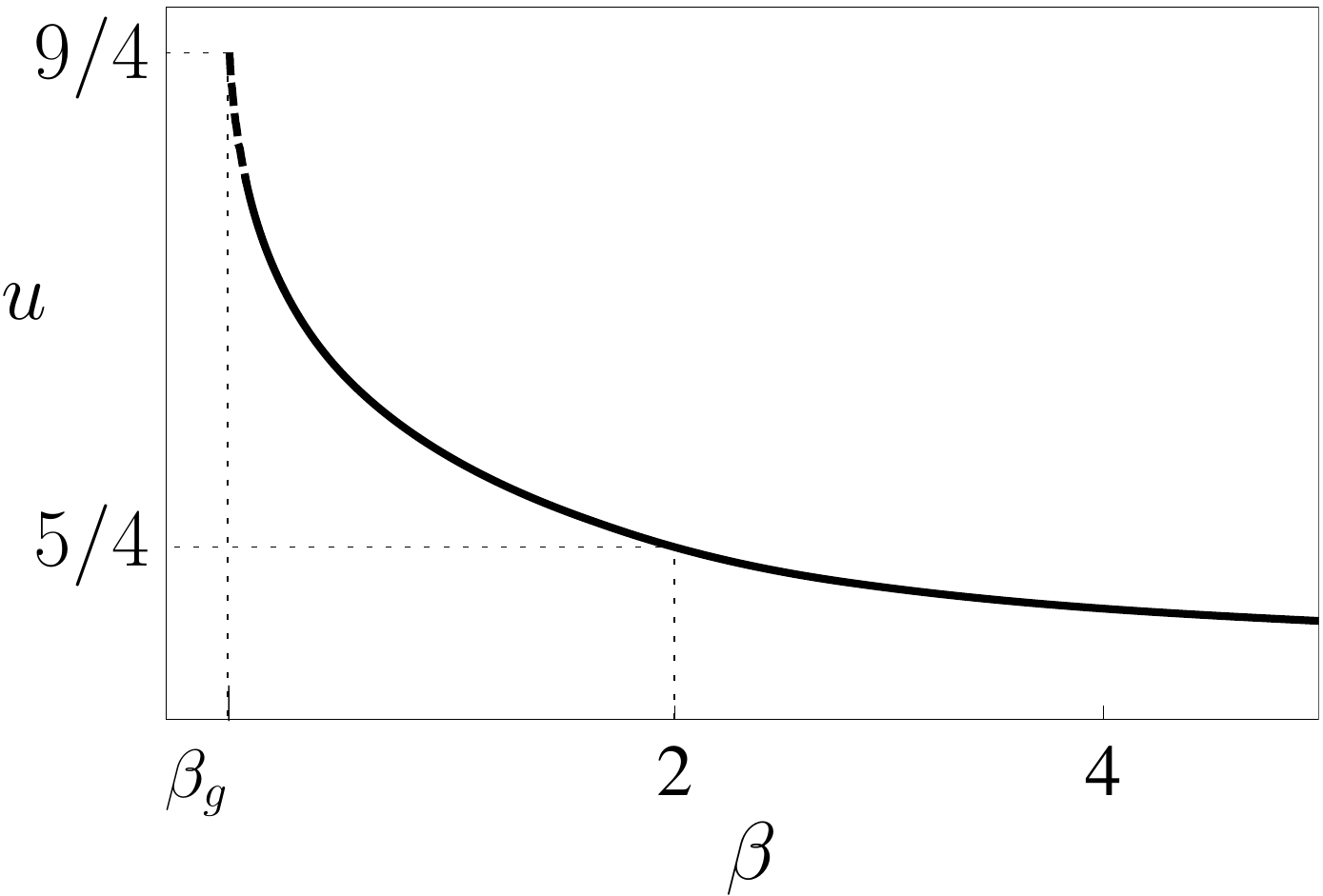}
\caption{$u=N\langle\pi_{AB}\rangle$ as a function of (inverse)
temperature. Notice that $u=2$ at $\beta=0$ (typical states).
In the $\beta\to\infty$ limit we find the minimum $u=1$. The
critical points described in the text are at $\beta_g=-2/27,u=9/4$
(left point) and $\beta_+=2,u=5/4$ (right point). However this phase
is unstable (dashed line) towards another phase as soon as $\beta<0$ (in this
scaling of $\beta$).} \label{fig: pi}
\end{figure}

In more details, the eigenvalue density (\ref{eq:ansatz1}) at $\beta=\beta_g=-2/27$, i.e.\ $\delta=3$
[see between Eqs.\ (\ref{eq:betasss}) and (\ref{eq:adibeta}), and Fig.~\ref{fig:betadeltapos}] reads
\begin{equation}
\rho(\lambda)=\frac{2}{27\pi} \sqrt{\frac{(6-\lambda)^{3}}{\lambda}},
\end{equation} 
and from (\ref{eq:puri1}) $u=9/4$ (see Fig. \ref{fig:entrophy at beta+} and \ref{fig: pi}). The derivative at the right edge of eigenvalue density
in Fig.\ \ref{fig:Wisharts} vanishes.

By expanding (\ref{eq:betasss})  for $\delta \to 3$ 
\begin{eqnarray} 
\beta &=& -\frac{2}{27} + \frac{2}{81}(\delta-3)^2 -\frac{16}{729}(\delta-3)^3 + \frac{10}{729}(\delta-3)^4 \nonumber \\
& & - \frac{16}{2187}(\delta-3)^5+\Ord{(\delta-3)^6},
\end{eqnarray}
that is, by setting $x=\sqrt{2(\beta-\beta_g)/9} \, \to 0$,
\begin{equation}
\delta = 3 \left(1+x+\frac{4}{3} x^2 +\frac{35}{18} x^3+\frac{80}{27} x^4 + \frac{1001}{216} x^5+\Ord{x^6}\right),
\end{equation}
and therefore for $\beta\to\beta_g$
\begin{eqnarray}
\beta f &=& \frac{3}{4}- \log \frac{3}{2} + \frac{9}{4} (\beta-\beta_g) - \frac{81}{16} (\beta-\beta_g)^2 \nonumber \\ 
& & 
- \frac{81 \sqrt{2}}{5} (\beta-\beta_g)^{5/2}+\Ord{(\beta-\beta_g)^3}.
\end{eqnarray}
In fact, if one relaxes the unit trace condition,
our partition function $\cZ$ has been studied in the context of
random matrix theories \cite{Morris91} before. The objects generated
in this way correspond to chequered polygonations of surfaces. Our
calculations show that the constraint $\Tr\ \rho_A=1$ is irrelevant
for the critical exponents in this region.

However, this is not a real critical point of our Coulomb gas. As this is an analytic continuation of the solution obtained for $\beta>0$, we are not assured that this is indeed a stable branch. In the next section we will show that a first order phase transition occurs at a lower value of $\beta$, namely at $\beta\simeq -2.455/N$ in this scaling (and therefore the exponent $\alpha$ needs to be lowered from 3 to 2 for negative $\beta$). The new stable phase will take over for all negative $\beta$, where $\beta=-\infty$ corresponds to separable states.
However the analytic continuation described here, although metastable, exists for all negative $\beta$ and we can study in more detail the behavior of the eigenvalue density (\ref{eq:phi(x)}) and of its free energy (\ref{eq:freeenergy}).
The solution is straightforward but lengthy and is given in the following. It is of interest in itself because, as we shall see, it entails a restoration of the $\mathbb{Z}_2$ symmetry that was broken at the phase transition at $\beta=2$ described in the previous section.

\begin{figure}
\hspace{-0.2cm}\includegraphics[width=0.72\columnwidth]{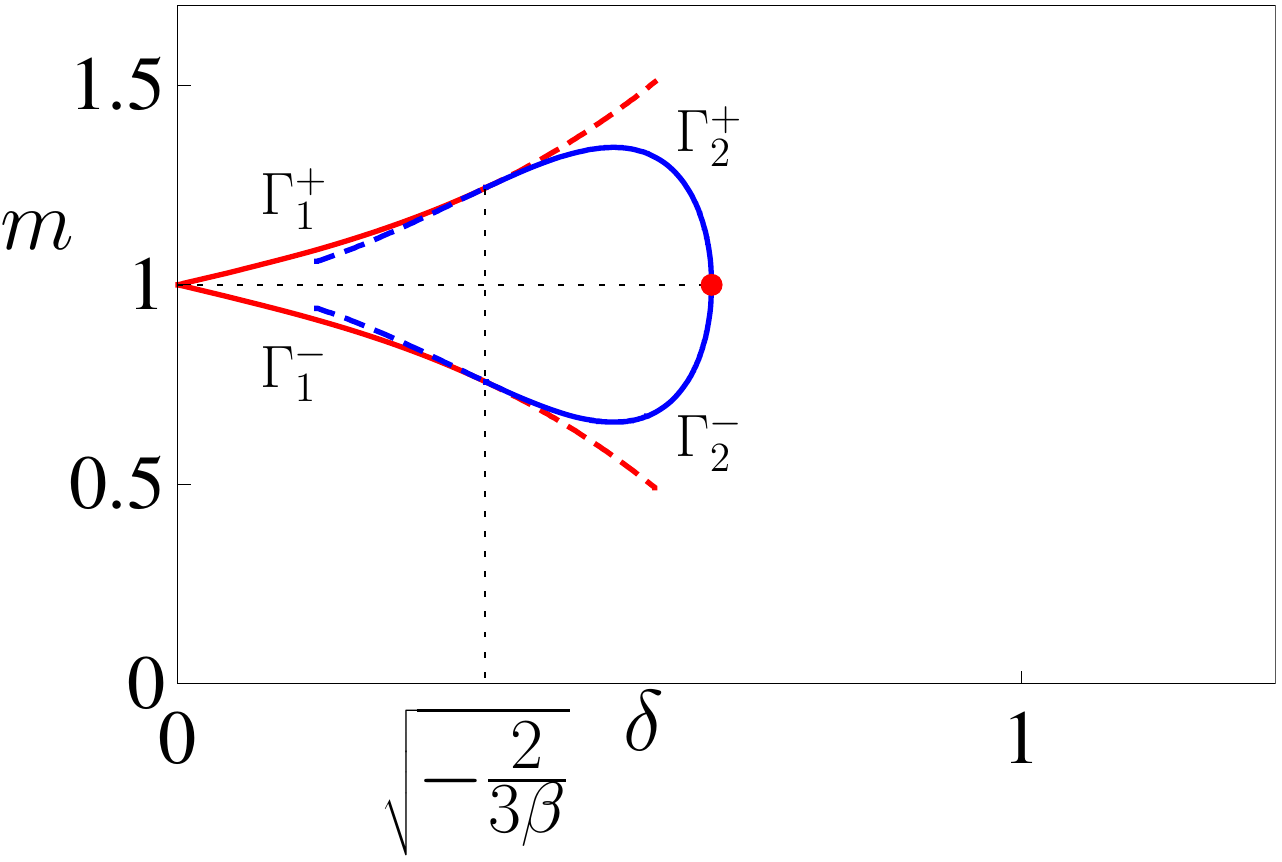}\\ \vspace{0.1cm}
\includegraphics[width=0.7\columnwidth]{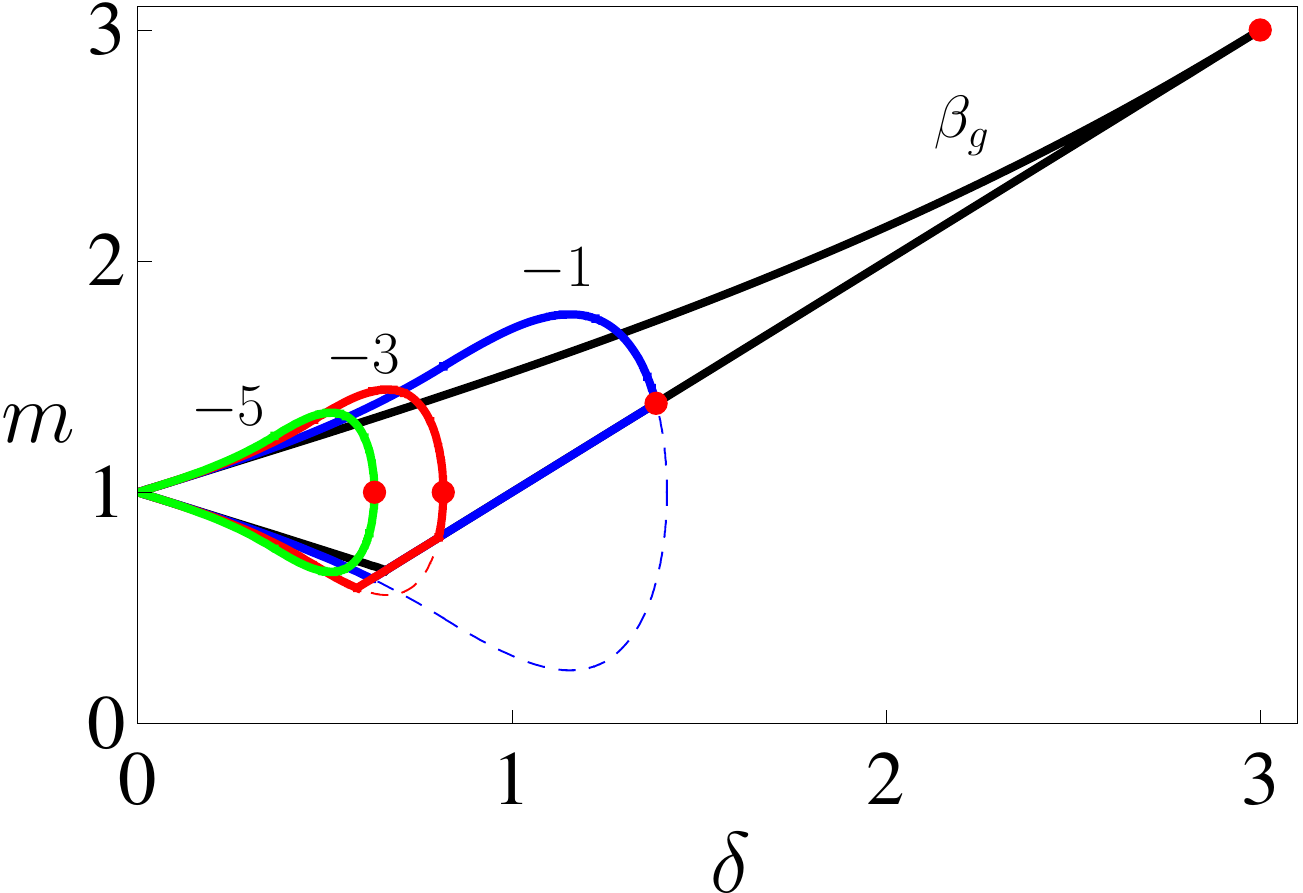}
\caption{Metastable branch. Domain of existence for the solution $(m,\delta)$ for negative temperatures.}
\label{fig:domneg}
\end{figure}

Recall that the density $\phi(x)$ must be nonnegative for all $x\in(-1,1)$. This condition  for
$\beta < 0$ gives $x\notin (x_+,x_-)$, with $x_\pm$ given by (\ref {eq:roots})-(\ref {eq:discriminant}).
The level curves $x_{\pm}=\pm 1$ are given by $m=\Gamma_1^{\pm} (\delta,\beta)$, with
$\Gamma_1^{\pm}$ in (\ref {Gamma1}), while the level curves  $\Delta=0$ are given by $m=\Gamma_2^{\pm} (\delta,\beta)$ with
\begin{equation}
\label{Gamma2}
\Gamma_2^{\pm} (\delta,\beta) = 1\pm\delta^2 \sqrt{-\beta\left(1+\frac{\beta\delta^2}{2}\right)},
\end{equation}
They are symmetric with respect to the line $m=1$ and intersect at $\delta=0$ and at $\delta=\sqrt{-2/\beta}$.
Moreover, they are tangent to $\Gamma_1^{\pm}$ at the points
\begin{equation}
(\delta,m)=\left(\sqrt{-\frac{2}{3\beta}},\; 1\pm\frac{2}{3}\sqrt{-\frac{2}{3\beta}}\right),
\end{equation}
as shown in Fig.~\ref{fig:domneg}.
Therefore, the condition $(-1,1)\cap[x_1,x_2]=\emptyset$ implies that the points $(\delta, m)$ should be restricted to a (possibly cut) eye-shaped domain given by
\begin{equation}\label{eq:negative_temp_eyesShapeDomain}
\max\left\{ \delta, \; h_{-}(\beta,\delta)\right\} \leq m \leq  h_{+}(\beta,\delta),
\end{equation}
where
\begin{equation}
h_{\pm}(\delta,\beta) =
\begin{cases}
\Gamma_1^{\pm} (\delta,\beta), &  0\leq \delta\leq \sqrt{-\frac{2}{3\beta}}, \\
\\
\Gamma_2^{\pm} (\delta,\beta), &  \delta > \sqrt{-\frac{2}{3\beta}} .
\end{cases}
\end{equation}
The right corner of the eye is given by
\begin{equation}
\label{eq:rightcornerneg}
(\delta, m)=\left(\sqrt{-\frac{2}{\beta}},  1\right)
\end{equation}
and belongs to the boundary as long as $\beta\leq -2$. For $\beta\geq-4$ the eye is cut by the line $m=\delta$. See Fig.\ \ref{fig:domneg}.

\begin{figure}
\includegraphics[width=0.6\columnwidth]{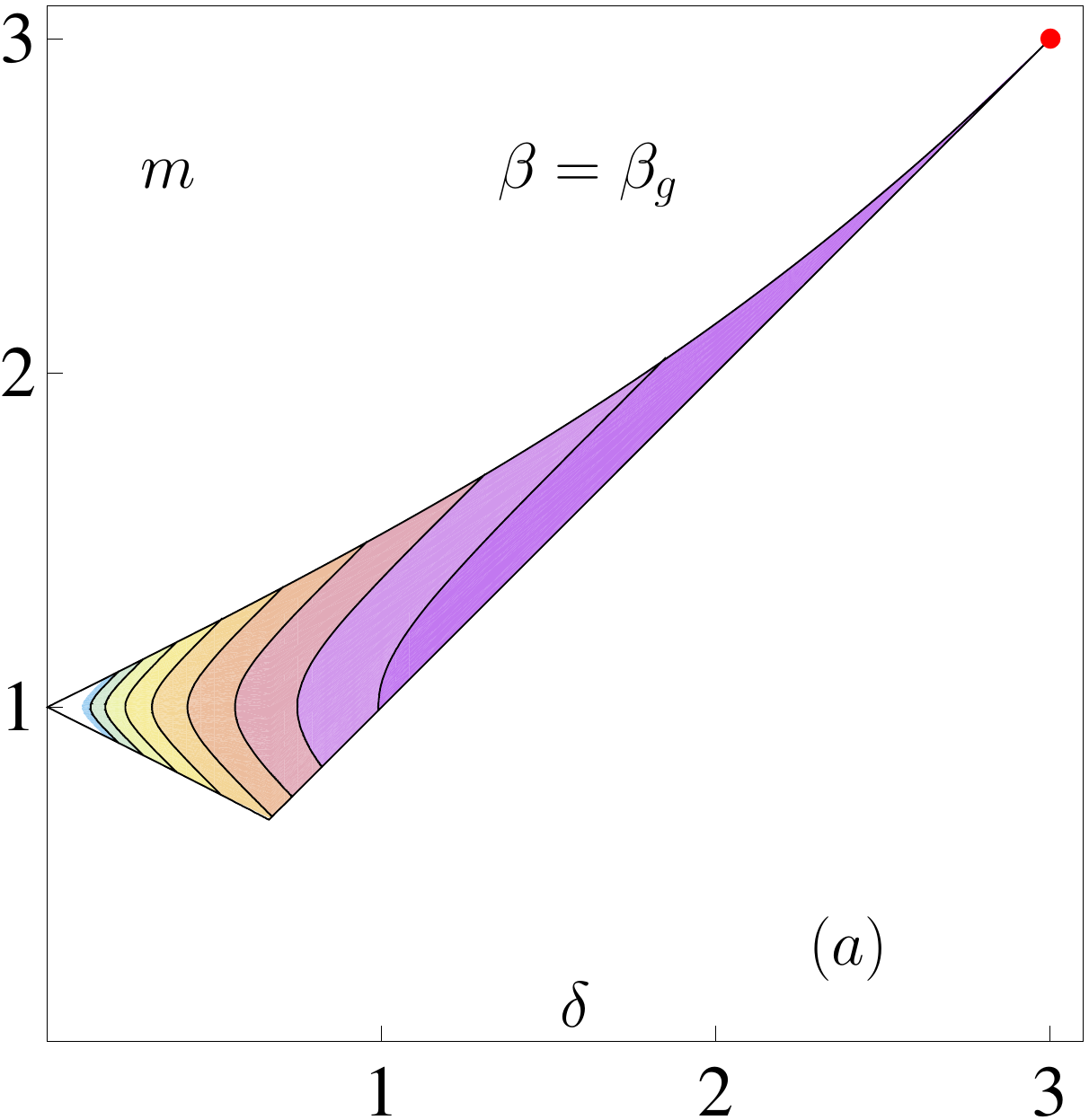} \quad
\includegraphics[width=0.6\columnwidth]{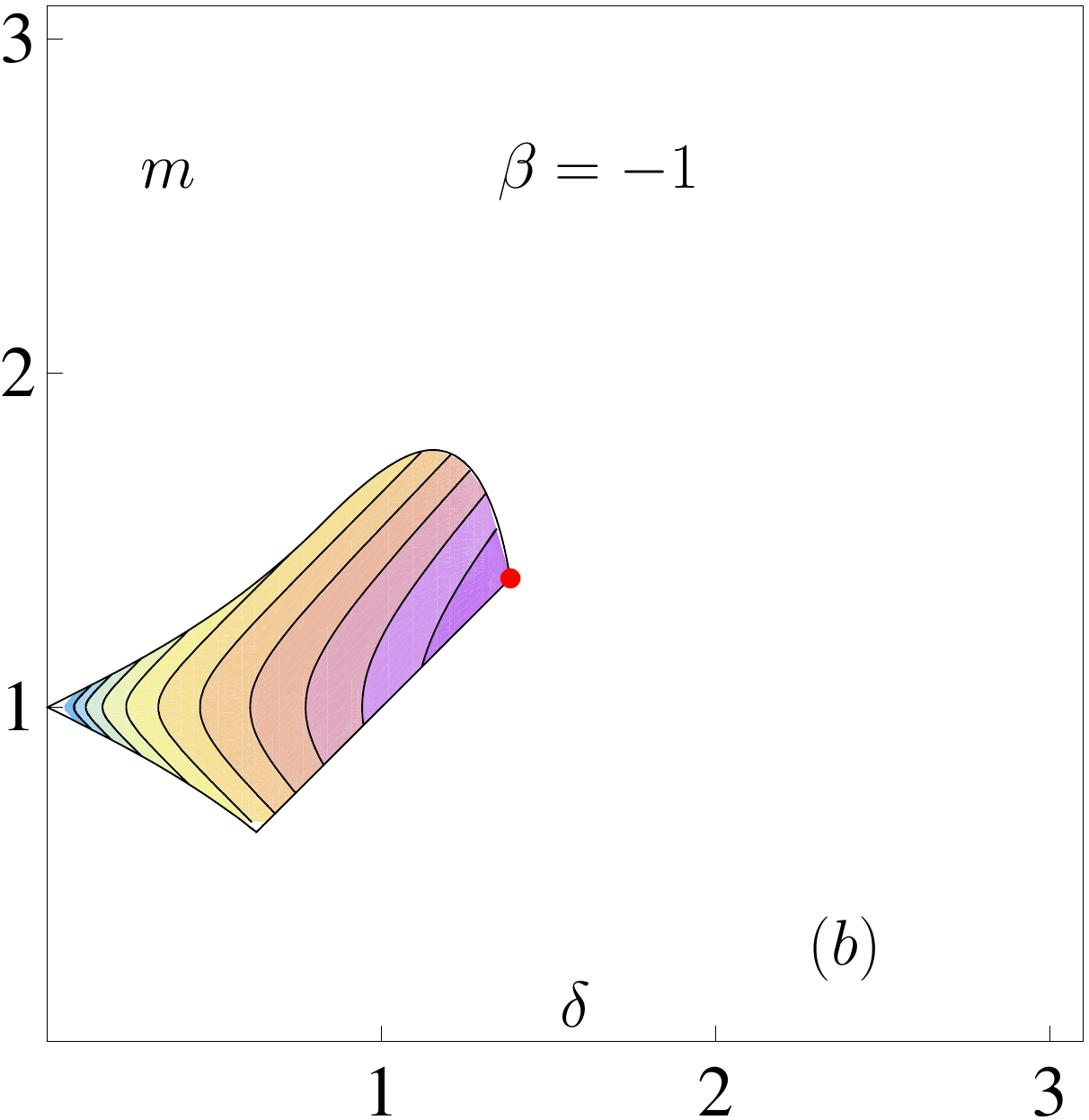}\quad
\includegraphics[width=0.6\columnwidth]{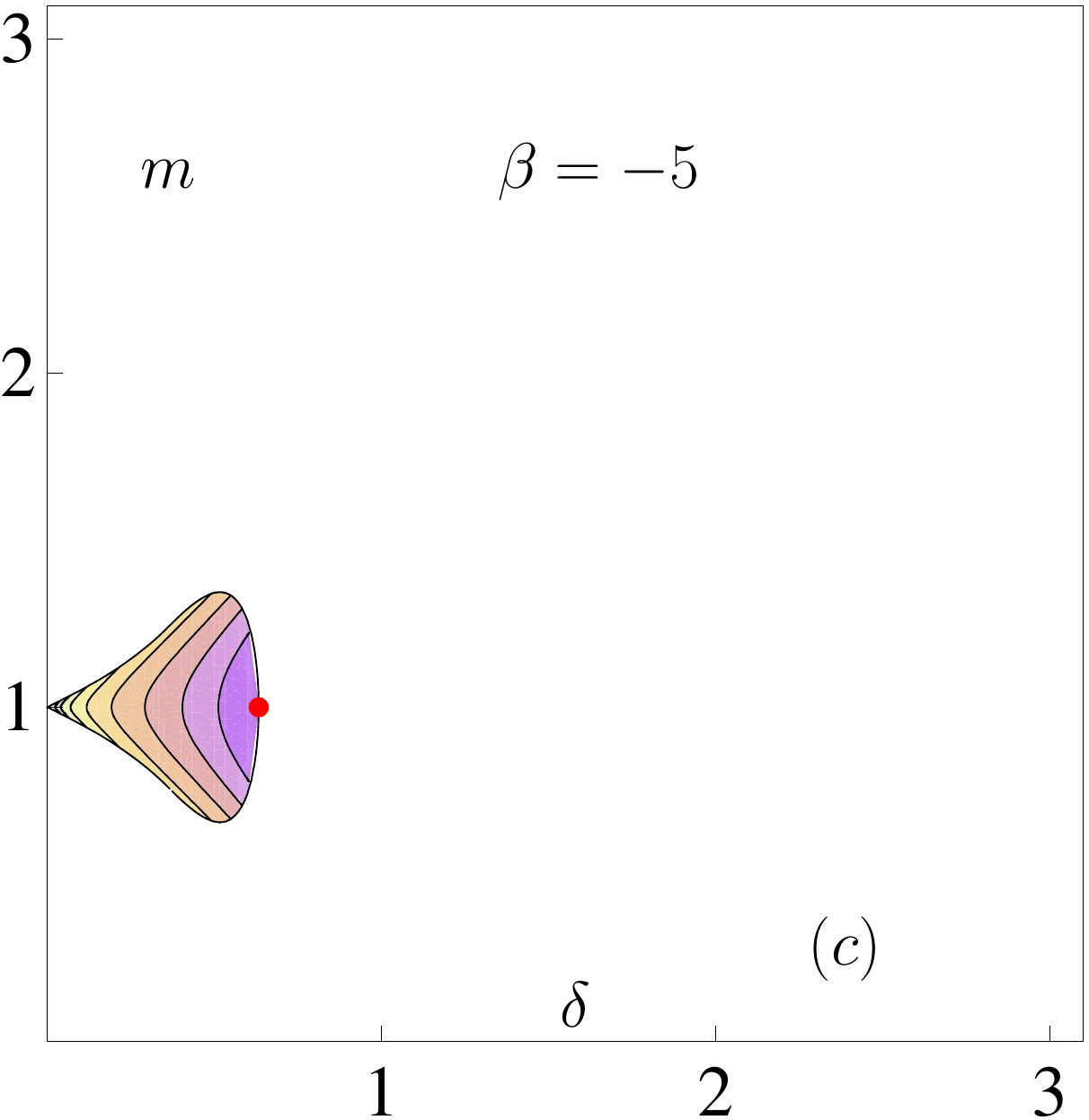}
\caption{Metastable branch. Contour plots of the free energy in regions of the
parameter space such that $\phi(x)\geq0$, for (a) $\beta=\beta_g$,
(b) $\beta=-1$ and (c) $\beta=-5$. Darker regions have lower free
energy.} \label{fig:contoursneg}
\end{figure}

\begin{figure}
\includegraphics[width=0.7\columnwidth]{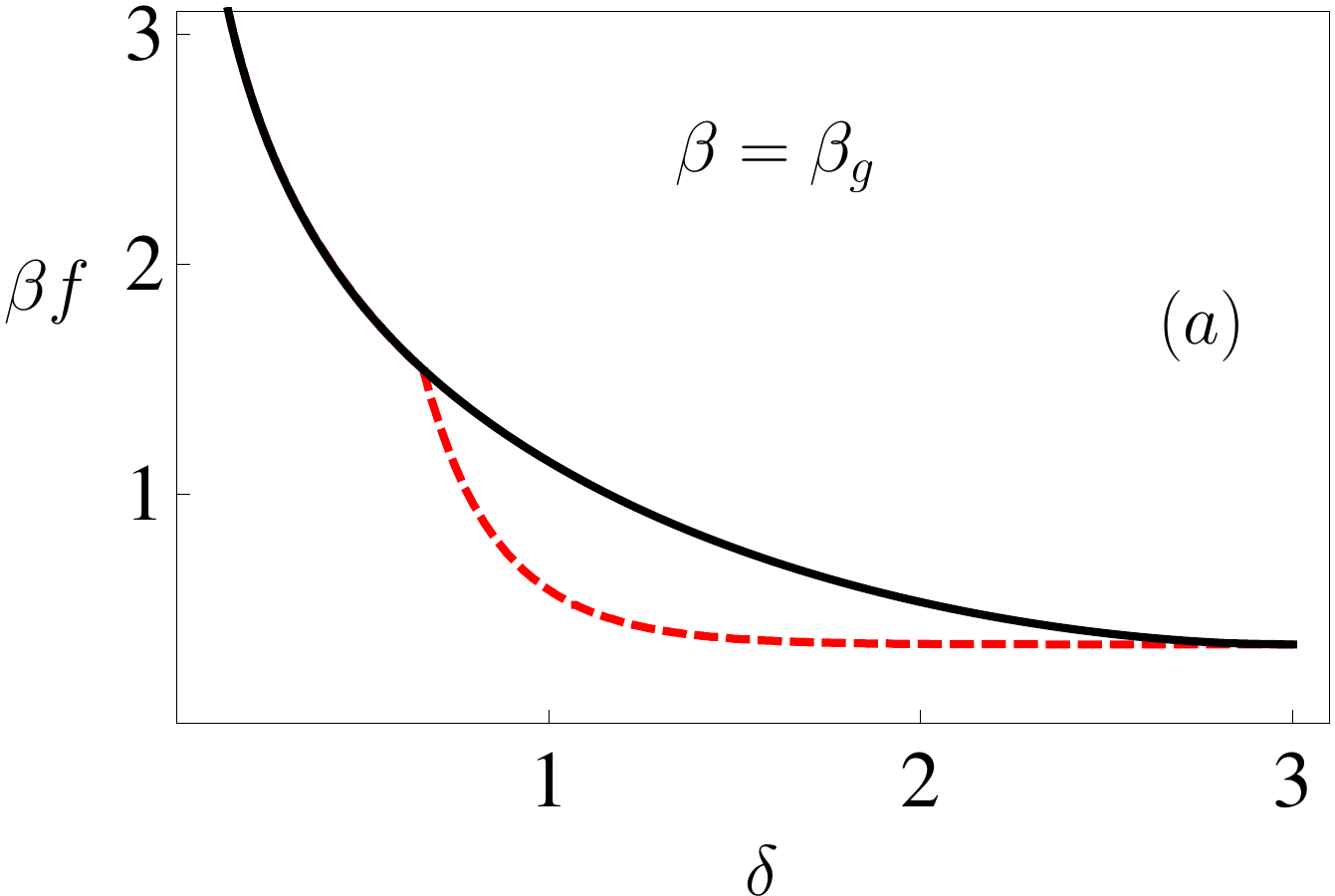}
\\  \vspace{0.1cm}
\hspace{-0.4cm} \includegraphics[width=0.74\columnwidth]{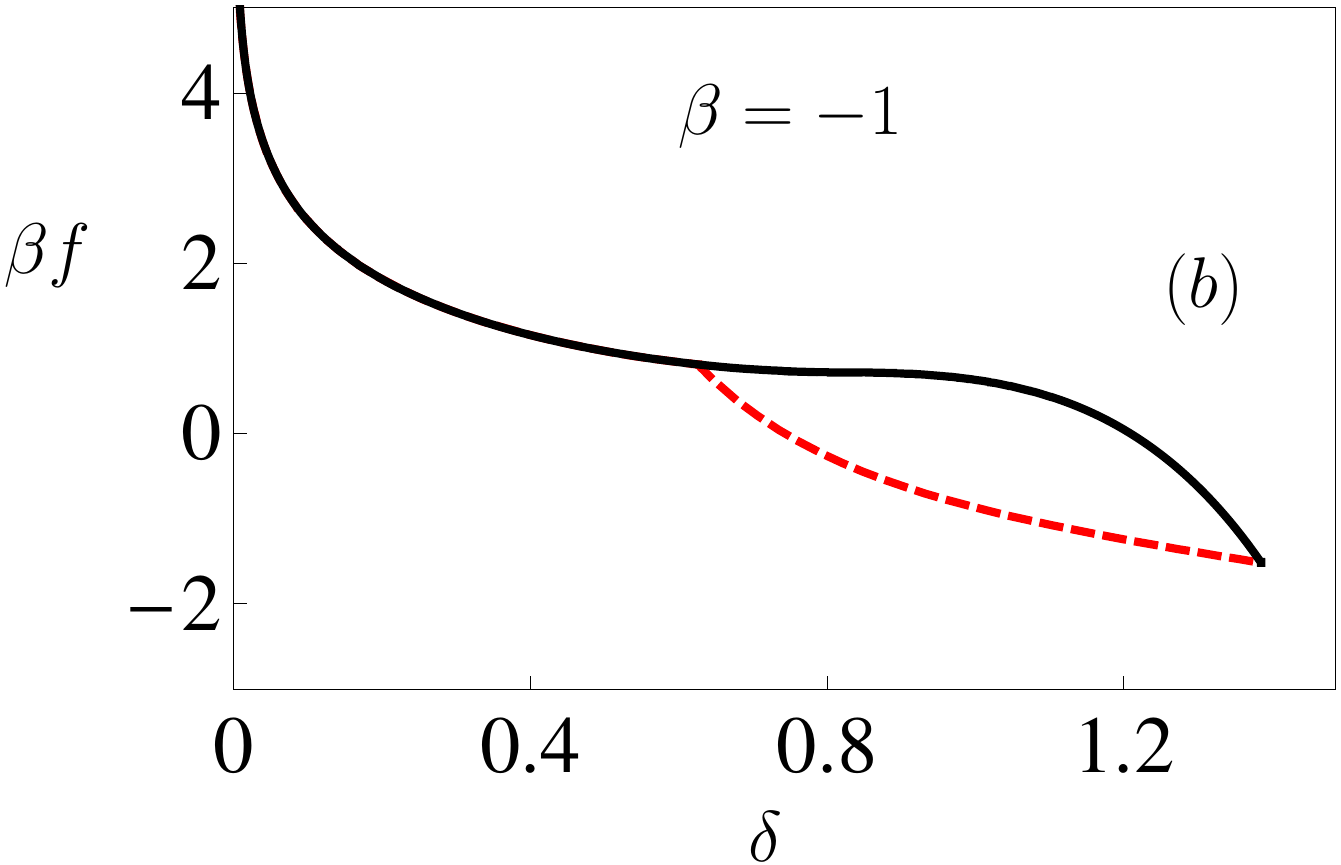}\\ \vspace{0.1cm}
\hspace{-0.4cm}  \includegraphics[width=0.74\columnwidth]{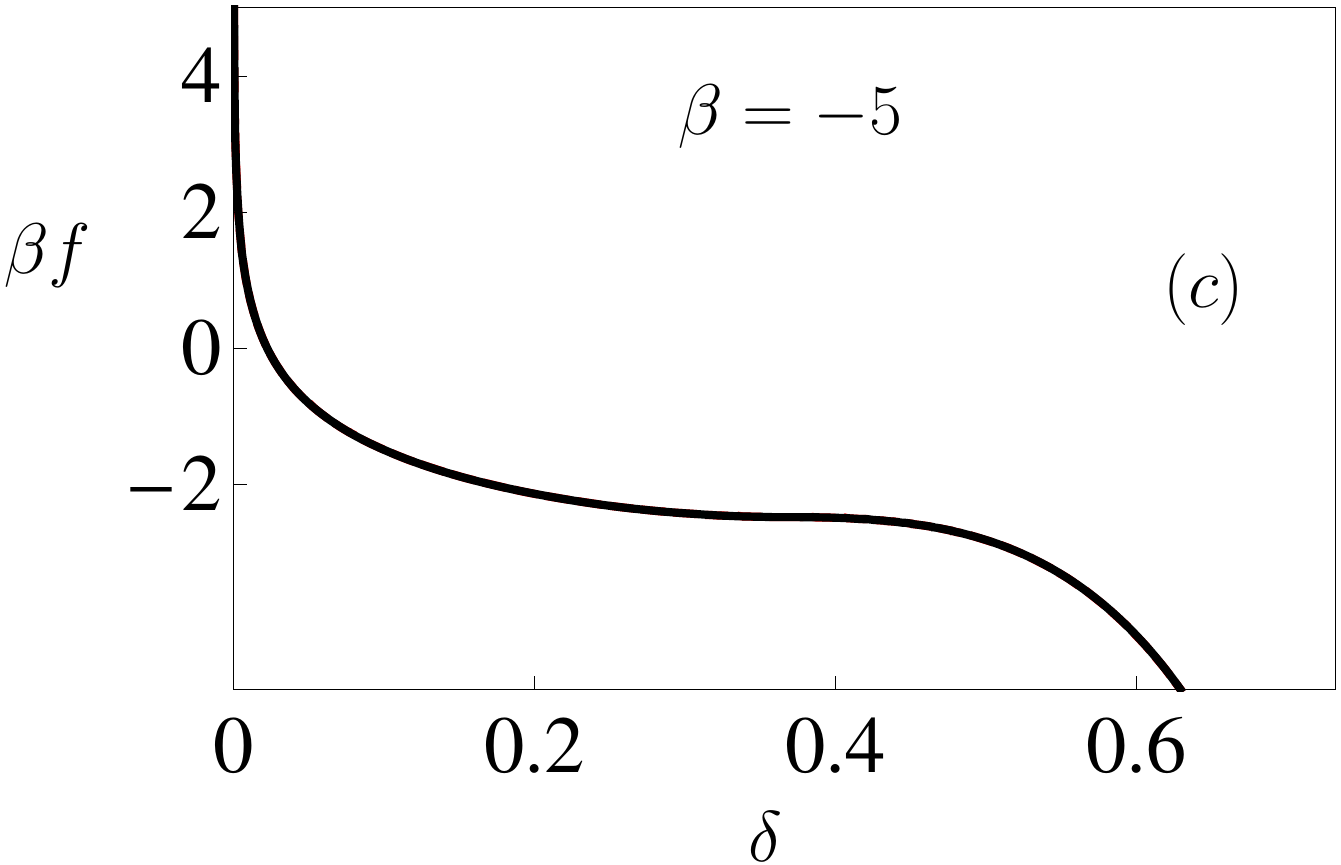}
\caption{Metastable branch. Free energy on the boundary of the region of the domain
where $\phi(x)\geq0$, for different temperatures (indicated).
Dashed line: free energy $\beta f$ on the lower boundary of the eye-shaped domain;
full line: free energy on the upper boundary. The sought minima of the free energy can be inferred from the graph and coincide with the dots in Figs.\ \ref{fig:domneg} and \ref{fig:contoursneg}.}
\label{fig:3freeneg}
\end{figure}

The contour plots of the free energy  (\ref{eq:freeenergy}) are shown in Fig.\ \ref{fig:contoursneg}.
The free energy density
$f$ has no stationary points for $\beta<0$.
The behavior of the free energy at the boundaries of the allowed domain is shown in Fig.\ \ref{fig:3freeneg} for different temperatures.
For $\beta\leq-2$ the right corner of the eye  is the absolute minimum,
whereas for $-2<\beta<0$ the absolute minimum is at the right upper corner of the allowed region, namely at
\begin{equation}
m=\delta, \quad\mbox{with} \quad \delta= h_{+}(\beta,\delta),
\label{eq:beta-2<0}
\end{equation}
that is
\begin{equation}
\beta \frac{\delta^3}{4} +\frac{\delta}{2} -1 =0, \quad \mbox{for} -\frac{2}{27}\leq \beta\leq0, 
\label{eq:prolong}
\end{equation}
and
\begin{equation}
 \delta -1 = \delta^2\sqrt{-\beta\left(1+\beta\frac{\delta^2}{2}\right)}, \quad \mbox{for} -2 \leq \beta \leq -\frac{2}{27}.
 \label{eq:prolong1}
\end{equation}
Note that (\ref{eq:prolong}) coincides with (\ref{eq:beta<2}) and thus is the prolongation of the curve (\ref{eq:betasss})
which runs monotonically from $\beta=0$ when $\delta=2$ to its minimum  $\beta_g=-2/27$ at $\delta=3$.

\begin{figure}
\includegraphics[width=0.7\columnwidth]{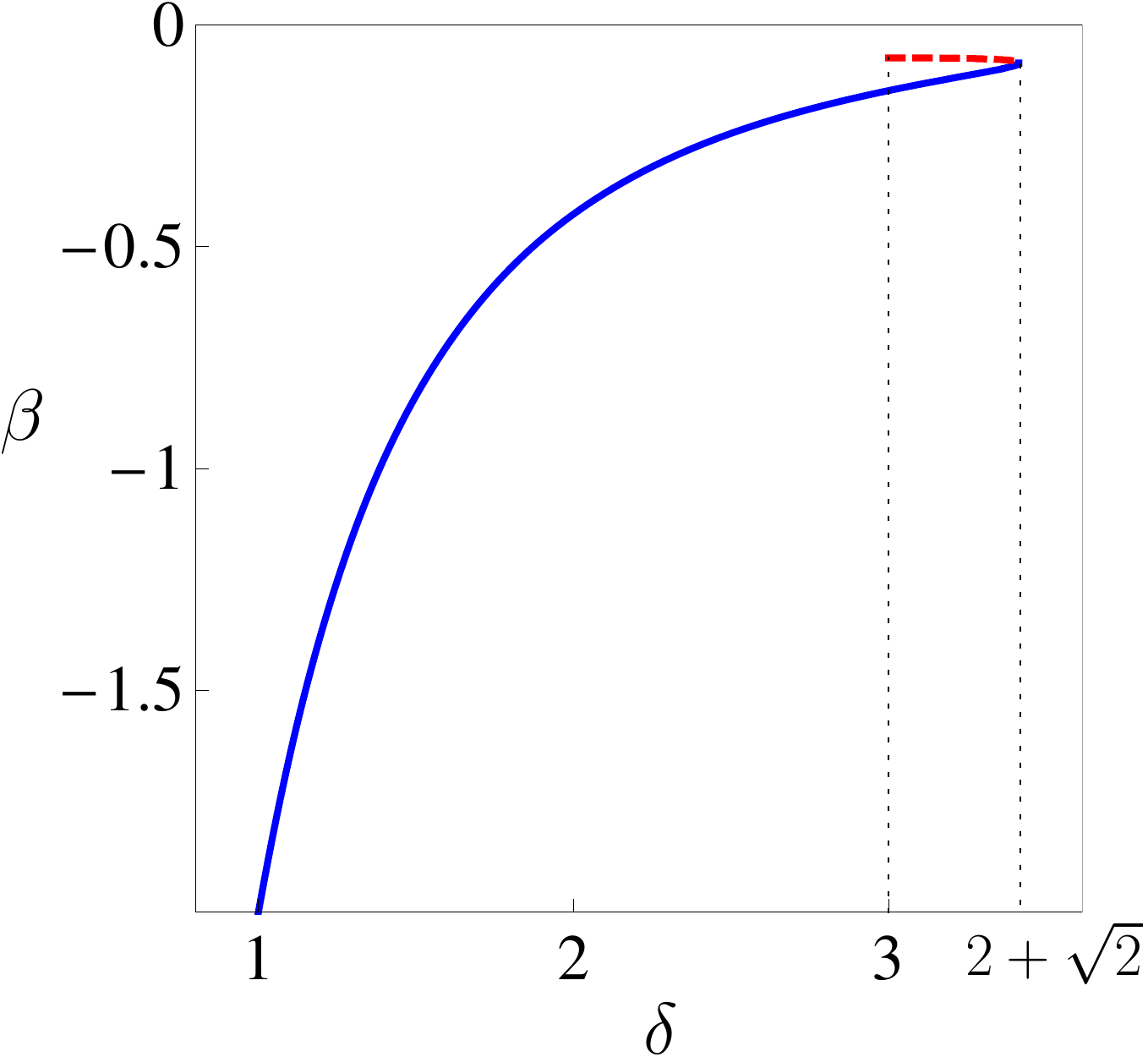}
\caption{Metastable branch. $\beta$ vs $\delta$. See Eq.\ (\ref{eq:2rami}).}
\label{fig: betadeltaneg}
\end{figure}

On the other hand, (\ref{eq:prolong1}) is given by the curves
\begin{eqnarray}
\beta &=& -\frac{1}{\delta^2}\pm\frac{1}{\delta^3}\sqrt{-\delta^2+ 4 \delta-2} \nonumber \\
&=& -\frac{1}{\delta^2}\pm\frac{1}{\delta^3}\sqrt{(2+\sqrt{2}-\delta)(\delta-2+\sqrt{2})},
\label{eq:2rami}
\end{eqnarray}
which run from $\beta= \beta_g$ when $\delta=3$ (with derivative zero) up to $\beta=-3/2 +\sqrt{2}$ when $\delta=2+\sqrt{2}$ (with derivative $-\infty$) and then  from $\beta=-3/2 +\sqrt{2}$ when $\delta=2+\sqrt{2}$ (with derivative $+\infty$) up to $\beta=-2$ when $\delta=1$. See Fig.~\ref{fig: betadeltaneg}.

Let us look at the eigenvalue density (\ref{eq:phi(x)}). When   $\beta_g\leq \beta \leq 0$ the solution is
obtained by plugging (\ref{eq:beta<2}) into (\ref{eq:phi(x)})
\begin{equation}
\phi(x)=\frac{2}{\pi\delta}\sqrt{\frac{1-x}{1+x}}\big(1+(2-\delta)x\big) ,
\label{eq:Whishprolong}
\end{equation}
with $2\leq\delta\leq3$, and is Wishart.
At $\beta_g$ one gets  $\delta=3$ and
\begin{equation}
\phi(x)=\frac{2}{3 \pi}\sqrt{\frac{(1-x)^3}{1+x}} ,
\end{equation}
whose derivative at the right edge $x=1$ vanishes. On the other hand,
when   $-3/2+\sqrt{2}\leq\beta\leq \beta_g$ by (\ref{eq:2rami}) one gets
\begin{widetext}
\begin{equation}
\phi(x)=\frac{1}{\pi\delta\sqrt{1-x^2}}\left[\frac{1}{2}\left(\delta + \sqrt{-\delta^2+ 4 \delta-2} \right)  +2(1-\delta)x  + \left(\delta - \sqrt{-\delta^2+ 4 \delta-2} \right) x^2 \right] ,
\end{equation}
with $3\leq\delta\leq2+\sqrt{2}$,
while for   $-2\leq \beta\leq -3/2+\sqrt{2}$
\begin{equation}
\phi(x)=\frac{1}{\pi\delta\sqrt{1-x^2}}\left[\frac{1}{2}\left(\delta - \sqrt{-\delta^2+ 4 \delta-2} \right)  +2(1-\delta)x  + \left(\delta + \sqrt{-\delta^2+ 4 \delta-2} \right) x^2 \right] ,
\end{equation}
with $1\leq\delta\leq2+\sqrt{2}$.
Note that this eigenvalue density diverge \emph{both} at the left edge $x=-1$ \emph{and} at the right edge $x=+1$.\begin{figure}
\includegraphics[width=0.24\columnwidth]{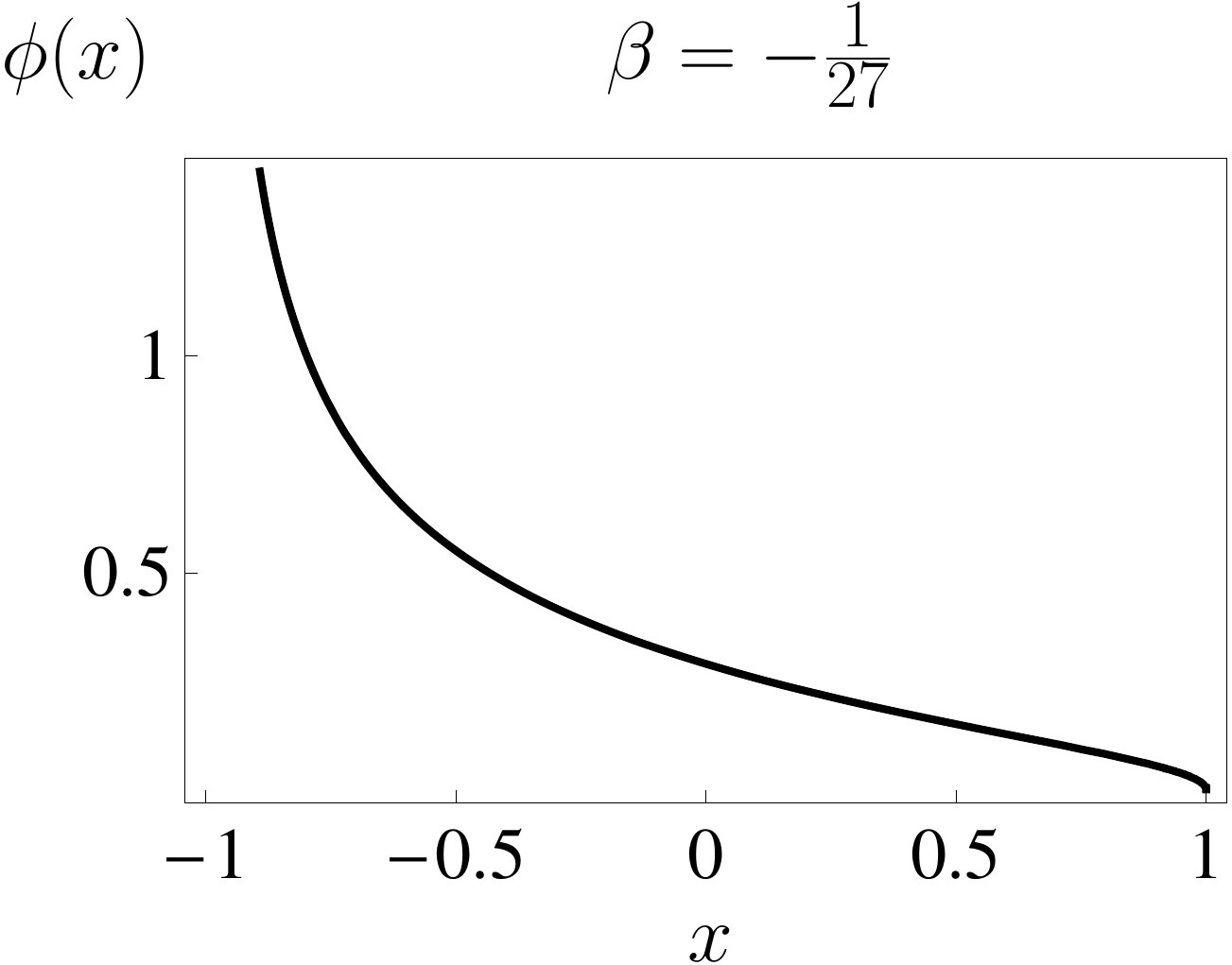}
\includegraphics[width=0.24\columnwidth]{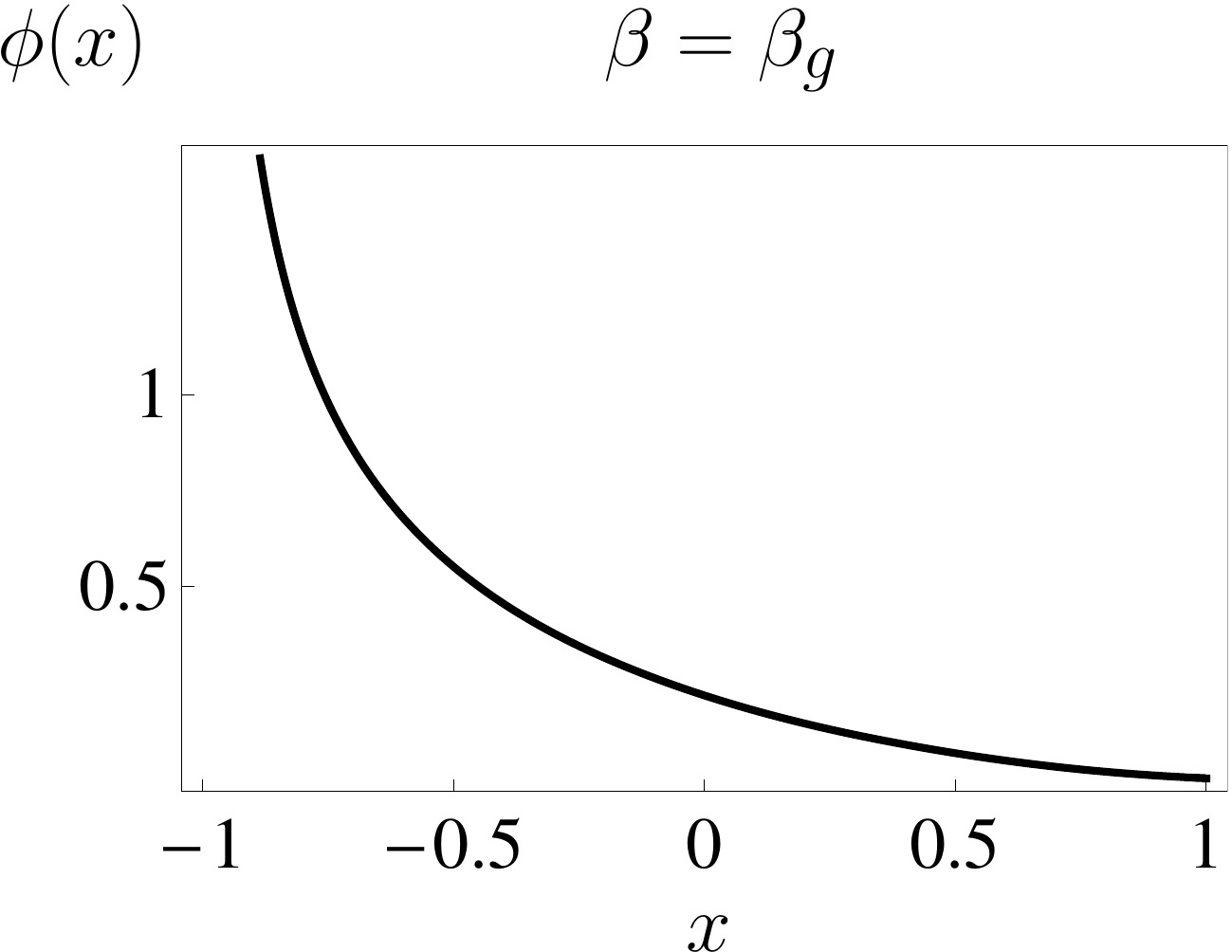}
\includegraphics[width=0.24\columnwidth]{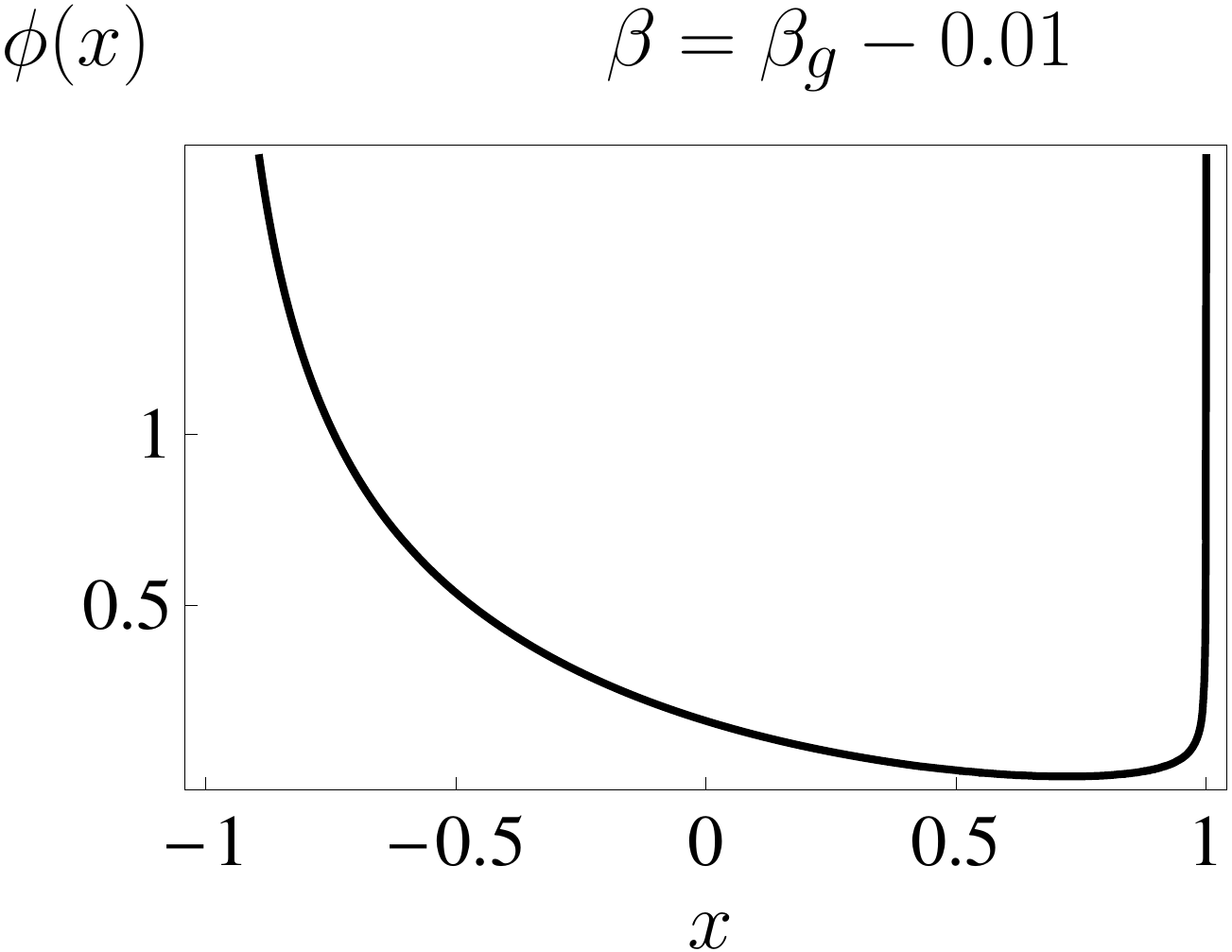}
\includegraphics[width=0.24\columnwidth]{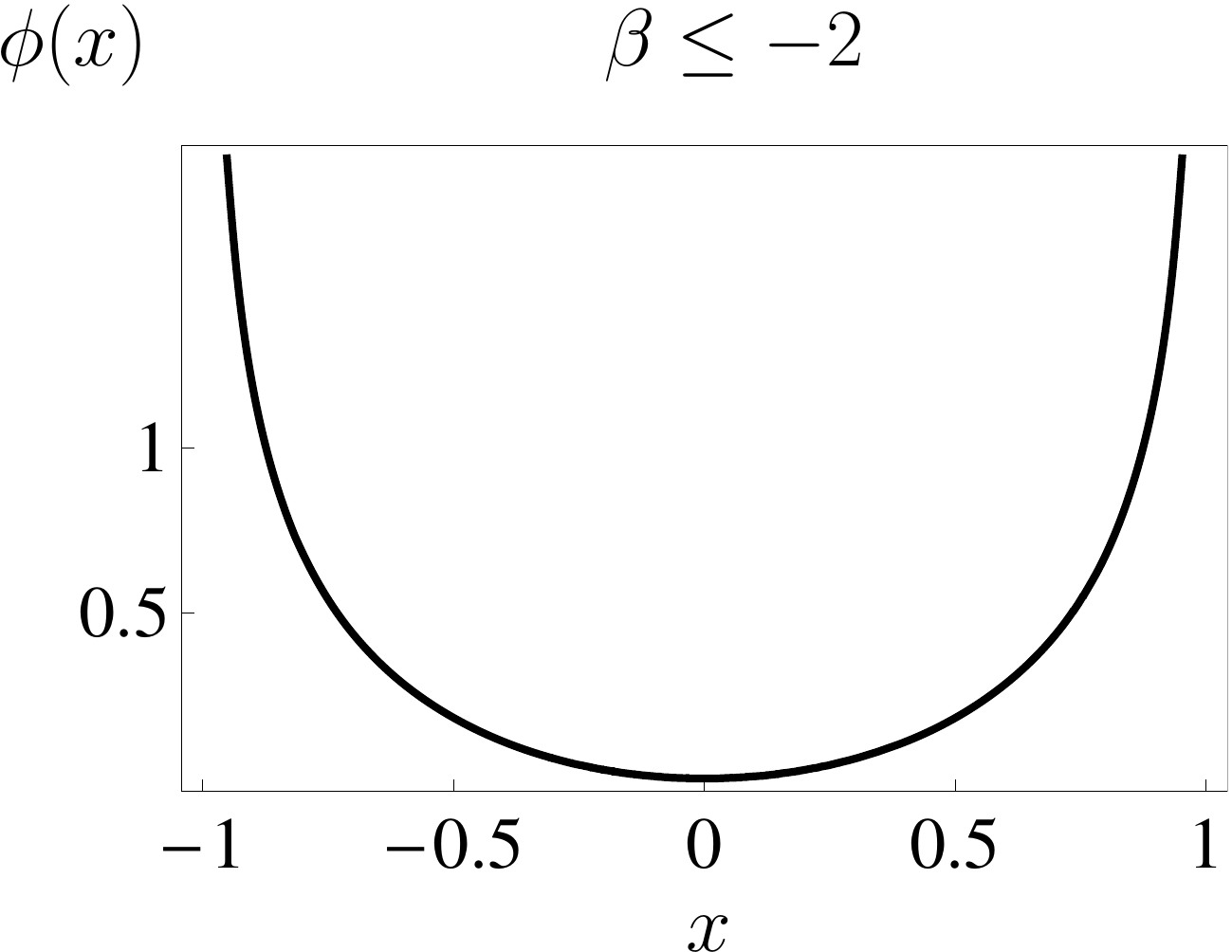}
\caption{Metastable branch. Eigenvalue density for $\beta=-1/27$, $\beta_g=-2/27$,
$\beta\lesssim \beta_g$ and $\beta< \beta_-=-2$. From left to right, notice how the distribution (initially Wishart, whose derivative at the rigth edge of the domain diverges) gets first a vanishing derivative at the right edge, then develops a singularity there and eventually restores the $\mathbb{Z}_2$ symmetry that was broken at the phase transition at $\beta=2$ described in the previous section. } 
\label{fig:3phineg}
\end{figure}
\end{widetext}

At $\beta=-2$ one obtains $\delta=1$ and
\begin{equation}
\phi(x)=\frac{2 x^2}{\pi\sqrt{1-x^2}}.
\end{equation}
One gets the above density for all $\beta\leq -2$, where the $\mathbb{Z}_2$ symmetry is restored.
The interesting behavior of the eigenvalue density as $\beta$ is varied is displayed in Fig.\ \ref{fig:3phineg}. 

\begin{figure}
\includegraphics[width=0.7\columnwidth] {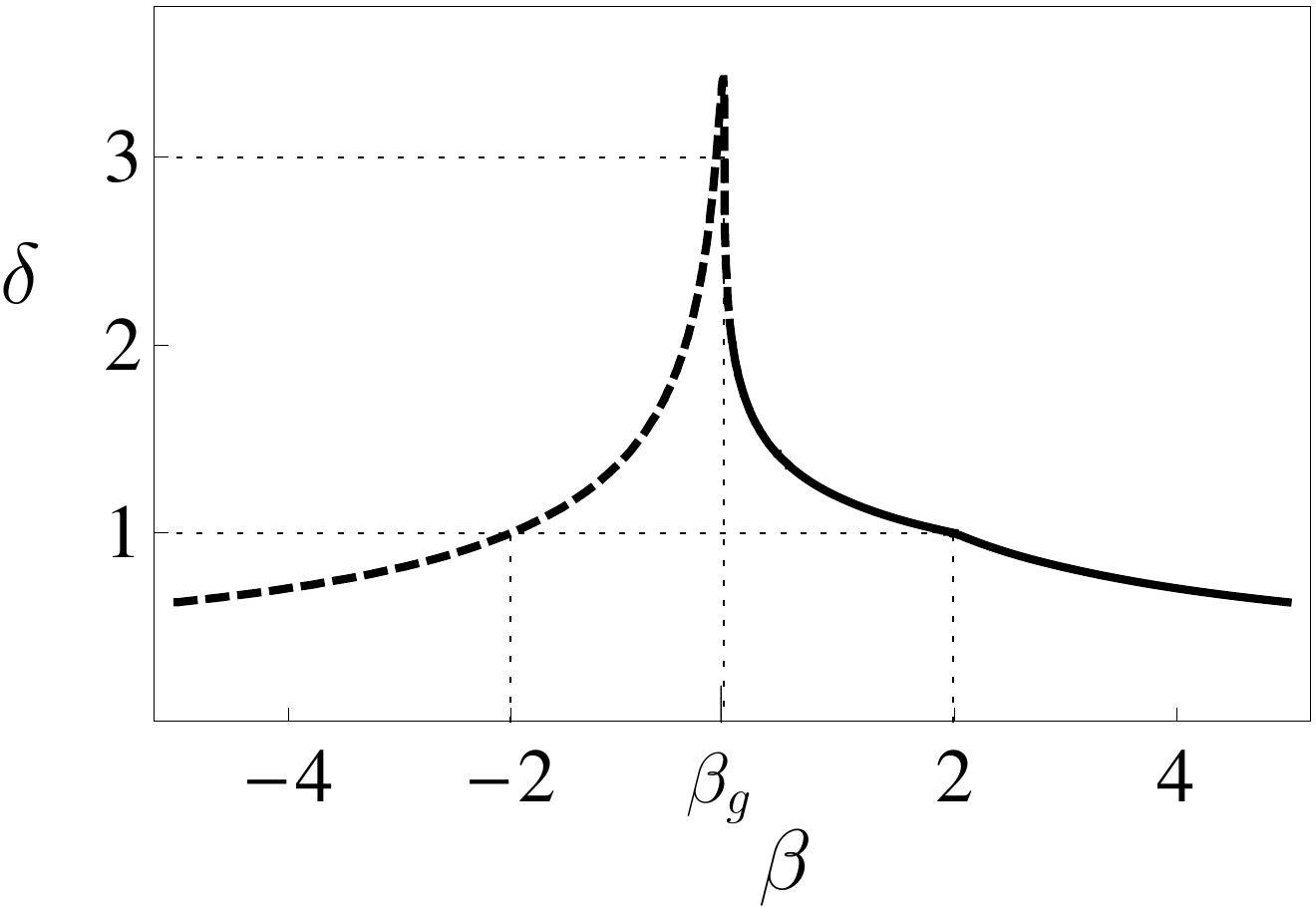}\\ \vspace{0.2cm}
\hspace{-0.2cm}\includegraphics[width=0.72\columnwidth] {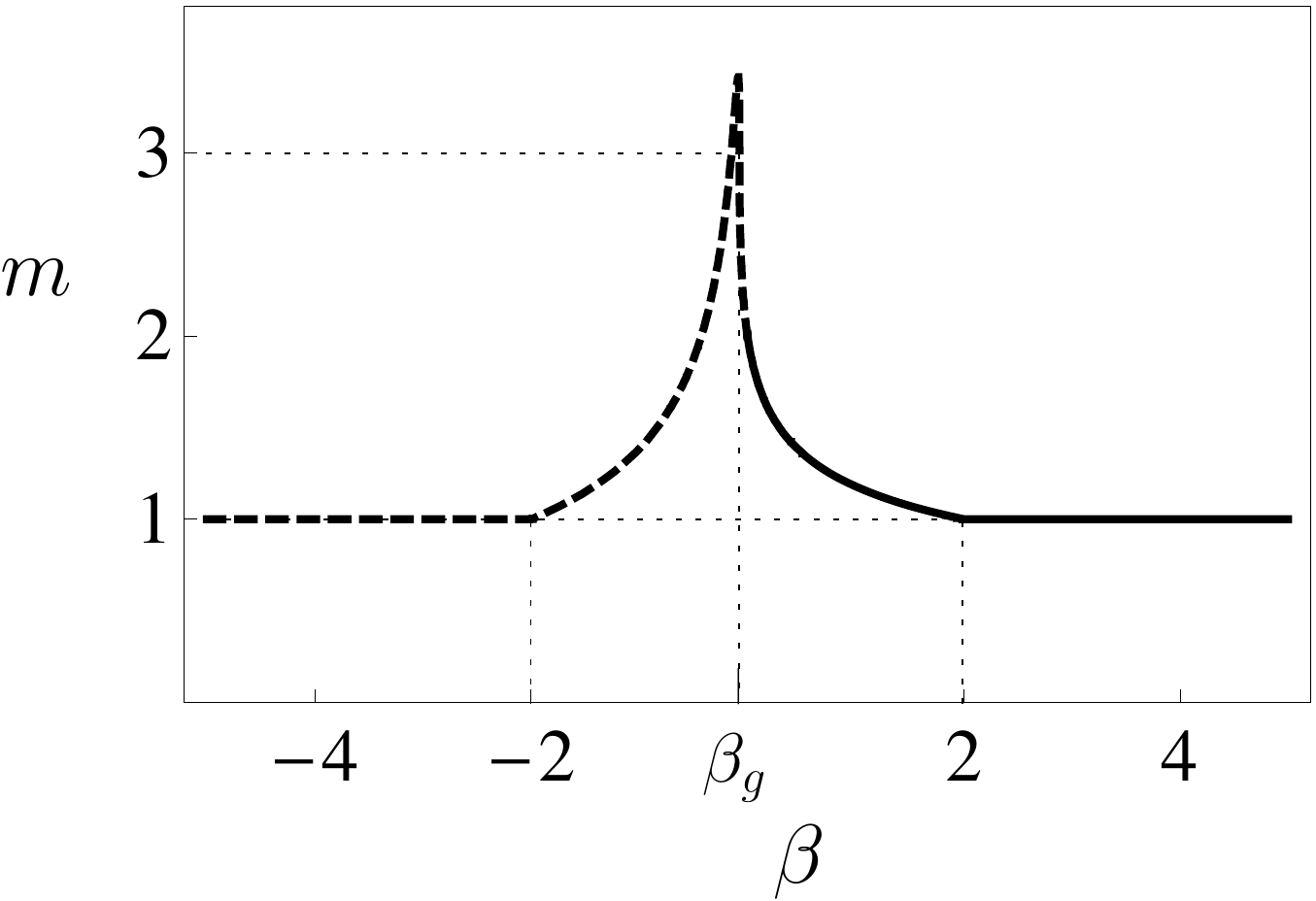}
\caption{Average and width of the solution domain $m$ and $\delta$  [Eq.\ (\ref{eq:mdeltadef})] as a function of $\beta$. Solid line: stable branch. Dotted line: metastable branch.}
\label{fig:deltavsbeta}
\end{figure}

\begin{figure}
\includegraphics[width=0.8\columnwidth]{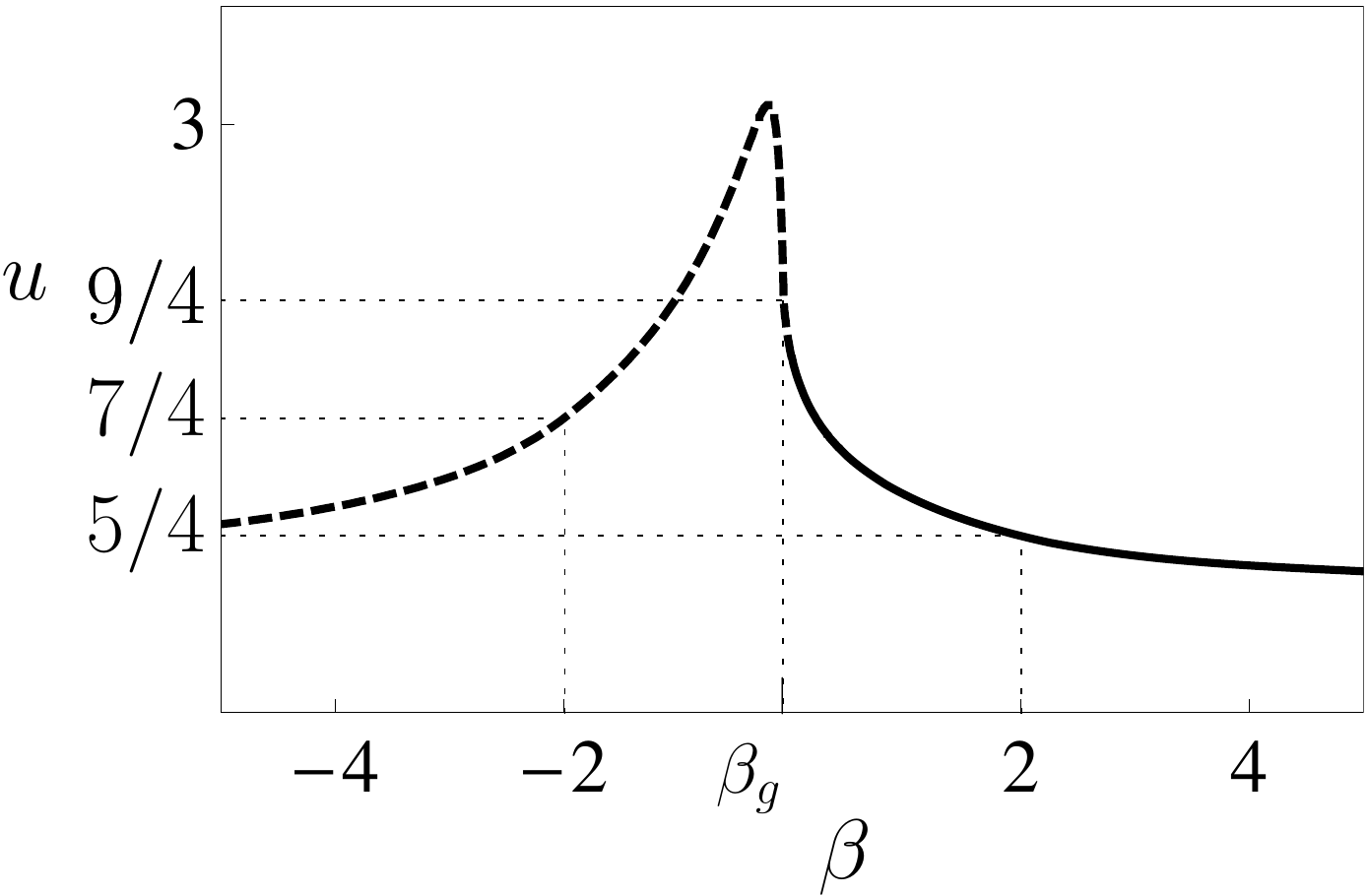}\\
\vspace{0.1cm}
\hspace{-0.3cm}\includegraphics[width=0.83\columnwidth]{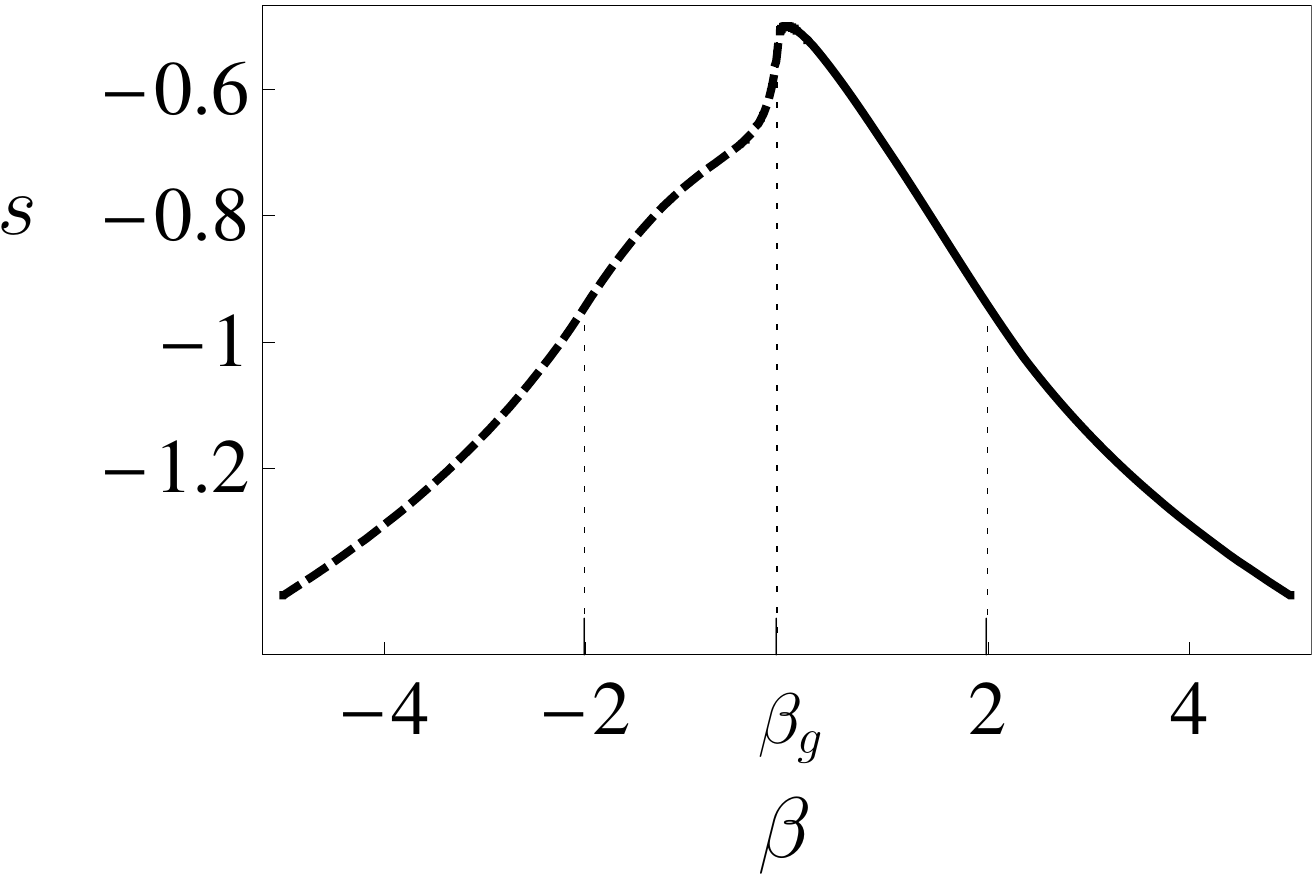}
\caption{Internal energy density $u$  
and entropy density $s$ 
versus $\beta$. Solid line: stable branch. Dotted line: metastable branch.}
\label{fig:uvsbeta}
\end{figure}

The values of $(m,\delta)$ [that define the eigenvalue domain, see Eq.\ (\ref{eq:mdeltadef})] and the 
thermodynamic functions $u$ (internal energy density) and $s$ (entropy density) are shown in Figs.\ \ref{fig:deltavsbeta}, 
\ref{fig:uvsbeta}, 
respectively.
Their explicit expressions are given for positive temperatures in Sec.\ \ref{sec:positivetemp}, while for negative temperatures are given in the following.

In the gravity branch, for $\beta_g \leq\beta\leq 0$ ($2\leq\delta\leq 3$) we get
\begin{eqnarray}
m&=&\delta,  \quad \beta=\frac{4}{\delta^3}-\frac{2}{\delta^2} , \nonumber\\
u &=& \frac{3}{2}\delta -\frac{\delta^2}{4}, \nonumber\\
s &=&  -\frac{9}{4} + \frac{5}{\delta} - \frac{3}{\delta^2}
+\ln\frac{\delta}{2} , \nonumber\\
\beta f &=& \frac{11}{4} -\frac{9}{\delta} + \frac{9}{\delta^2} - \ln \frac{\delta}{2} .
\label{eq:gravitysol}
\end{eqnarray}

Beyond a second order phase transition at the critical temperature $\beta_g$ we get that 
for $-3/2+\sqrt{2}\leq\beta\leq \beta_g$ ($3\leq\delta\leq2+\sqrt{2}$) 
both $u$ and $s$ increase together with the eigenvalue density halfwidth $\delta$,
\begin{eqnarray}
m&=&  \delta,  \quad \beta= -\frac{1}{\delta^2}+\frac{1}{\delta^3}\sqrt{-\delta^2+ 4 \delta-2}, \nonumber\\
u &=& 2\delta -\frac{3}{8} \delta^2 -\frac{\delta}{8} \sqrt{-\delta^2+ 4 \delta-2}, \nonumber\\
s &=&-2 + \frac{15}{4\delta} -\frac{15}{8\delta^2} +\frac{1}{8\delta} \sqrt{-\delta^2+ 4 \delta-2} + \ln \frac{\delta}{2},
\nonumber\\
\beta f &=&\frac{5}{2} -\frac{25}{4\delta} +\frac{17}{8\delta^2} -\left(\frac{3}{8\delta} -\frac{2}{\delta^2} \right) \sqrt{-\delta^2+ 4 \delta-2} \nonumber\\
&&- \ln \frac{\delta}{2},
\end{eqnarray}
and then decrease for $-2\leq \beta\leq -3/2+\sqrt{2}$ ($1\leq\delta\leq2+\sqrt{2}$),
\begin{eqnarray}
m&=&  \delta,  \quad \beta= -\frac{1}{\delta^2}-\frac{1}{\delta^3}\sqrt{-\delta^2+ 4 \delta-2}, \nonumber\\
u &=& 2\delta -\frac{3}{8} \delta^2 +\frac{\delta}{8} \sqrt{-\delta^2+ 4 \delta-2}, \nonumber\\
s &=& -2 + \frac{15}{4\delta} -\frac{15}{8\delta^2} -\frac{1}{8\delta} \sqrt{-\delta^2+ 4 \delta-2} + \ln \frac{\delta}{2}, \nonumber \\
\beta f &=&  \frac{5}{2} -\frac{25}{4\delta} +\frac{17}{8\delta^2} +\left(\frac{3}{8\delta} -\frac{2}{\delta^2} \right) \sqrt{-\delta^2+ 4 \delta-2}\qquad \nonumber\\
&&- \ln \frac{\delta}{2}.
\end{eqnarray}

Finally, beyond another  second order phase transition for $\beta\leq -2$ ($0\leq \delta\leq 1$), when the $\mathbb{Z}_2$ symmetry is restored, we get
\begin{eqnarray}
m&=&1,  \quad \beta = -\frac{2}{\delta^2}, \nonumber\\
u &=& 1+\frac{3}{4} \delta^2 =1-\frac{3}{2\beta},\nonumber\\
s &=& \ln \frac{\delta}{2} - \frac{1}{4} =  -\frac{1}{2} \ln \left(- 2 \beta \right) - \frac{1}{4} , \nonumber\\
\beta f &=& - \frac{5}{4} -\frac{2}{\delta^2} - \ln \frac{\delta}{2}= -\frac{5}{4}  +\beta + \frac{1}{2} \ln \left(- 2 \beta \right)  .
\end{eqnarray}

\begin{figure}
\includegraphics[width=0.8\columnwidth]{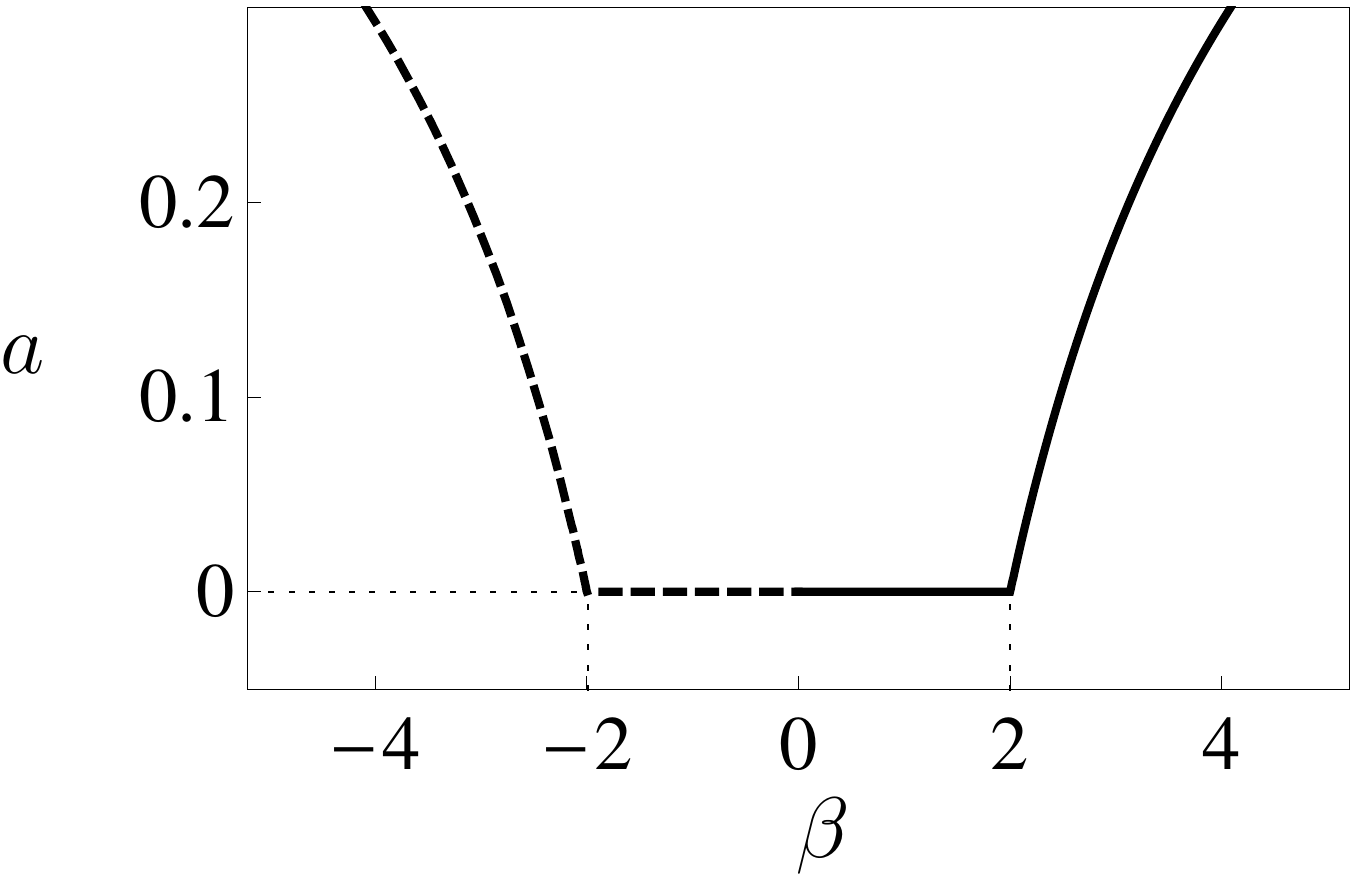}
\caption{Minimum eigenvalue $a=m-\delta$ versus $\beta$. Solid line: stable branch. Dotted line: metastable branch.}
\label{fig:a}
\end{figure}

Finally, we record the interesting behavior of the minimum eigenvalue $a=m-\delta$: see Fig.\ \ref{fig:a}.
For $-2<\beta<2$, $a$ coincides with the origin (left border of the solution domain).
This variable can be taken as an order parameter for both the second order phase transitions at $\beta=-2$ and at 
$\beta_+=2$. The $\mathbb{Z}_2$ symmetry is broken for
$-2<\beta<2$. Notice, however, that the gravity critical point at $\beta_g=-2/27$ remains undetected by $a$.

Let us briefly comment on the fact that the analytic continuation of the solution at positive $\beta$ described in the previous subsection has \emph{not} led us towards separable states. The eigenvalues remain of $\Ord{1/N}$ (and so does purity)  \emph{even} though the temperature can be (very) negative (as $\beta$ crosses $0$). In order to find separable states we will have to look at the stable branch in the following subsection.

\subsection{Stable branch of separable states}

In this section we will search the stable solution of the system at negative temperatures. As anticipated in Sec.\ \ref{sec:introductionstatisticalapproach},
from the definition (\ref{eq:partitionfunction}) of the partition
function one expects that, for any $N$, as $\beta \rightarrow
-\infty$ the system approaches the region of the phase space
associated to separable states: here the purity is $\Ord{1}$
and the right scaling in Eqs.\ (\ref{eq:partitionfunction})-(\ref{eq:piscale}) is $\alpha = 2$. In other words, by adopting the scaling $N^2$ for the exponent of the partition function, we will
explore the region $\beta=\Ord{1/N}$ of the scaling $N^3$
introduced for positive temperatures. Notice that the critical point
$\beta=-2 / 27$ for the solution at negative temperatures now reads
$\beta=-(2/27) N$ and escapes to $-\infty$ in the thermodynamic
limit.

\begin{figure}
\includegraphics[width=0.8\columnwidth]{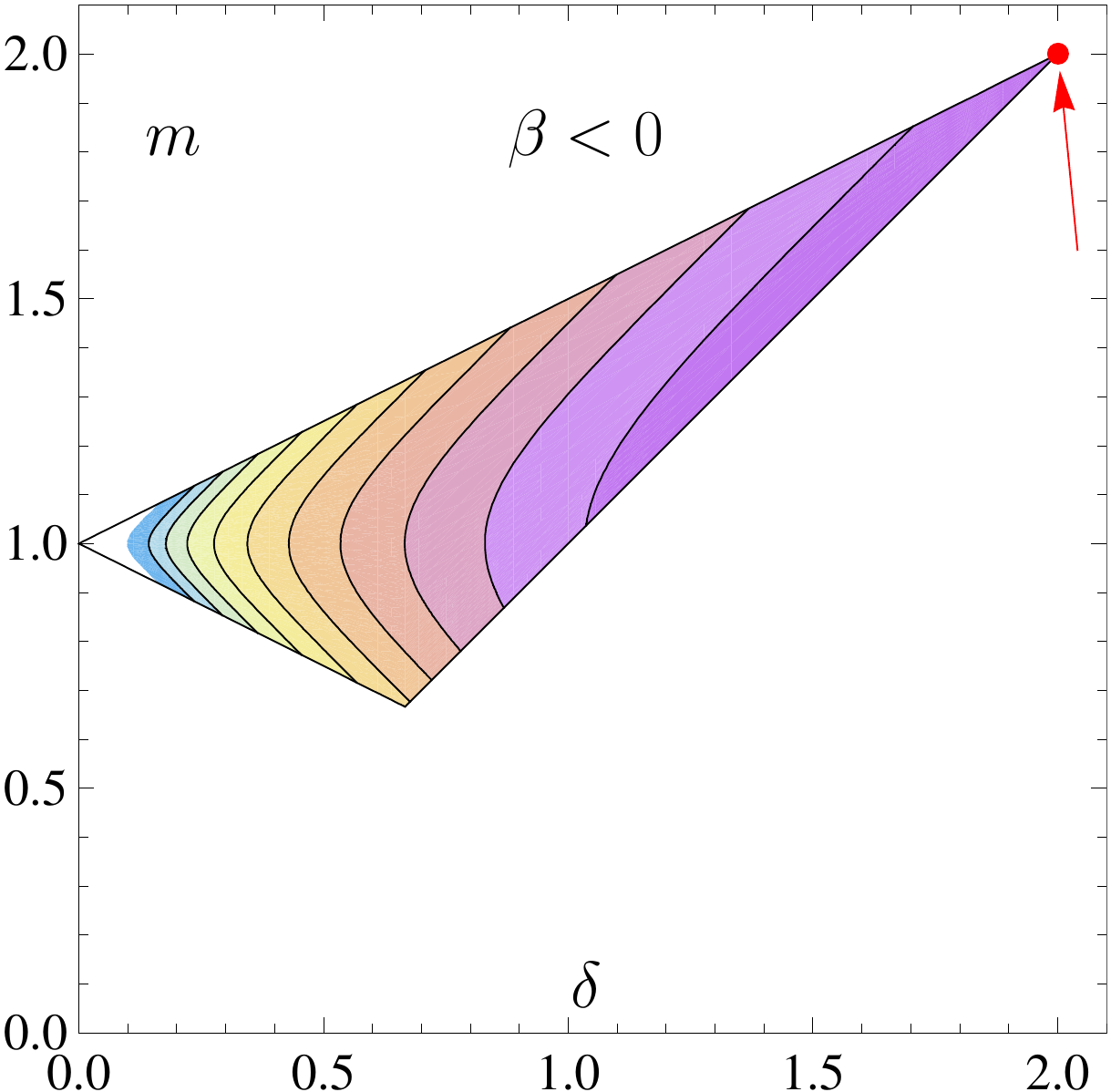}
\caption{Contour plot of the reduced free energy $\beta f_{\mathrm{red}}(\delta,m)$ of the sea for negative
temperatures. The arrow points at the minimum.}
\label{fig:contourPlotForNegativeTemp}
\end{figure}

We will show that the solution (\ref{eq:ansatz1}), according to which all the eigenvalues are $\Ord{1/N}$, becomes metastable in the region of negative
temperatures, and the distribution of the eigenvalues minimizing the
free energy is such that one eigenvalue is $\Ord{1}$: this
solution in the limit $\beta \rightarrow -\infty$ will correspond to the
case of separable states. By following an approach similar to that adopted for positive temperatures, we will first look for
the set of eigenvalues $\{\lambda_1, \ldots, \lambda_N\}$ satisfying
the saddle point equations (\ref{eq:stat1})-(\ref{eq:normal}) with
$\alpha=2$, getting as in Sec.\ \ref{sec:positivetemp} a continuous
family of solutions. We will select among them the set maximizing
($\beta<0$) the free energy (\ref{eq:freeF}), with $\alpha =2$:
\begin{eqnarray}\label{eq: free energy}
f_{N}= \sum_{i=1}^N \lambda_i^2-\frac{2}{N^2 \beta} \sum_{1\leq
i<j\leq N} \ln |\lambda_j-\lambda_i|.
\end{eqnarray}
As emphasized at the beginning of this section, since we are
approaching the limit $\beta \rightarrow -\infty$ the states
occupying the largest volume in phase space are separable; we
then define $\lambda_N= \mu$ as the maximum eigenvalue and
conjecture it to be of order of unity, whereas the other eigenvalues
are $\Ord{1/N}$:
\begin{equation}\label{eq:eigenvalues for negative temperatures}
\lambda_N=\mu= \Ord{1}, \quad \sum_{1 \leq i \leq N-1}\lambda_i=1-\mu .
\end{equation}
From this it follows that we need to introduce the natural scaling
only for the first $N-1$ eigenvalues in order to solve the saddle
point equations in the continuous limit and then estimate the
thermodynamic quantities:
\begin{eqnarray}
\label{eq:naturalRescaling for Negative temeperatures}
\lambda_i &=& (1-\mu)\frac{\lambda(t_i)}{N-1}, \\ 
& & \quad 0<t_i=\frac{i}{N-1}\leq1, \quad \forall i = 1, \ldots, N-1. \nonumber
\end{eqnarray}
In particular we will separately solve the saddle point equations
(\ref{eq:normal0})-(\ref{eq:normal}), corresponding to the minimization of the exponent of the partition
function with respect to the first $N-1$ eigenvalues and the
Lagrange multiplier $\xi$, given, in the limit $N \to \infty$, by
\begin{eqnarray}
\label{eq:neg_temp_saddle point equations for the sea}
& & P \int_0^\infty \frac{\bar{\rho}(\lambda')d
\lambda'}{\lambda-\lambda'} - i \frac{\xi}{2}(1-\mu)=0, \\
& & \quad \mathrm{with} \qquad \int_0^\infty \lambda \bar{\rho}(\lambda)d
\lambda=1, \nonumber 
\end{eqnarray}
and we will then consider the condition deriving from the saddle
point equation associated to $\mu$
\begin{equation}\label{eq:saddle point eq for mu}
 2\mu \beta + i\xi= 0.
\end{equation}
The function $\bar{\rho}$ introduced in (\ref{eq:neg_temp_saddle
point equations for the sea}) is the density of the eigenvalues
associated to $\lambda_1, \ldots \lambda_{N-1}$ in
(\ref{eq:naturalRescaling for Negative temeperatures})
and has the same form (\ref{eq:rhodef}) introduced for $\rho(\lambda)$ in the
regime of positive temperatures. By the same change of variables
introduced in Sec.\ \ref{sec:positivetemp}, Eqs.\
(\ref{eq:mdeltadef})-(\ref{eq:new variable}), the solution
of the integral equations (\ref{eq:neg_temp_saddle point equations
for the sea}) can be expressed in terms of $\bar{\phi}(x)=
\bar{\rho}(\lambda)\delta$:
\begin{equation}
\bar{\phi}(x)=\frac{1}{\pi
\sqrt{1-x^2}}\left(1-\frac{2x(m-1)}{\delta}\right).
\end{equation}
and the Lagrange multiplier is
$\xi=-i 4 (m-1)/(\delta^2(1-\mu))$.
The region of the parameter space $(m,\delta)$ such that the density
of eigenvalues $\bar{\phi}$ is nonnegative reads
\begin{equation}\label{eq:negative_temp_DomainForPhi}
\max\left\{ \delta, \;1-\frac{\delta}{2}\right\} \leq m \leq 1+
\frac{\delta}{2},
\end{equation}
which is the same expression of the domain found for the range of
positive temperatures (\ref{eq:positive_temp_eyesShapeDomain}) when
$\beta=0$, namely $\Gamma_1^\pm(\delta,0)=1\pm\delta/2$
(see Fig.\ \ref{fig:contourPlotForNegativeTemp}, which is the analog of Fig.\
\ref{fig:positive_eye}). This is consistent with the change in temperature scaling from $N^3$ in the case of positive
temperatures to $N^2$ in the case of negative temperatures: we are
``zooming'' into the region near $\beta \rightarrow 0^-$ of the range of
temperatures analyzed in \cite{paper1} and Sec.\
\ref{sec:positivetemp}. Summarizing, as could be expected from what
we have shown for positive temperatures, the solution of the
saddle point equations is a two parameter continuous family of
solutions. We now have to determine the density of eigenvalues that maximizes the free energy of the system.
From Eqs.\ (\ref{eq:eigenvalues for negative temperatures}) and
(\ref{eq: free energy}) we get
\begin{equation}
f_N=\mu^2-\frac{2}{N^2 \beta} \sum_{1\leq i<j\leq N} \ln
|\lambda_j-\lambda_i|+ \Ord{\frac{1}{N}}
\end{equation}
and by applying the scaling
(\ref{eq:naturalRescaling for Negative temeperatures}) 
\begin{eqnarray}
f_N &=& \mu^2 -\frac{1}{\beta}
\ln{(1-\mu)} + f_{\mathrm{red}}(\delta,m,\beta)
\nonumber\\
& &
+\frac{1}{\beta}\ln N
+\Ord{\frac{\ln N}{N}}
\nonumber\\
&=& f +\frac{1}{\beta}\ln N+\Ord{\frac{\ln N}{N}},
\end{eqnarray}
where
\begin{equation}
f= \lim_{N \to \infty}\left(f_N-\frac{1}{\beta}\ln N\right) =
\mu^2 -\frac{1}{\beta} \ln{(1-\mu)} + f_{\mathrm{red}},
\end{equation}
is the free energy density in the thermodynamic limit,
and
\begin{eqnarray}
 f_{\mathrm{red}}(\delta,m,\beta)&=&-\frac{1}{\beta}\int_{-1}^{1} \ dx \bar \phi(x) \int_{-1}^{1} \ dy  \bar \phi(y) \ln ( \delta |x-y| ) \nonumber \\
&=& \frac{2(m-1)^2}{\beta \delta^2} -\frac{1}{\beta} \ln\left(\frac{\delta}{2}\right)
\end{eqnarray}
is the reduced free energy density of the sea of eigenvalues.

It is easy to see that $\beta  f_{\mathrm{red}} (m,\delta)$,
has no stationary
points, but only a global minimum $\beta  f_{\mathrm{red}}=1/2$ at $(\delta,m)=(2, 2)$, see arrow in Fig.\
\ref{fig:contourPlotForNegativeTemp}; this point yields the Wishart distribution
found at $\beta=0$ for the case of positive temperature (see also
\cite{paper1}):
\begin{equation}\label{eq:whistart in zero}
\bar \phi(x)=\frac{1}{\pi}\sqrt{\frac{1-x}{1+x}} , \quad
\bar \rho(\lambda)=\frac{1}{2\pi} \sqrt{\frac{4-\lambda}{\lambda}},
\end{equation}
where one should remember that the $\lambda$'s are also scaled by $1-\mu$, see Eq.\ (\ref{eq:naturalRescaling for Negative temeperatures}).

\begin{figure}
\includegraphics[width=0.8\columnwidth]{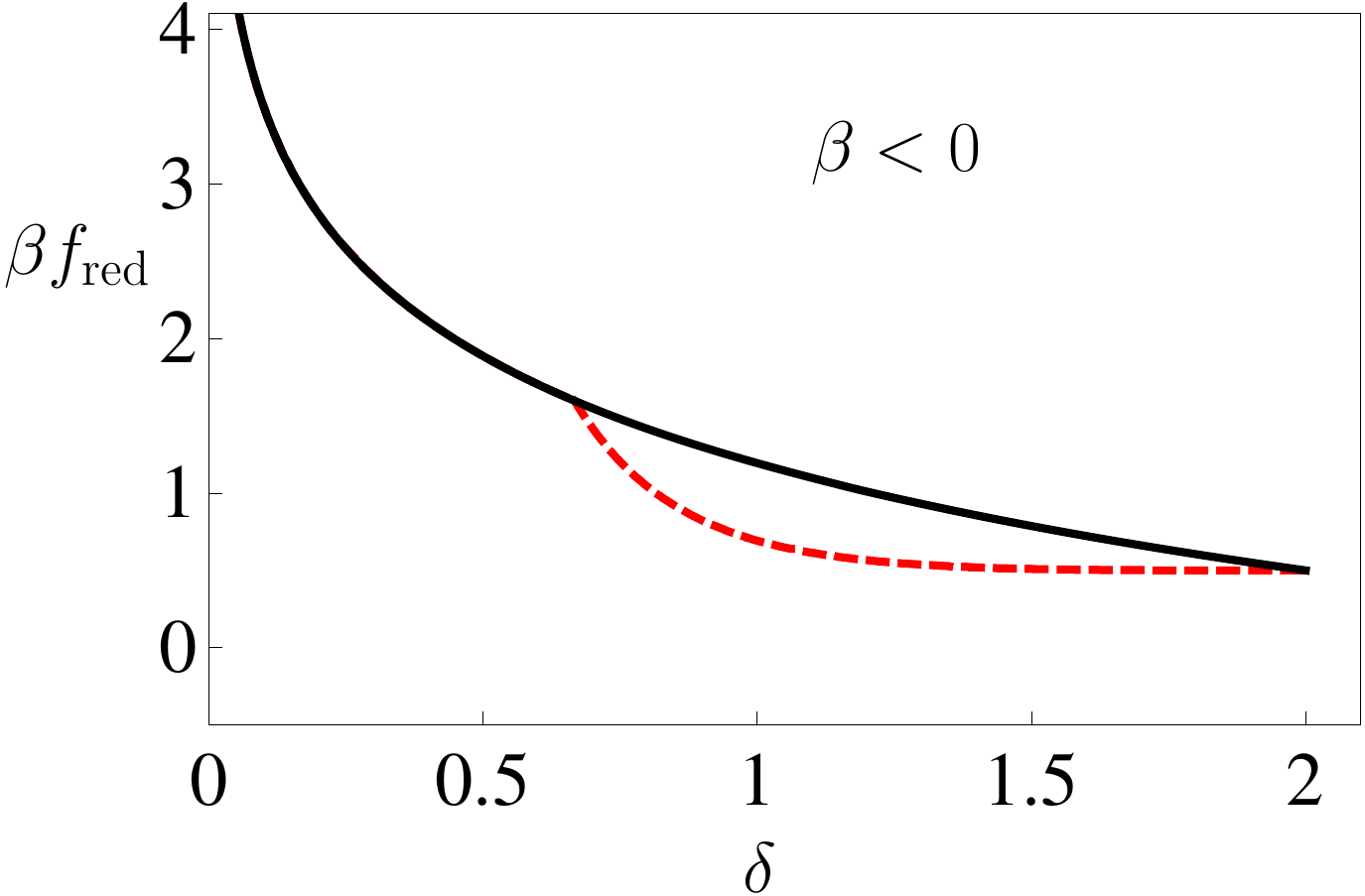}
\caption{Reduced free energy $\beta f_{\mathrm{red}}$ on the boundary of the triangular domain in Fig.\ \ref{fig:contourPlotForNegativeTemp} for
the case of negative temperatures. Solid line: upper boundary; dashed line: lower boundary.}
\label{fig: free energy su frontiera temp negative}
\end{figure}

We stress that this result is valid for all $\beta <0$. In order
to check this solution one has to compute the free energy on the
boundary of this domain, see Fig. \ref{fig: free energy su frontiera temp negative}
(which is the analog of Fig.\ \ref{fig:3free}). One gets
 that the free energy density is given by
\begin{eqnarray}
f(\mu, \beta)=\mu^2 -\frac{1}{\beta}
\ln{(1-\mu)}+\frac{1}{2\beta}.
\end{eqnarray}

\begin{figure}
\includegraphics[width=0.8\columnwidth]{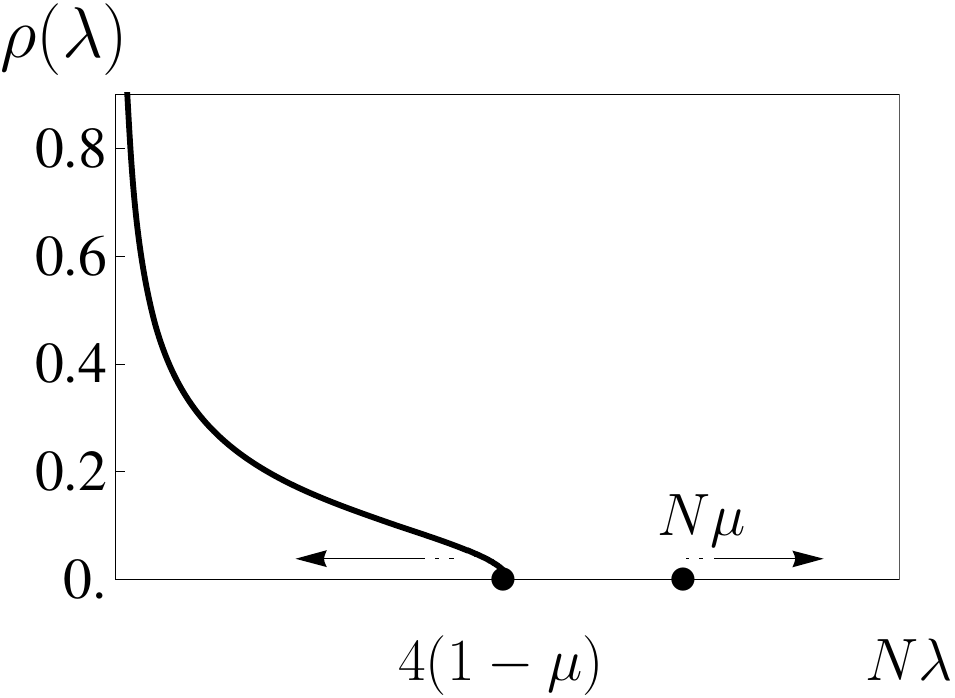}
\caption{Evaporation of the eigenvalue $\mu = \Ord{1}$ from the sea of eigenvalues $\Ord{1/N}$.}
\label{fig:evaporation}
\end{figure}

A new stationary solution, in which the largest isolated eigenvalue
$\mu$ becomes $\Ord{1}$, can be found by minimizing the free energy
density and yields
\begin{equation}
\label{eq:isolated eigenvalue}
\mu(\beta)=\frac{1}{2}+\frac{1}{2}\sqrt{1+\frac{2}{\beta}} ,
\end{equation}
being defined only for $\beta<-2$; this expression can also be obtained directly by
the saddle point equation (\ref{eq:saddle point eq for mu}) corresponding to the isolated eigenvalue $\mu$.
This eigenvalue, $\Ord{1}$, evaporates from the sea of eigenvalues $\Ord{1/N}$, as pictorially represented in Fig.\ \ref{fig:evaporation}.
The isolated eigenvalue moves at a speed $-d \mu /d\beta = 1/(2\sqrt{\beta^4+2\beta^3})$, which diverges at $\beta=-2$: another symptom of criticality. However, this new solution,  when it appears at $\beta=-2$, is not the global minimum of $\beta f$: as we shall see it eventually becomes stable at a lower value of $\beta$.
We get for  $\beta <-2$ (i.e.\ $0<\mu<1$)
\begin{eqnarray}
\label{eq:unegmu}
u &=& \mu^2=\frac{1}{2}+\frac{1}{2\beta}+\frac{1}{2}\sqrt{1+\frac{2}{\beta}},
\\
\label{eq:snegmu}
s &=& 
\ln(1-\mu) -\frac{1}{2}= \ln \left(\frac{1}{2}-\frac{1}{2}\sqrt{1+\frac{2}{\beta}}\right) -\frac{1}{2}, \\
\label{eq:fnegmu}
\beta f &=& \beta u -s = \frac{1-2\mu}{2(1-\mu)} - \ln{(1-\mu)} \nonumber \\
&=& 1+ \frac{\beta}{2}+\frac{\beta}{2}\sqrt{1+\frac{2}{\beta}}-\ln \left(\frac{1}{2}-\frac{1}{2}\sqrt{1+\frac{2}{\beta}}\right),
\end{eqnarray}

\begin{figure}
\includegraphics[width=0.8\columnwidth]{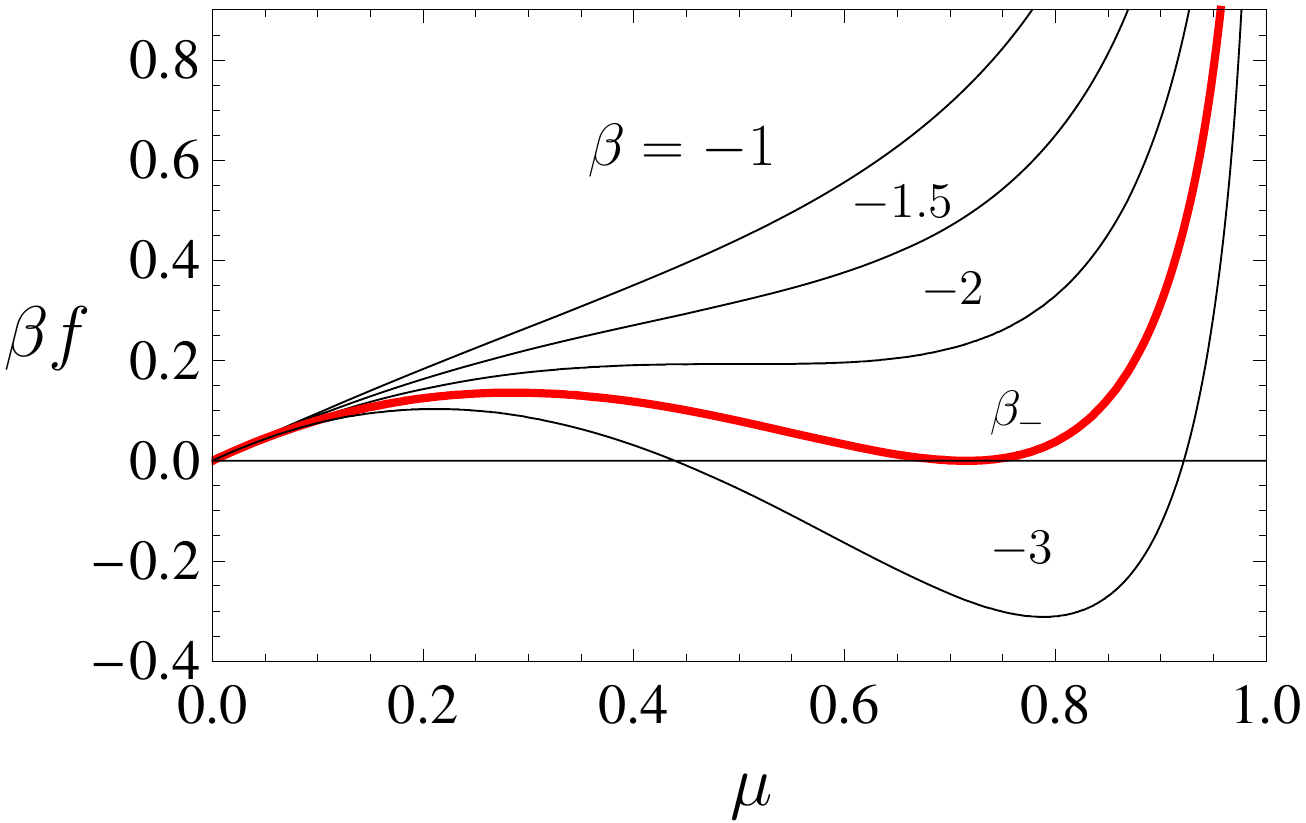}
\caption{Reduced free energy as a function of $\mu$ for different
values of $\beta(<0)$. Notice the birth of a stationary point for $\beta=-2$ that becomes the global minimum for $\beta\leq\beta_-$.}
\label{fig:neg_temp_free_energy}
\end{figure}

\begin{figure}
\hspace{-0.5cm}\includegraphics[width=0.75\columnwidth]{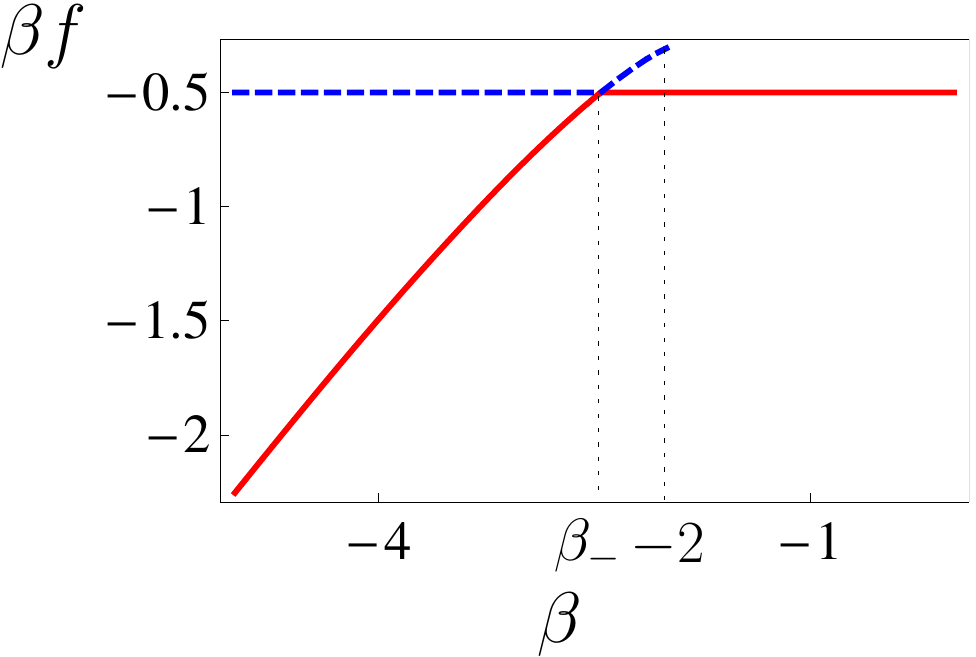}
\\ \vspace{0.1cm}
\includegraphics[width=0.7\columnwidth]{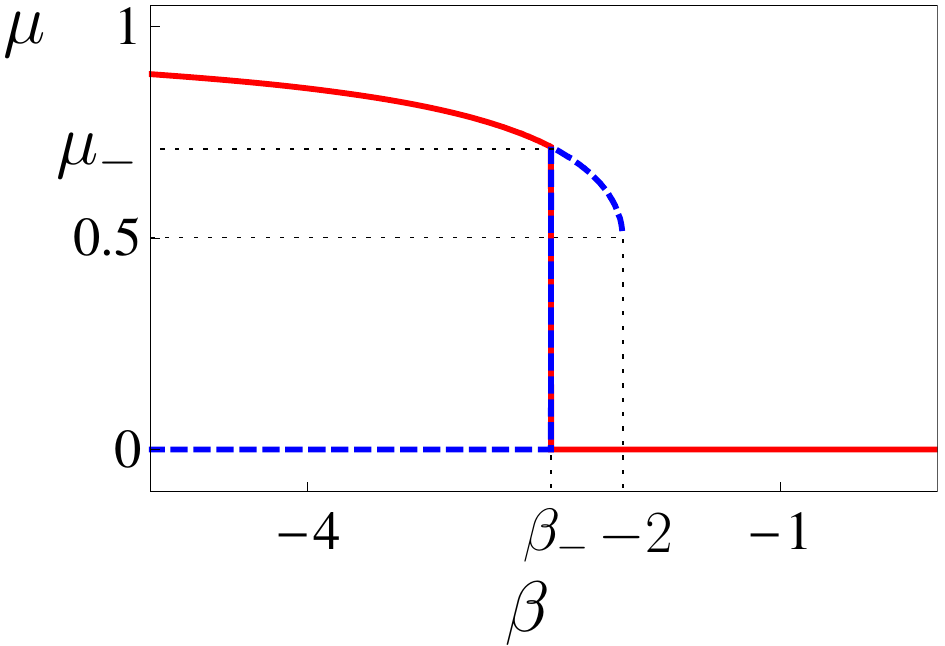}
\caption{Free energy and maximum eigenvalue at negative
temperatures. The two solutions are exchanged at $\beta_- \simeq
2.45541$, where there is a first order phase transition. Full line:
solution of mimimal free energy; dashed line: solution of higher
free energy.} \label{fig:BETAfreeEnergyTeoricaNegTemp1}
\end{figure}

We are now ready to unveil the presence of a \emph{first order}
phase transition in the system. In Fig.\ \ref{fig:neg_temp_free_energy} we plot the free energy density as a function of $\mu$ for different values of $\beta$. For $\beta>-2$ there is a global minimum of $\beta f$  at
$\mu=0$; $\mu$ is still in the sea of the eigenvalues $\Ord{1/N}$ and
the stable solution is given by the Wishart distribution (\ref{eq:Whishprolong}) with the potentials (\ref {eq:gravitysol}) (remember that, in the zoomed scale considered here, $\beta_g$ corresponds to the very large inverse temperature $N \beta_g$). At
$\beta=-2$ there appears a stationary point for the free energy density
corresponding to $\mu = \Ord{1}$ [see (\ref{eq:isolated
eigenvalue})]; notice however that $\beta f$ at this point remains larger than its
value at the global minimum, until $\beta$
reaches $\beta_-$. Finally, for $\beta<\beta_-$ the global
minimum of $\beta f$ moves to the right, to the solution containing $\mu = \Ord{1}$. Summarizing, for
$\beta>\beta_-$ the solution of saddle point equations maximizing the
free energy of the system is such that all eigenvalues are $\Ord{1/N}$, at $\beta=-2$ there appears a metastable solution for the
system with one eigenvalue $\Ord{1}$, and for $\beta\leq
\beta_-$ this becomes the stable solution, that maximizes the free
energy, whereas the distribution of the eigenvalues found in Sec.\ \ref{sec:positivetemp} becomes now metastable. The maximum eigenvalue is then a discontinuous function of the temperature at $\beta=\beta_-$
and in the limit $\beta \rightarrow -\infty$, $\mu$
approaches $1$: the state becomes separable. This
critical temperature $\beta_- $
is the solution of the transcendental
equation $f(\beta_-,0)=f(\beta_-,\mu_-)$, that is
\begin{equation}
\frac{\mu_-}{2(1-\mu_-)} =-\ln(1-\mu_-),
\label{eq:mu-}
\end{equation}
which yields
\begin{equation}
\mu_-\simeq 0.71533, \qquad \beta_- = -\frac{1}{2\mu_-
(1-\mu_-)} \simeq -2.45541.
\label{eq:beta-}
\end{equation}
Therefore, the branch (\ref{eq:snegmu})-(\ref{eq:fnegmu})
is stable for $\beta < \beta_-$ while it becomes metastable for
$\beta_-<\beta<-2$. On the other hand, the solution $\mu=0$,
corresponding to
\begin{eqnarray}
\label{eq:unegmu0}
u&=&\mu^2=0, \\
\label{eq:snegmu0}
s &=& \beta(u-f)= -\frac{1}{2},\\
\label{eq:fnegmu0}
\beta f &=& \frac{1}{2},
\end{eqnarray}
has a lower free energy for $\beta_-<\beta<0$, and a higher one for $\beta<\beta_-$.
See Fig.\ \ref{fig:BETAfreeEnergyTeoricaNegTemp1}.

At $\beta_{-}$ there is a first order phase transition. At this fixed temperature  the internal energy of the system goes from $u_r=0$ up to $u_l = \mu_-^2\simeq0.5117$, while the entropy goes from $s_r=-1/2$ down to $s_l =  -1/2 + \ln(1-\mu_-)\simeq -1.75643$. One gets $\Delta s/ \Delta u = \beta_{-}$. Therefore, the entropy density as a function of the internal energy density reads
\begin{equation}
s(u) =
\begin{cases}
\beta_- u -\frac{1}{2}, &  0<u<\mu_-^2, \\
\\
\ln(1-\sqrt{u})-\frac{1}{2}, &  \mu_-^2 \leq u <1 .
\end{cases}
\label{eq:sumu}
\end{equation}
It is continuous together with its first derivative at $u=\mu_-^2$, while its second derivative is discontinuous. 
Notice that $\Delta u = \Delta s/\beta_{-} $ is the specific latent heat of the evaporation of the largest eigenvalue from the sea of the eigenvalues, from $\Ord{1/N}$ up to $\mu_{-}$.

A few words of interpretation are necessary. As we have seen, it has been necessary to follow the stable branch of the solution in order to obtain separable states at negative temperatures. The analytic continuation of the stable solution for positive temperatures would yield an unstable branch in which all eigenvalues remain $\Ord{1/N}$. By contrast, the new stable solution consists in a sea of $N-1$ eigenvalues $\Ord{1/N}$ plus one isolated eigenvalue $\Ord{1}$. 

Let us discuss this result in terms of purity, like at the end of Sec.\ \ref{sec:positivetemp} (we stress again that $\beta$ is a Lagrange multiplier that fixes the value of the purity of the reduced density matrix of our  $N^2$-dimensional
system). Assume that we pick a given isopurity manifold in the original Hilbert space, defined by a given \emph{finite} value $\pi_{AB}$ of purity. If we randomly select a vector belonging to this isopurity manifold, its reduced density matrix (for the fixed bipartition) will have one finite eigenvalue $\mu \simeq \sqrt{\pi_{AB}}$ and many small eigenvalues $\Ord{1/N}$ (yielding a correction $\Ord{1/N}$ to purity). In this sense, the quantum state is largely separable. 
The probability of finding in the afore-mentioned manifold a vector whose reduced density matrix has, say, \emph{two} (or more) finite eigenvalues $\mu_1$ and $\mu_2$ (such that $\mu_1^2+\mu_2^2 \simeq \pi_{AB}$, modulo corrections $\Ord{1/N}$) is vanishingly small. By contrast, remember (from the results of Sec.\ \ref{sec:positivetemp}) that if the isopurity manifold is characterized by a \emph{very small} value $\Ord{1/N}$ of purity, the eigenvalues of a randomly chosen vector on the manifold are \emph{all} $\Ord{1/N}$ (being distributed according to the semicircle or Wishart, depending on the precise value of purity, as seen in Sec.\ \ref{sec:positivetemp}).
This is the significance of the statistical mechanical approach adopted in this article. We will come back to this point in Sec.\ \ref{sec:QI}.

\section{Finite size systems}
\label{sec:finitesize}

The results of the previous section refer to $N\to\infty$. In order to understand how finite-$N$ corrections affect our conclusions we have numerically minimized the free energy for various temperatures. The two phases of the system discussed in the previous section correspond to the two solutions
obtained by minimizing the free energy $\beta f_N$  (\ref{eq: free energy}) on the $N$
dimensional simplex of the normalized eigenvalues. Indeed, we have
numerically proved that $\beta f_N(\beta)$ presents two local minima at negative temperatures: for $0 \geq \beta > \beta_-^{(N)}$ the minimum
giving the lower value of $\beta f_N(\beta)$ corresponds to the
distribution of eigenvalues (\ref{eq:ansatz1}), found in the last
section; the other minimum is reached when the highest eigenvalue is $\Ord{1}$. The point $\beta=\beta_-^{(N)}$ is a crossing point for these two solutions, and for $\beta \leq \beta_-^{(N)}$ these two solutions are inverted, see Fig.\ \ref{fig:BETAfreeEnergyTeoricaNegTemp1} (and \ref{fig:neg_temp_free_energy}). Summarizing, there exists a negative temperature at which the system undergoes a first-order phase transition, from typical to
separable states.

The first thing to notice is that qualitatively the phase transition remains of first order even for finite $N$. The second is that the finite $N$ corrections are quite relevant for the location of the phase transition and the value of the maximum eigenvalue as a function of $\beta$.
For example, for $N=30$, the negative critical temperature $\beta_{-}^{(30)}= -1.935$ instead of $-2.455$.
This is evinced from Fig.\ \ref{fig:betaF}, which is the finite size version of Fig.\ \ref{fig:BETAfreeEnergyTeoricaNegTemp1}.
This can be understood, as the corrections to $f(\mu)$ around $\mu=0$ are quite large. In the limit $\mu=1/N$ there is a hard wall for the maximum eigenvalue $\mu$, as the condition $\sum_i\lambda_i=1$ cannot be satisfied if $\mu<1/N$. It is therefore likely that all sorts of large corrections occur as $\mu$ tends to $1/N$, probably yielding an effective size to the corrections which is a lower power of $1/N$ (or even possibly $1/\ln N$). The limits $\mu\to 0$ and $N\to\infty$ do not commute.

\begin{widetext}

\begin{figure}
\includegraphics[width=0.41\columnwidth]{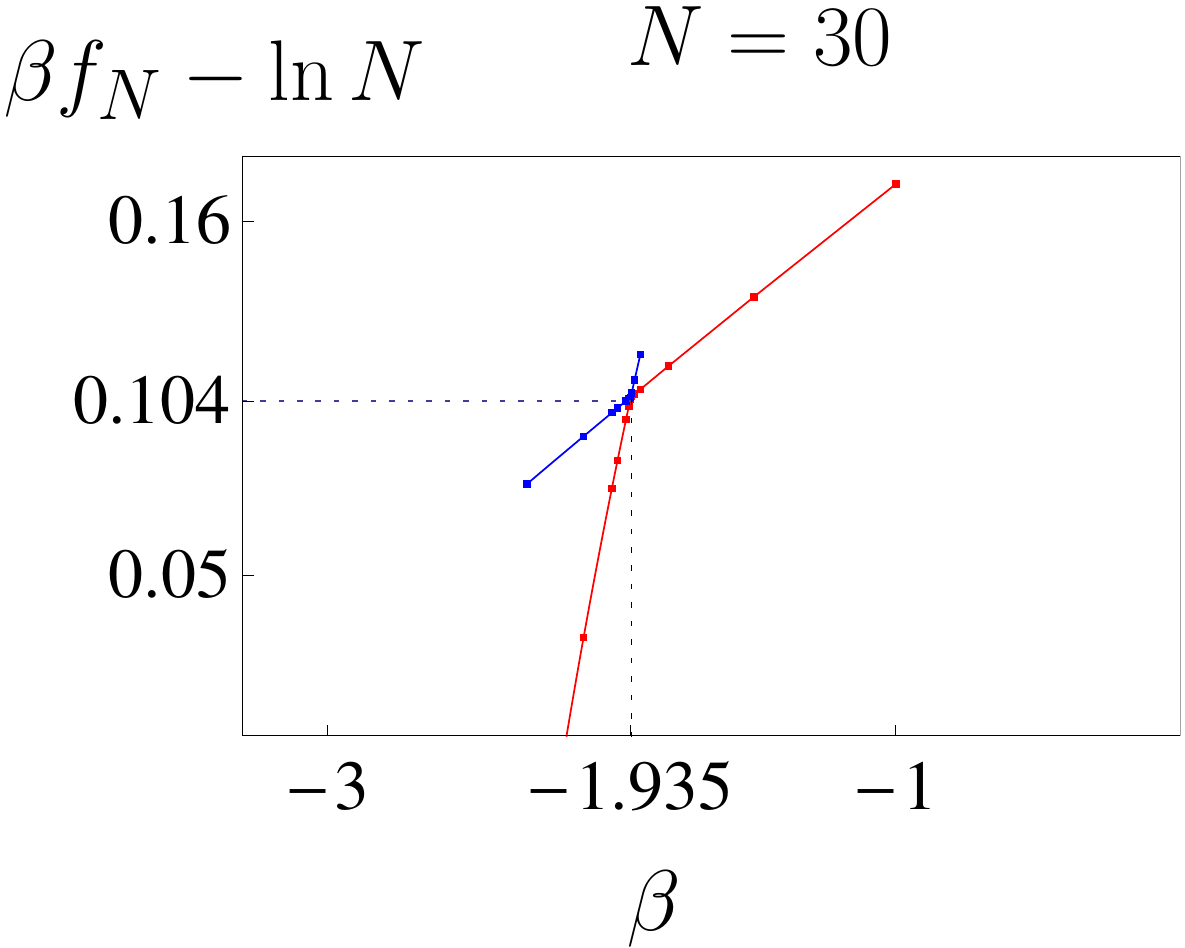}
\quad
\includegraphics[width=0.4\columnwidth]{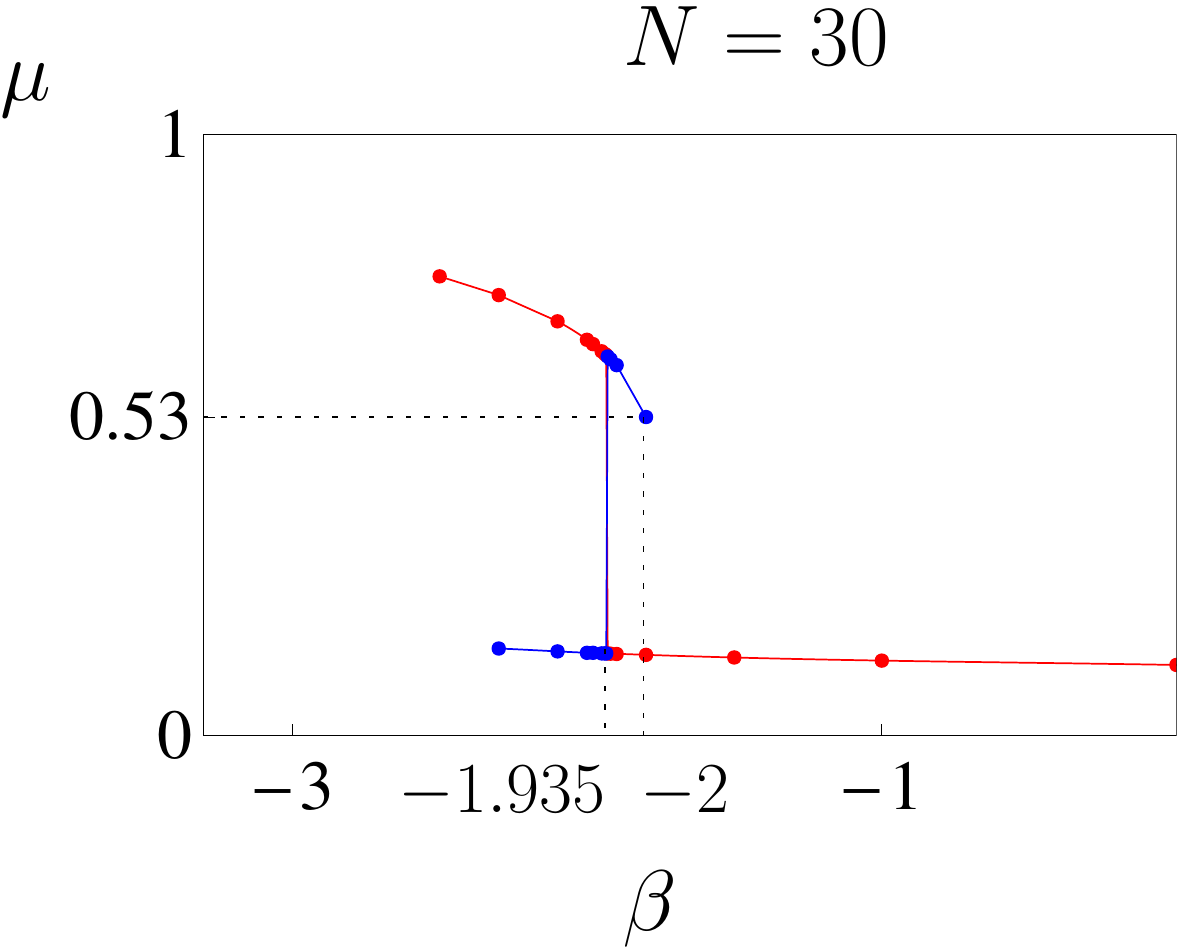}
\caption{Finite $N$ version of Fig.\ \ref{fig:BETAfreeEnergyTeoricaNegTemp1}. Free energy and maximum eigenvalue in the saddle point
approximation as function of $\beta$ at $N=30$. The local minimum is
in blue, the global one in red. The two minima swap stability at $\beta=-1.935$.
Notice the birth of the new local minimum at $\beta=-1.8$ (for $N=\infty$ this takes place at $\beta=-2$) and the exchange of stability at $\beta=-1.93$ (for $N=\infty$, $\beta=-2.45$). } 
\label{fig:betaF}
\end{figure}

\begin{figure}
\includegraphics[width=0.7\columnwidth]{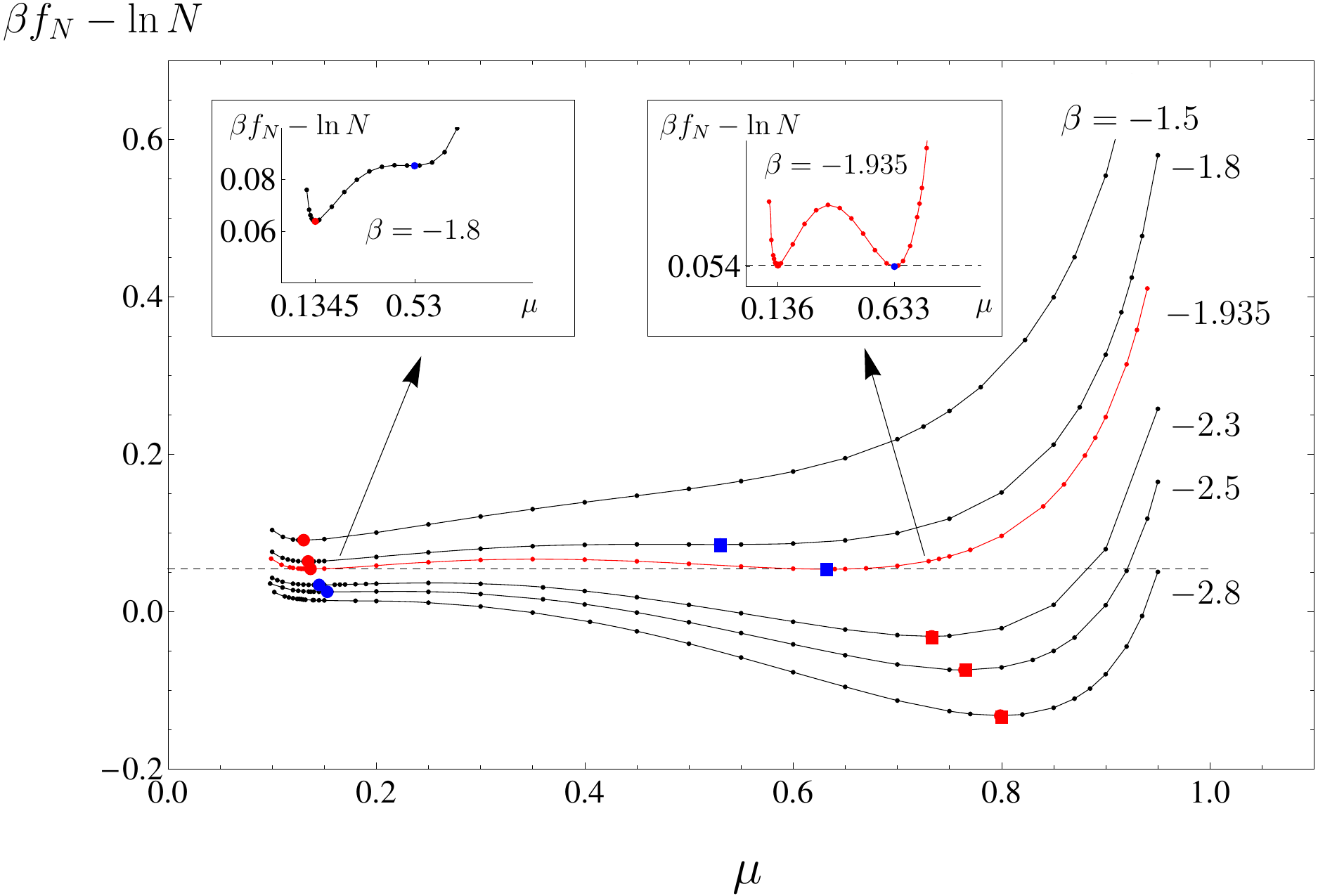}
\caption{Finite $N$ version of Fig.\ \ref{fig:neg_temp_free_energy}. $\beta f_N-\ln N$ as a function of the maximum eigenvalue $\mu$, obtained by numerical minimization over the remaining $N-1$ eigenvalues for various $\beta$. Observe the formation of a new minimum and the exchange of stability, although the critical values of $\beta$ at which these phenomena occur differ from the theoretical ones, due to large finite $N$ corrections. However, it is clear that at small $\mu$, $1/N$ corrections tend to increase the value of $\beta f_N$, making the critical value $\beta_-$ move towards 0, as observed in the numerics.}\label{fig:FreeEnergyAtFixedX}
\end{figure}
\end{widetext}

\begin{figure}
\includegraphics[width=0.8\columnwidth]{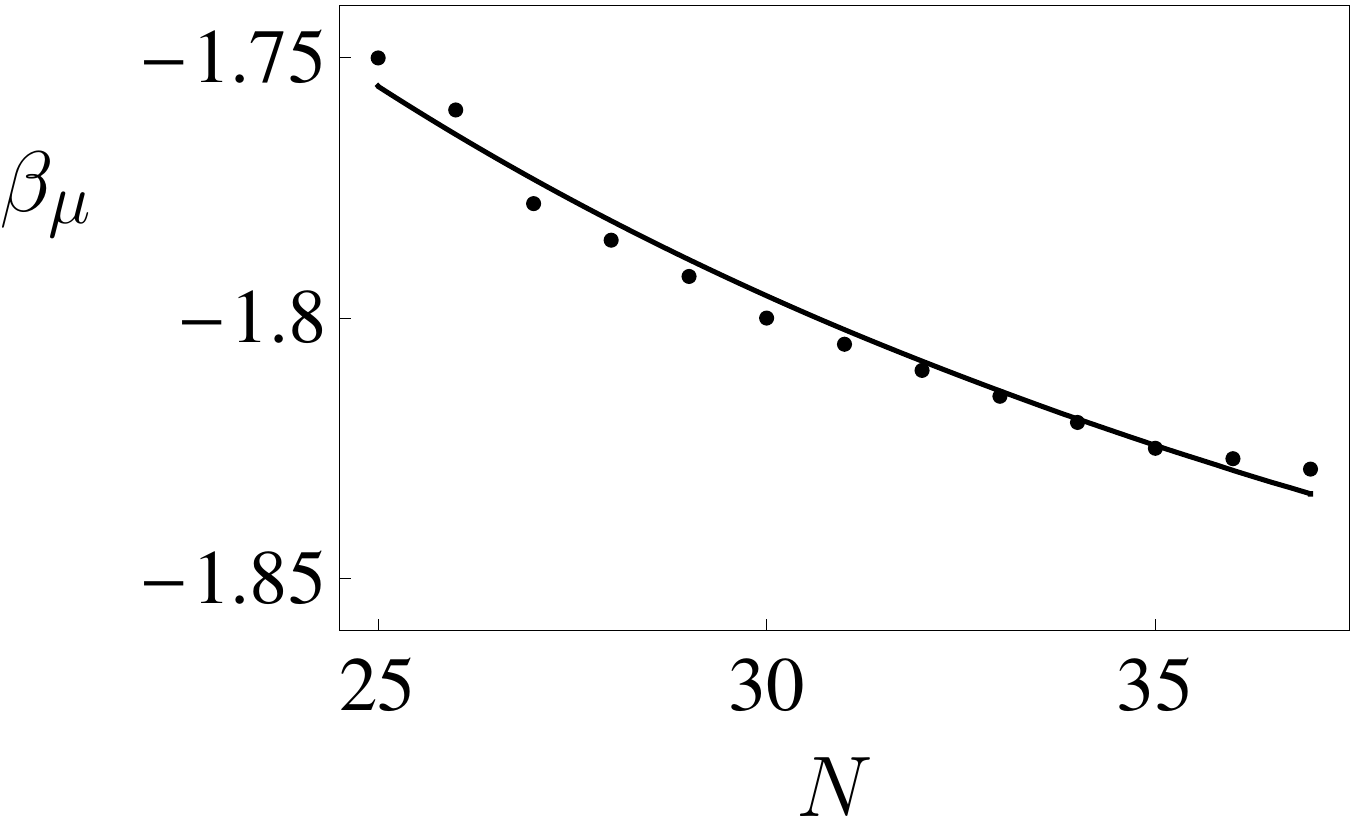}
\caption{Value of $\beta$ at which the second minimum is born as a function of $N$. The solid curve is a best fit returning $\beta_\mu^{(N)}=-1.997-6.04/N$. The asymptotic value should be 2 and is in good agreement with the constant of the fit.}
\label{fig:primabeta}
\end{figure}

\begin{figure}
\includegraphics[width=0.81\columnwidth]{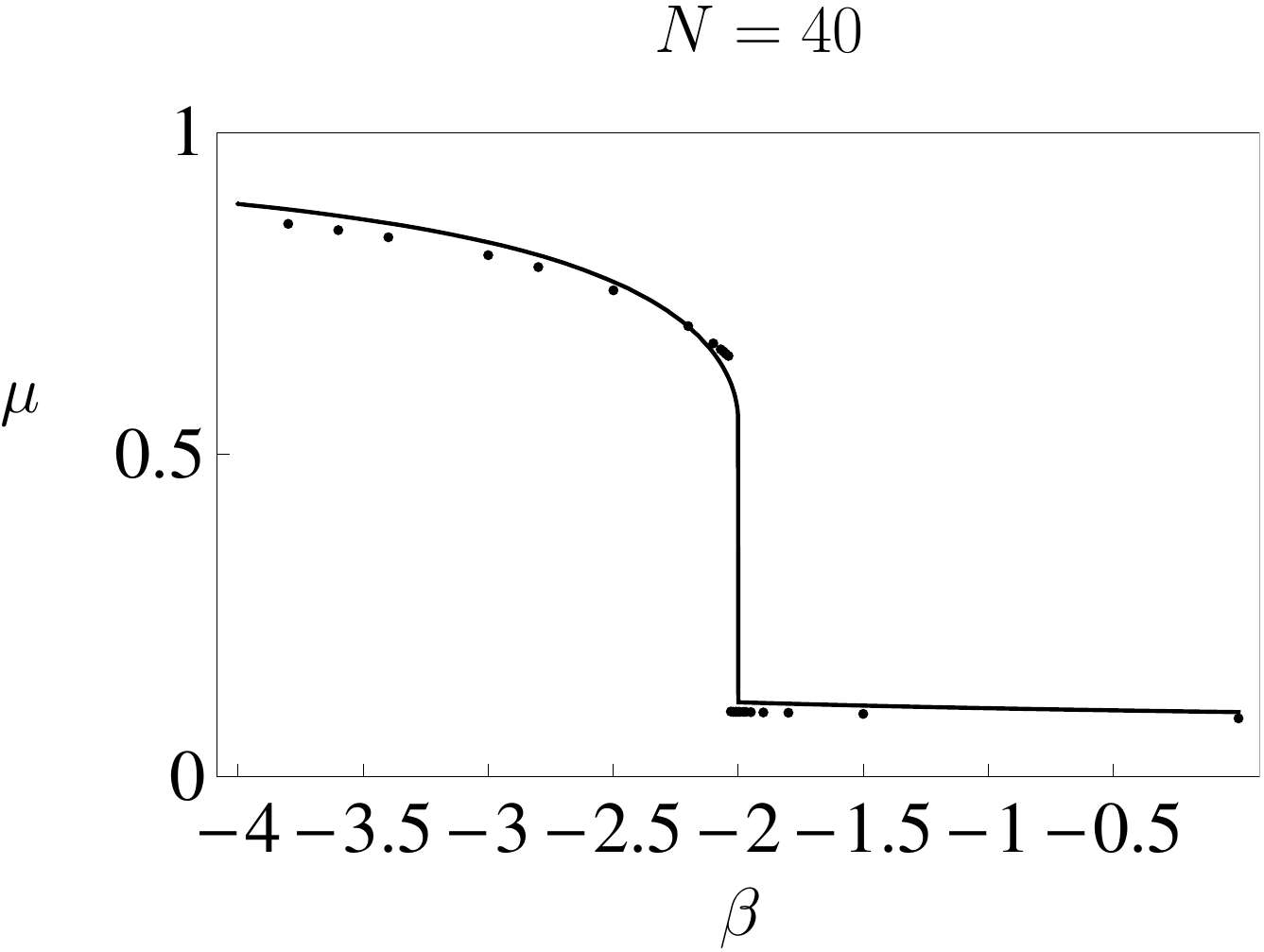}
\caption{Maximal eigenvalue in the saddle point approximation as function of $\beta$. The points are the result of a numerical evaluation for $N=40$, while the full line is the expression in Eq.\ (\ref{eq:Nfinfin}).}
\label{fig:betaF40}
\end{figure}

To further explore this effect we minimized $\beta f_N$ with respect to $\lambda_1, \ldots, \lambda_{N-1}$,
for fixed values of the largest eigenvalue $\lambda_N=\mu$ and for different temperatures. The results for $N=30$ are shown in Fig.\ \ref{fig:FreeEnergyAtFixedX}. One can see that between $\beta=-1.8$ and $\beta=-2.5$ there is a competition between two well defined local minima, corresponding to the two solutions discussed above. At
$\beta=\beta_{-}^{(30)}= -1.935$ their free energies are equal. For higher $\beta$ the global minimum corresponds to the solution (\ref{eq:ansatz1}), whereas on the other side of $\beta_{-}^{(30)}$ the solution with $\mu = \Ord{1}$ minimizes $\beta f_N$.
Similar corrections are observed for the value of $\beta=\beta_\mu^{(N)}$ at which the second minimum is born. See Fig.\ \ref{fig:primabeta}.

We have seen that for $\beta>\beta_{-}$ the stable solution has no detached eigenvalue. By taking into account the scaling $\beta\to\beta/N$ we get that the solution is given by the very first part of the gravity branch (\ref{eq:gravitysol}). In particular, the maximum eigenvalue is given by $b/N=(m+\delta)/N=2\delta/N$. On the other hand for $\beta<\beta_{-}$ the maximum eigenvalue is given by (\ref{eq:isolated eigenvalue}). Therefore, we get
\begin{eqnarray}
\mu = 
\begin{cases}
\frac{1}{2}+\frac{1}{2}\sqrt{1+\frac{2}{\beta}}  , &  \beta \leq\beta_-, \\
\\
\frac{2}{N} \delta(\beta/N), &   \beta_-<\beta<0 ,
\end{cases}
\label{eq:Nfinfin}
\end{eqnarray}
with $\delta(\beta)$ given by (\ref{eq:betasss})-(\ref{eq:adibeta}). The numerical results for $N=40$ are compared with the expressions in Eq.\ 
(\ref{eq:Nfinfin}) in Fig.\ \ref{fig:betaF40}. The agreement is excellent.

The corresponding free energy follows from (\ref{eq:fnegmu}) and (\ref{eq:gravitysol}) with the appropriate scaling 
\begin{widetext}
\begin{eqnarray}
\beta f = 
\begin{cases}
1+ \frac{\beta}{2}+\frac{\beta}{2}\sqrt{1+\frac{2}{\beta}}-\ln \left(\frac{1}{2}-\frac{1}{2}\sqrt{1+\frac{2}{\beta}}\right), &  \beta \leq\beta_- , \\
\\
\frac{11}{4} -\frac{9}{\delta(\beta/N)} + \frac{9}{\delta(\beta/N)^2} - \ln \frac{\delta(\beta/N)}{2} , &  \beta_- <\beta<0 .
\end{cases} 
\end{eqnarray}
\end{widetext}
Notice that in order to have a finite size scaling of the critical temperature $\beta_{-}^{(N)}$ one should take into account $\Ord{1/N}$ corrections  to the expression of $\beta f$ and then evaluate the intersection between the two branches, but this analysis goes beyond our scope.

\section{Overview}
\label{sec:QI}

Let us summarize the main results obtained in this article in more intuitive terms, by focusing on those quantities that are more directly related to physical intuition.
In the statistical mechanical approach adopted in this article, the temperature plays the usual role of a Lagrange multiplier, whose only task is to fix the value of energy (purity in our case). A given value of $\beta$ determines a set of vectors in the projective Hilbert space whose reduced density matrices have a given purity (isopurity manifold of quantum states). The distribution of the eigenvalues of (the reduced density matrices associated to) these vectors is that investigated in this article and yields information on the separability (entanglement) of these quantum states. 
The distribution of eigenvalues is the most probable one \cite{mehta} (in the same way as the Maxwell distribution of molecular velocities is the most probable one at a given temperature). Let us therefore abandon temperature in the following and fully adopt purity as our physical quantity.

Entropy counts the number of states with a given value of purity and is in this sense proportional to the logarithm of the volume in the projective Hilbert space.
The explicit expressions of the entropy density $s$, which is the logarithm of the volume of the isopurity manifold, as a function of the purity $\pi_{AB}$ of the state vectors in that volume, can be read directly from  Eqs.\ (\ref{eq:112}) and (\ref{eq:sumu}) by taking into account the correct scaling:
\begin{widetext}
\begin{eqnarray}
s(\pi_{AB})  = \begin{cases}
\frac{1}{2} \ln ( N \pi_{AB} -1) -\frac{1}{4},  & \frac{1}{N} < \pi_{AB} \leq  \frac{5}{4 N} ,  \\
\\
 \ln \left( \frac{3}{2}  - \sqrt{\frac{9}{4} - N \pi_{AB}}
\right) - \frac{9}{4} + \frac{5}{2\left( \frac{3}{2}  - \sqrt{\frac{9}{4} - N \pi_{AB}}
\right)} -
\frac{3}{4\left( \frac{3}{2}  - \sqrt{\frac{9}{4} - N \pi_{AB}}
\right)^2},
& \frac{5}{4 N} < \pi_{AB} \leq \frac{2}{N},
 \\
\\
\beta_- 
\pi_{AB}
-\frac{1}{2}, &  \frac{2}{N} <\pi_{AB} \leq \mu_-^2 
, \\
\\
\ln\left(1-\sqrt{ 
\pi_{AB} 
}\right)-\frac{1}{2}, &  \mu_-^2 
< \pi_{AB} <1 ,
\end{cases}
\label{eq:1121}
\end{eqnarray}
\begin{figure}
\includegraphics[width=0.41\columnwidth]{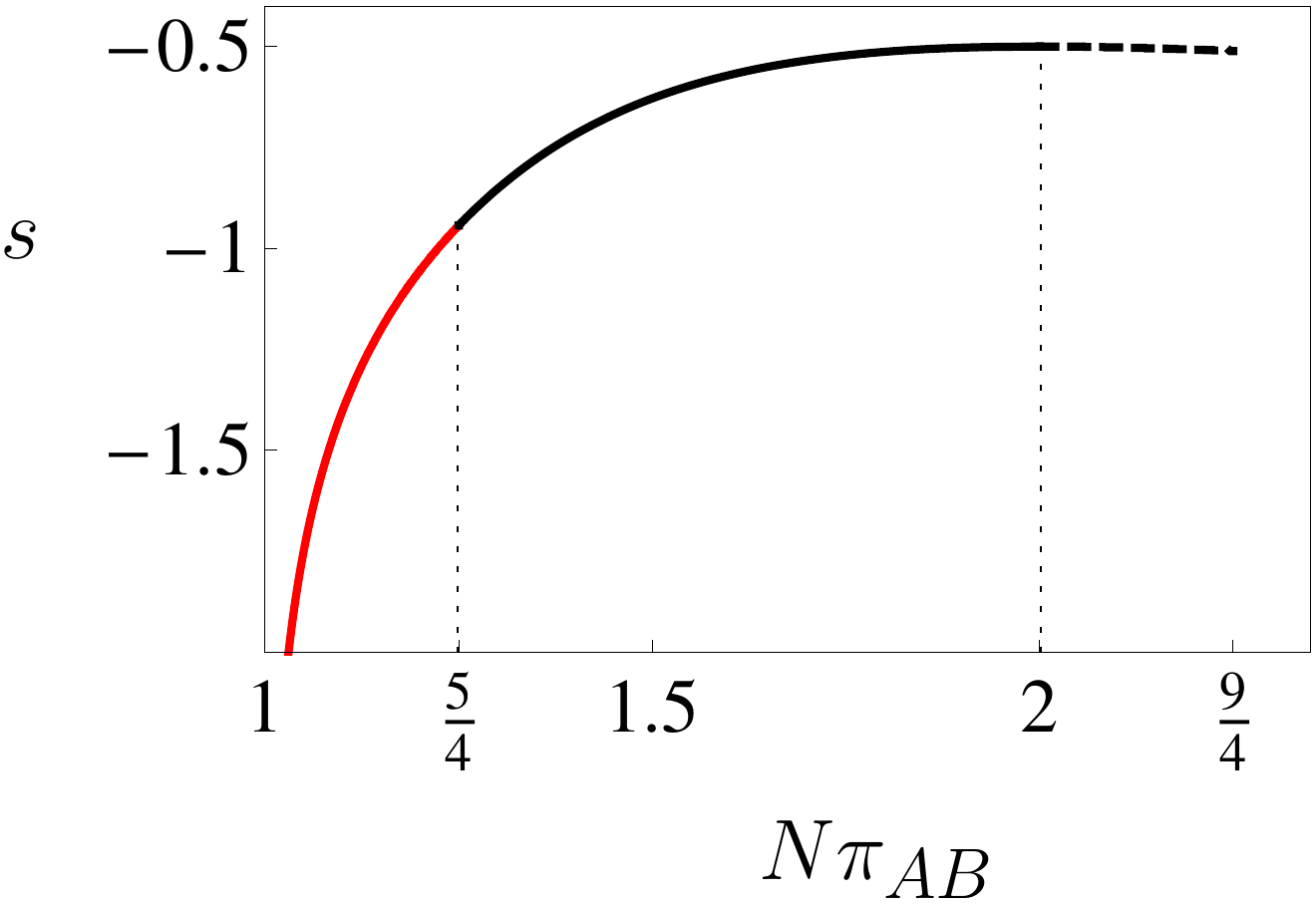} \quad
\includegraphics[width=0.405\columnwidth]{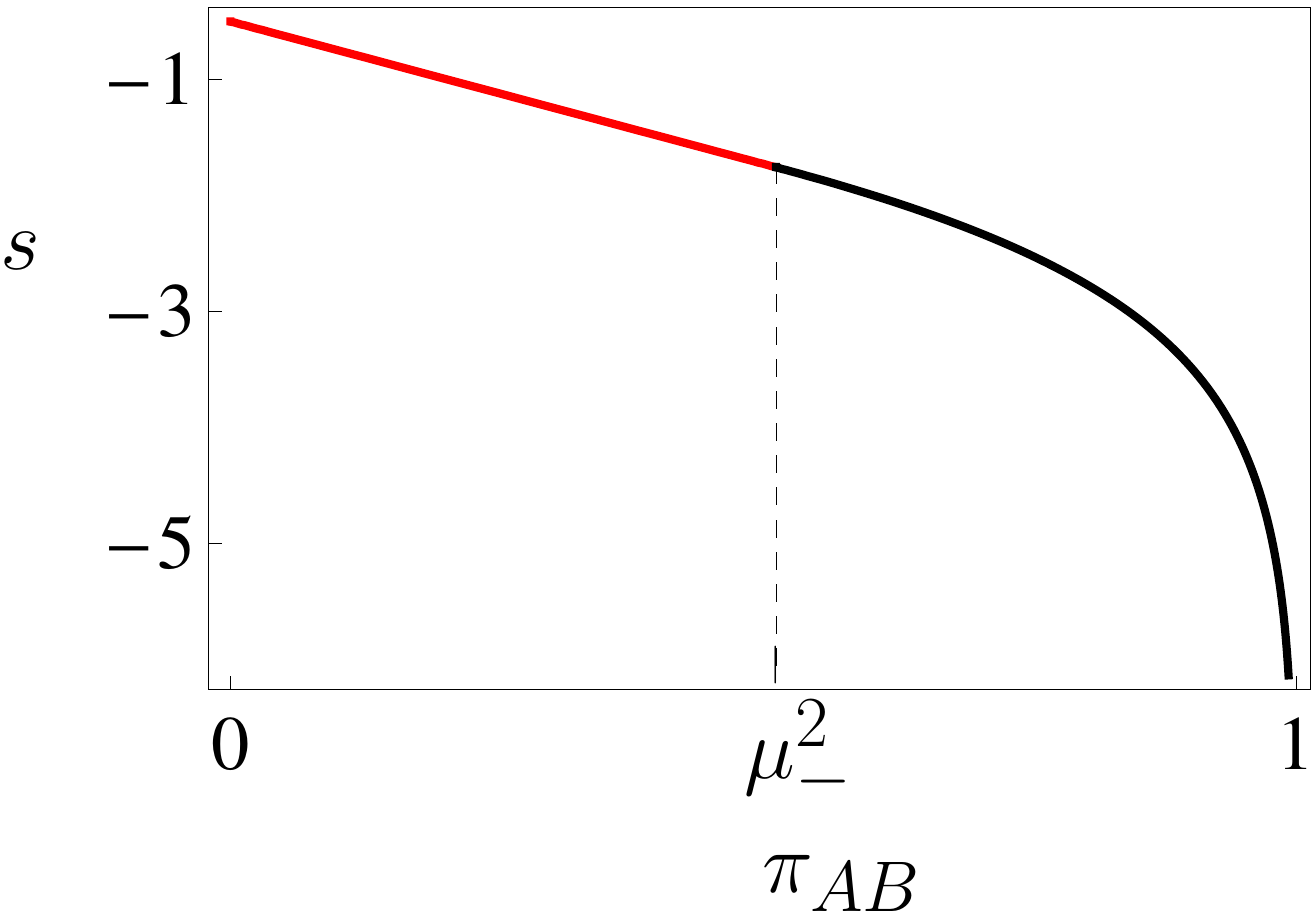}
\caption{Entropy density $s$ versus internal energy density $u$. 
Notice that the unit on the abscissae is $1/N$ in the left panel.}
\label{fig:svsuN1}
\end{figure}
\end{widetext}
with $\mu_{-}^2\simeq 0.512$ and $\beta_{-} \simeq -2.455$ given by (\ref{eq:mu-})-(\ref{eq:beta-}).
The plot of $s$ vs $\pi_{AB}$ in the two regions  $\pi_{AB}=\Ord{1/N}$ and $\pi_{AB}=\Ord{1}$ is shown in  Fig.~\ref{fig:svsuN1}.
By exponentiating the expression (\ref{eq:1121}) we get the volume $V=\exp{(N^2 s)}$ (i.e.\ the probability) of the isopurity manifolds
\begin{widetext}
\begin{eqnarray}
V(\pi_{AB})  \propto \begin{cases}
e^{-\frac{N^2}{4}} (N \pi_{AB} -1)^{N^2/2},  & \frac{1}{N} < \pi_{AB} \leq  \frac{5}{4 N} ,  \\
\\
 \left( \frac{3}{2}  - \sqrt{\frac{9}{4} - N \pi_{AB}}
\right)^{N^2} \exp \left[ N^2\left(- \frac{9}{4} + \frac{5}{2\left( \frac{3}{2}  - \sqrt{\frac{9}{4} - N \pi_{AB}}
\right)} -
\frac{3}{4\left( \frac{3}{2}  - \sqrt{\frac{9}{4} - N \pi_{AB}}
\right)^2}\right)\right] ,
& \frac{5}{4 N} < \pi_{AB} \leq \frac{2}{N},
 \\
\\
\exp \left[ N^2 \left(\beta_- \pi_{AB}
-\frac{1}{2}\right) \right], &  \frac{2}{N} <\pi_{AB} \leq \mu_-^2 
, \\
\\
e^{-\frac{N^2}{2}} \left(1-\sqrt{\pi_{AB} }\right)^{N^2}, &  \mu_-^2 
< \pi_{AB} <1 .
\end{cases}
\label{eq:1122}
\end{eqnarray}
\end{widetext}

\begin{figure}
\includegraphics[width=1.0\columnwidth]{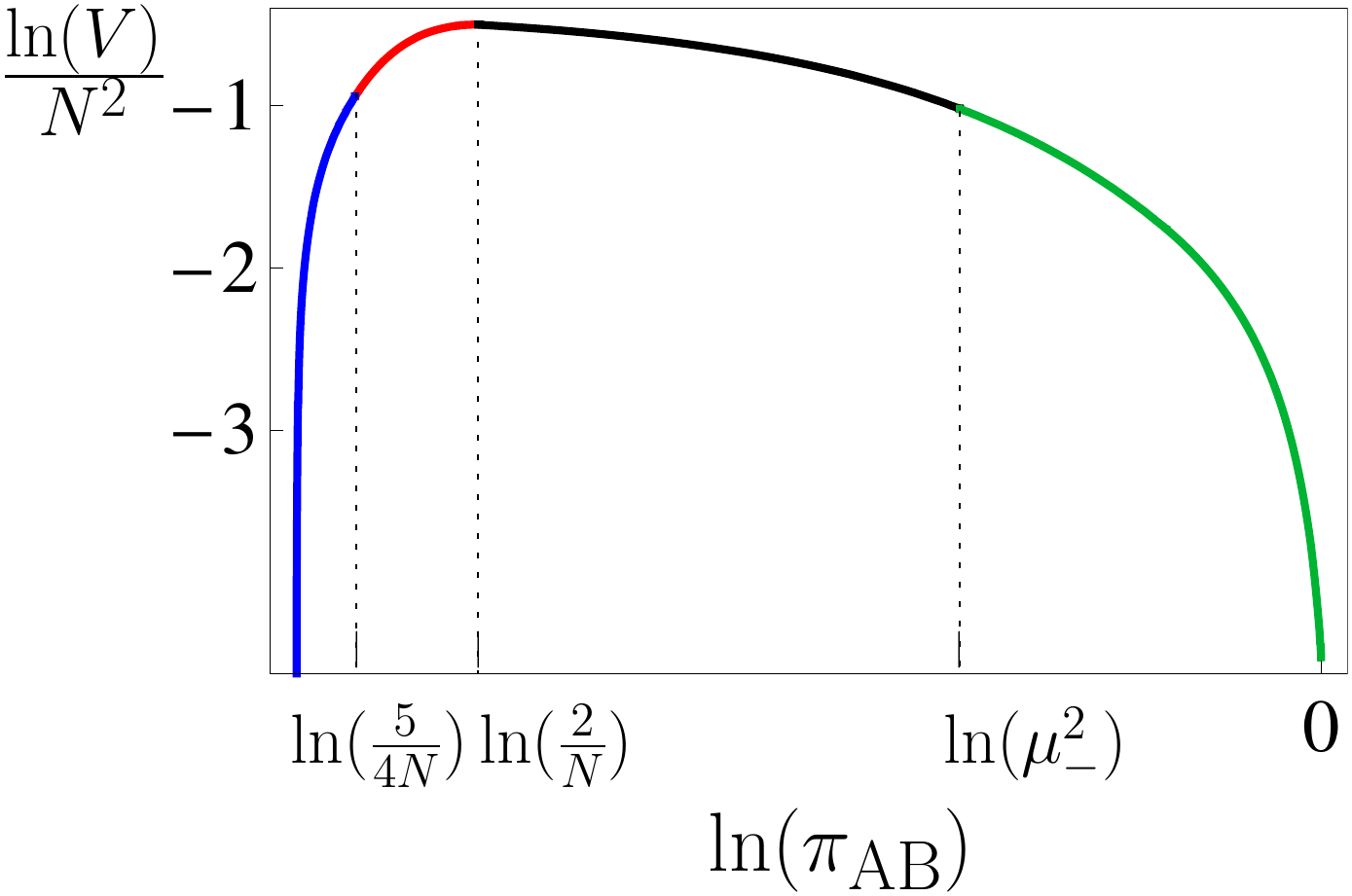}
\caption{Volume $V=\exp(N^2s)$ of the isopurity manifolds versus their purity $\pi_{AB}$ for $N=50$.}
\label{fig:volume}
\end{figure}
\begin{figure}
\includegraphics[width=0.9\columnwidth]{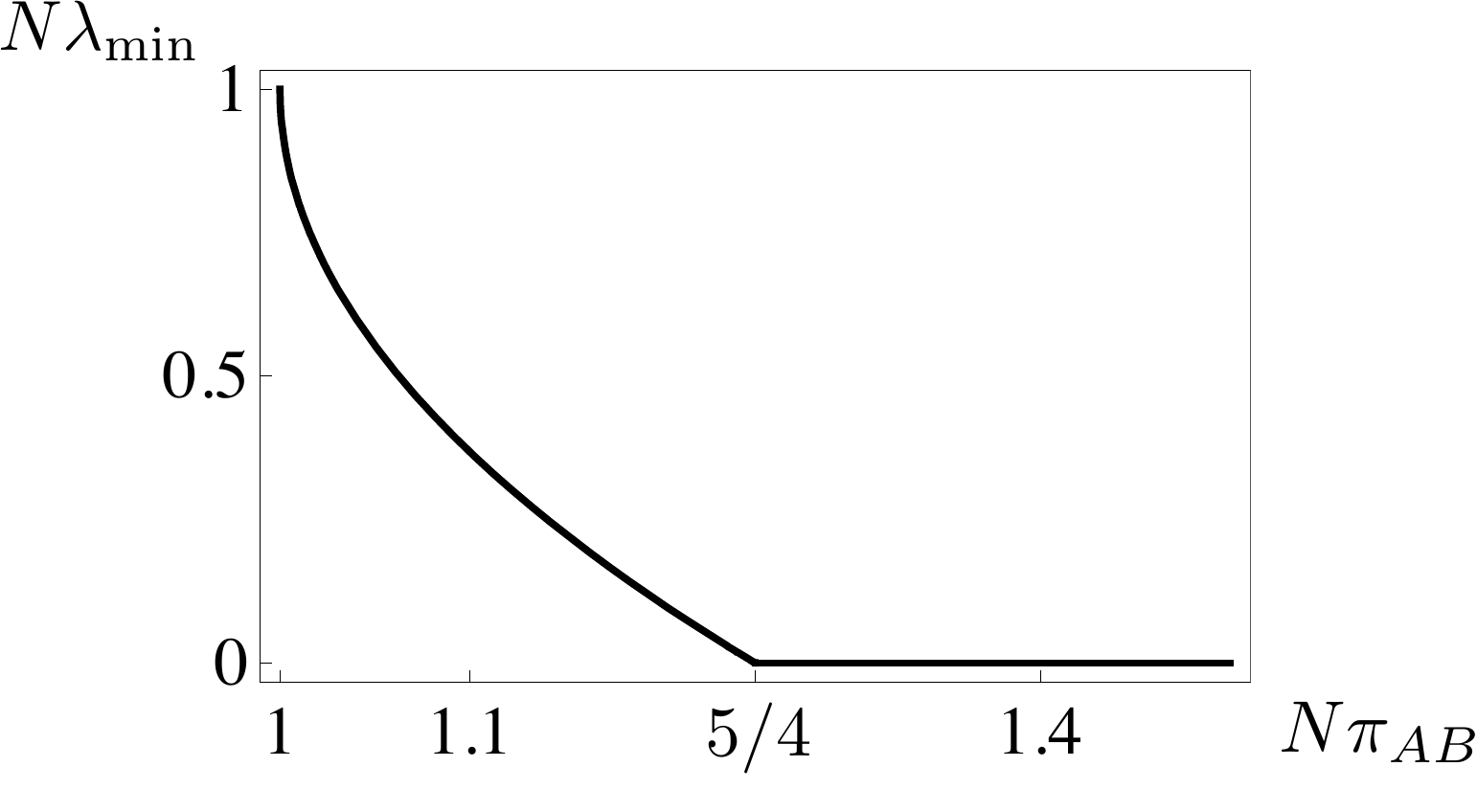}
\caption{Minimum eigenvalues as a function of $\pi_{AB}$.  At $\pi_{AB}=5/4N$ the gap vanishes.}
\label{fig:minmax}
\end{figure}
\begin{figure}
\hspace{0.1cm} \includegraphics[width=0.84\columnwidth]{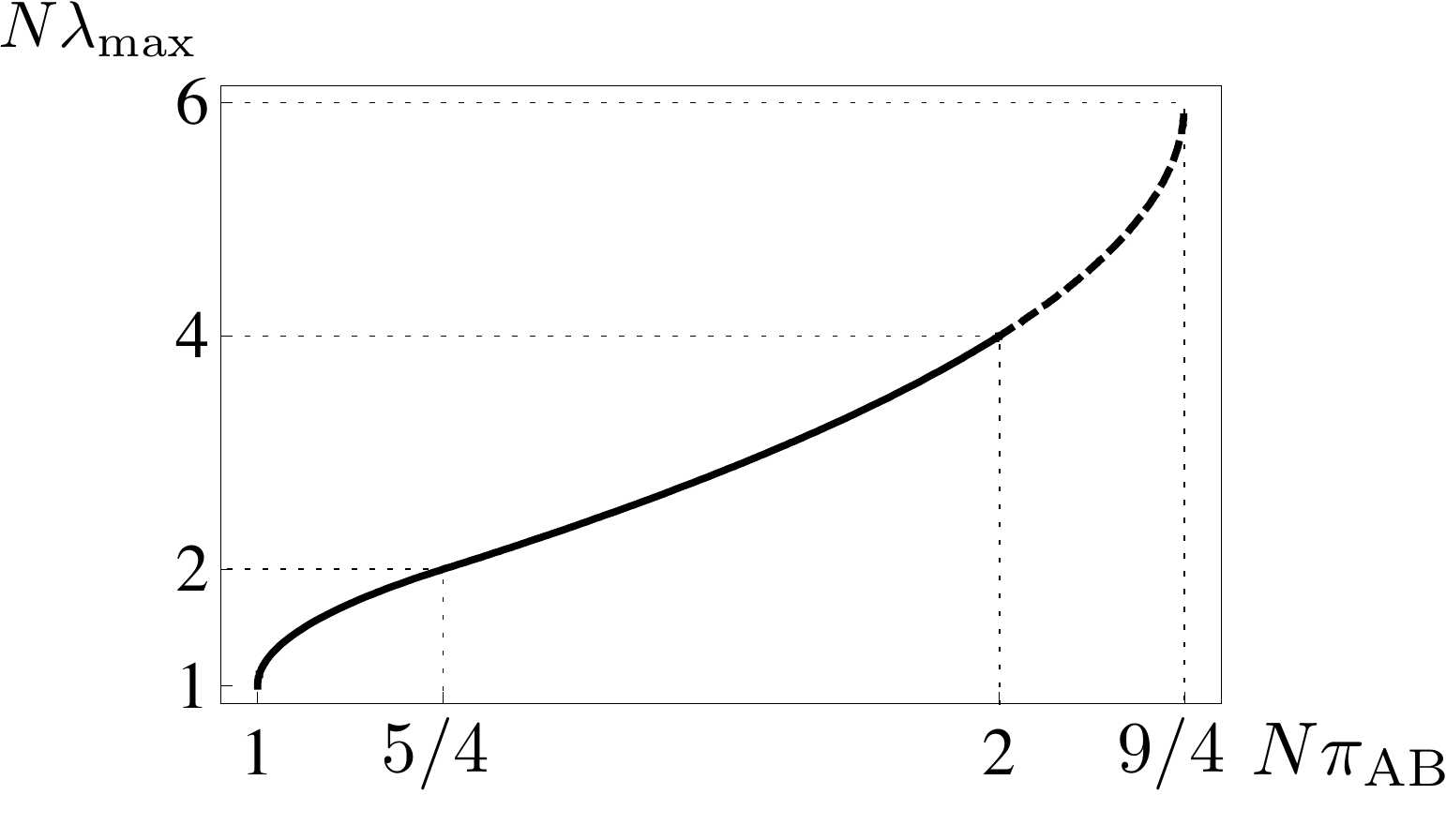}
\\ \vspace{0.1cm}
\includegraphics[width=0.8\columnwidth]{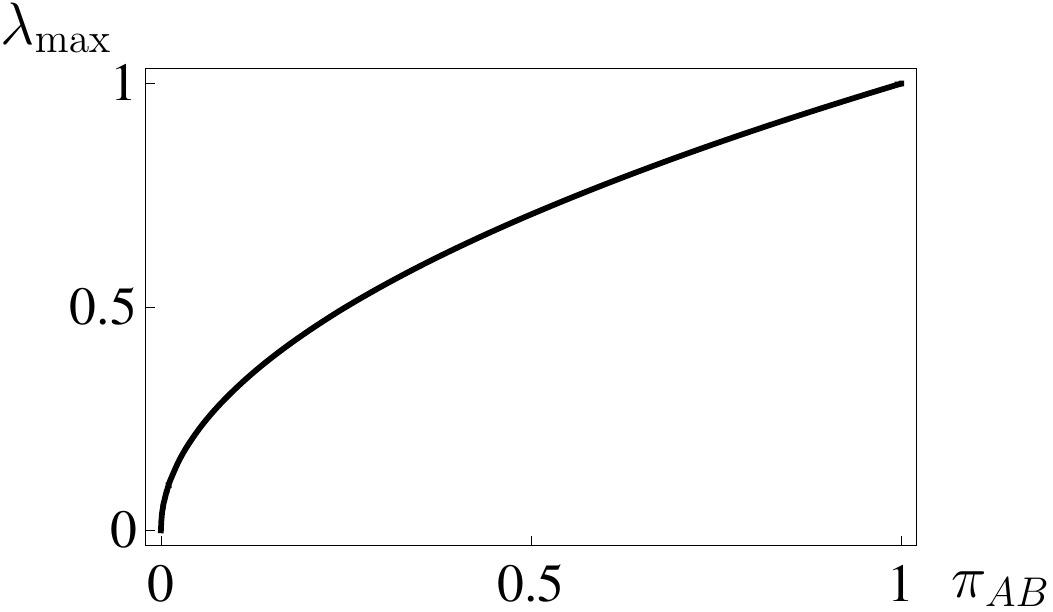}
\caption{Maximum eigenvalue as a function of $\pi_{AB}$. }
\label{fig:maximum}
\end{figure}
This  is plotted in  Fig.~\ref{fig:volume} for $N=50$.
The presence of discontinuities in some derivatives of entropy detects the two phase transitions. At $\pi_{AB}=5/4N$ there is a second order phase transition signaled  by a discontinuity in the third derivative. Indeed, in general, if $s, u$ and $T$ are entropy, energy and temperature, respectively, and $C=du/dT$ is the specific heat, one gets $ds/du =\beta=1/T$ and $d^2s/du^2= -1/(T^2 C)= -(1/T^3)(ds/dT)^{-1}.$ Discontinuities of the $n$th derivative of $ds/dT$ translate therefore in discontinuities of the $(n+1)$-th derivative of $ds/du$.  The first order phase transition, which takes place between $\pi_{AB}=2/N$ and $\pi_{AB}=\mu_{-}^2\simeq 0.512$ is signaled by discontinuities in the second derivative of the entropy at those points.
Observe that entropy is unbounded from below: 
at both endpoints of the range of purity, $\pi_{AB}=1/N$ (maximally entangled states) and $\pi_{AB}=1$ (separable states), when the isopurity manifold shrinks to a vanishing volume in the original Hilbert space, the entropy, being the logarithm of this volume, diverges, and the number of vector states goes to zero (compared to the number of typical vector states). See Fig.\  \ref{fig:volume}.

The presence of the phase transitions can be easily read out from the behavior of the distribution of the Schmidt coefficients, i.e.\ the eigenvalues of the reduced density matrix $\rho_{A}$ of one subsystem. From Eq.~(\ref{eq:deltavsu}) we get the expression of the minimum eigenvalue  $\lambda_{\mathrm{min}}=a = m-\delta$
as a function of $\pi_{AB}$
\begin{eqnarray}
\lambda_{\mathrm{min}} = \begin{cases}
\frac{1}{N}\left(1-2 \sqrt{N \pi_{AB} -1 }\right),  & \frac{1}{N} < \pi_{AB} \leq  \frac{5}{4 N} ,  \\
\\
 0 ,
& \frac{5}{4 N} < \pi_{AB} \leq 1.
\end{cases}
\label{eq:lambdamin}
\end{eqnarray}
which is shown in Fig.~\ref{fig:minmax}. The second order phase transition at $\pi_{AB}=5/4N$, associated to a ${\mathbb Z}_2$ symmetry breaking, is detected by a vanishing gap. On the other hand, the maximum eigenvalue $\lambda_{\mathrm{max}}$ coincides with the upper edge of the sea of eigenvalues $b = m+\delta$, as given by (\ref{eq:deltavsu}), until it evaporates according to Eq.~(\ref{eq:unegmu}). Thus,

\begin{widetext}
\begin{eqnarray}
\lambda_{\mathrm{max}} =  \begin{cases}
\frac{1}{N}\left(1+2 \sqrt{N \pi_{AB} -1 }\right),  & \frac{1}{N} < \pi_{AB} \leq  \frac{5}{4 N} ,  \\
\\
\frac{2}{N}\left( 3  - 2 \sqrt{\frac{9}{4} - N \pi_{AB}} \right),
 & \frac{5}{4 N} < \pi_{AB} \leq \frac{2}{N}\qquad\,
 \\
\\
\sqrt{\pi_{AB} }, &  \frac{2}{N} < \pi_{AB} <1 ,
\end{cases}
\label{eq:lambdamax}
\end{eqnarray}
\begin{figure}
\includegraphics[width=0.33\columnwidth] {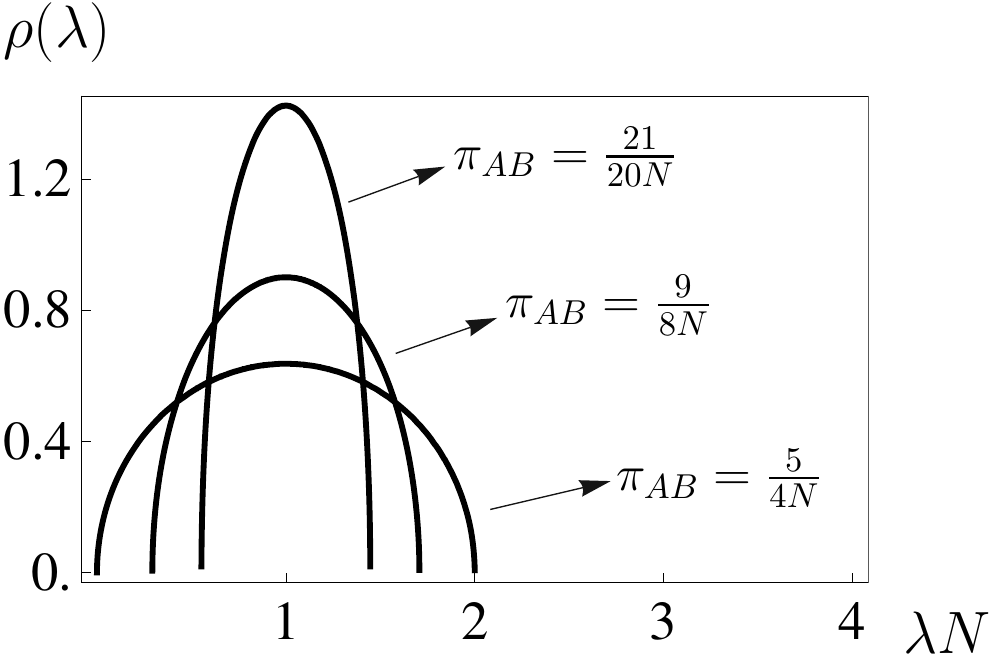}
\includegraphics[width=0.33\columnwidth] {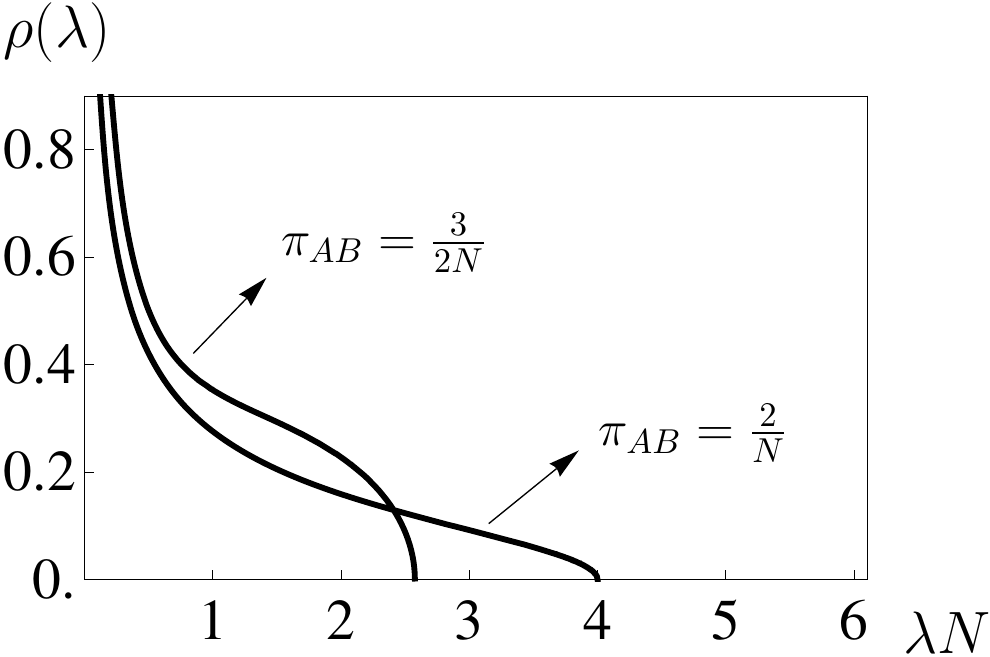} 
\includegraphics[width=0.30\columnwidth] {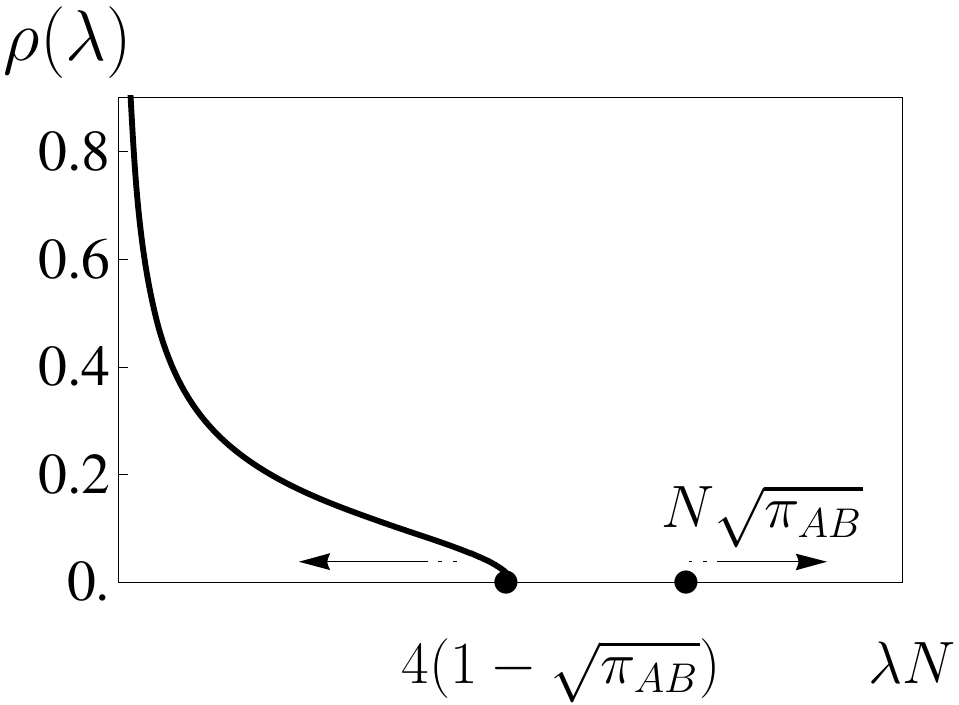}
\caption{Summary (stable branch): profiles of the eigenvalue density for $1 \leq \pi_{AB} \leq \frac{5}{4N}$, $\frac{5}{4N} \leq \pi_{AB} \leq \frac{2}{N}$, and $\frac{2}{N} \leq \pi_{AB} \leq 1$.}
\label{fig:distributions}
\end{figure}
\begin{figure}
\includegraphics[width=0.7\columnwidth] {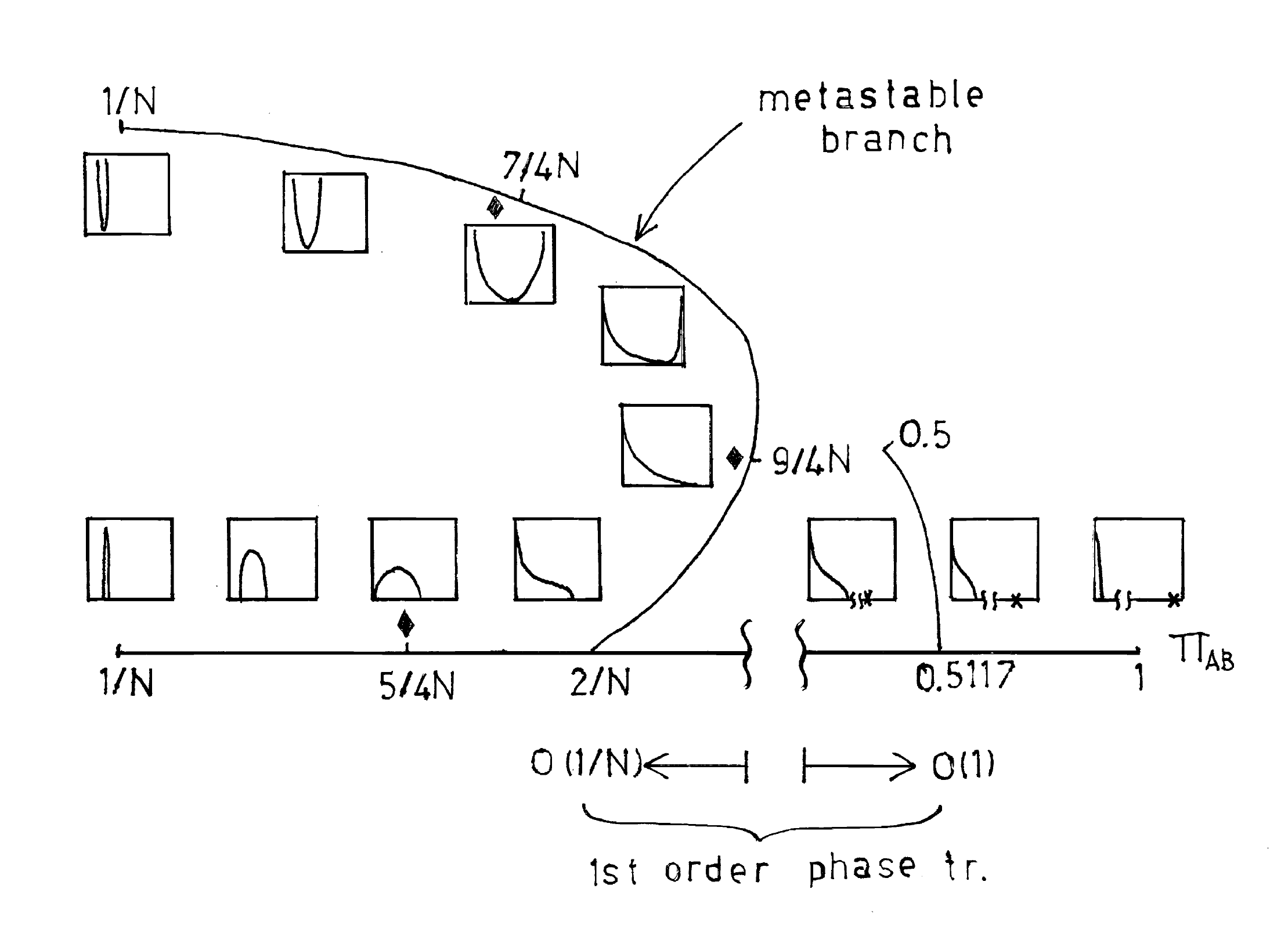}
\caption{Overview of the ``evolution" of the eigenvalue densities as a function of purity $(1/N \leq \pi_{AB} \leq 1)$.
The straight segment represents the stable branch: the distribution starts as a delta function, evolves into a semicircle, undergoes a second-order phase transition at $\pi_{AB}=5/4N$ (${\mathbb Z}_2$ symmetry breaking), becomes Wishart and undergoes a first-order phase transition between $2/N \leq \pi_{AB} \leq 0.5117$, during which one eigenvalue evaporates from the sea of the others  $\Ord{1/N}$ and becomes  $\Ord{1}$.
A metastable branch is born at $\pi_{AB}=2/N$: it starts as Wishart, undergoes a second-order phase transition at  $\pi_{AB}=9/4N$ (2-D gravity), develops a singularity at its right edge through a second order phase transition at $\pi_{AB}=9/4N$, then its support starts decreasing, undergoes a second-order phase transition at $\pi_{AB}=7/4N$ (${\mathbb Z}_2$ symmetry restoration) and eventually becomes sharply peaked (with two singularities). The diamonds indicate the three second order phase transitions.}
\label{fig:masterpiece}
\end{figure}
\end{widetext}
as shown in Fig.\ \ref{fig:maximum}.

In the different phases the distribution of the eigenvalues of $\rho_A$ have very different profiles. 
See Fig.\ \ref{fig:distributions}.
While for $1 \leq \pi_{AB} \leq \frac{5}{4N}$ the eigenvalues (all $\Ord{1/N}$) follow Wigner's semicircle law, they become distributed according to Wishart for larger purities, $\frac{5}{4N} \leq \pi_{AB} \leq \frac{2}{N}$, across the second order phase transition. This is a first signature of separability: some eigenvalues vanish and the Schmidt rank decreases.
For even larger  values of purity, $\frac{2}{N} \leq \pi_{AB} \leq 1$, across the first order phase transition, one
eigenvalue evaporates, leaving the sea of the other eigenvalues
$\Ord{1/N}$ and becoming $\Ord{1}$. This is the signature of factorization, fully attained when the eigenvalue becomes $1$ at $\pi_{AB} = 1$.

We tried to give an overview of the phenomenology of these phase transitions in Fig.\ \ref{fig:masterpiece}, where we also showed the presence of the metastable branches discussed  in Sec.~\ref{sec:negativetemp}. The global picture is both rich and involved and it would not be surprising if additional features would be unveiled by future investigation.

\section{Conclusions}
\label{sec:concl}

We have obtained a complete characterization of the statistical
features of the bipartite entanglement of a large quantum system 
in a pure state. The global picture is interesting as several locally stable solutions exchange stabilities. On the stable branch (solutions of minimal free energy) there is a second order phase transition,
associated to a ${\mathbb Z}_2$ symmetry breaking, and related to the vanishing of some Schmidt coefficients (eigenvalues of the reduced density matrix of one subsystem), followed by a first order phase transition, associated to the evaporation of the largest eigenvalue from the sea of the others.

In the different phases the distribution of the Schmidt coefficients
have very different profiles. While for large $\beta$ (small purity) the
eigenvalues (all $\Ord{1/N}$) follow Wigner's semicircle law,
they become distributed according to Wishart for smaller $\beta$ and larger purity,
across the first transition. For even smaller (and eventually
negative) values of $\beta$, when purity becomes finite, across the second phase transition, one eigenvalue evaporates, leaving the sea of the other eigenvalues
$\Ord{1/N}$ and becoming $\Ord{1}$. This is the signature of separability,
this eigenvalue being associated with the emergence of factorization
in the wave function (given the bipartition). This intepretation is
suggestive and hints at a profound modification of the distribution
of the eigenvalues as $\beta$, and therefore purity, are changed. Remember that $\beta$, viewed as a Lagrange multiplier in this statistical mechanical
approach, localizes the measure on set of states with a given entanglement (isopurity manifolds \cite{Kus01}).

It would be of great interest to understand whether the phase
transitions survive even in the multipartite entanglement scenario,
if one views the distribution of purity (over all balanced
bipartitions) as a characterization of the global entanglement of
the many-body wave function of the quantum system \cite{mmes}.
This description of multipartite entanglement displays the symptoms of frustration \cite{frust}, catapulting the problem into one of the most
fascinating arenas of modern statistical mechanics \cite{parisi}.

While we were completing this work a paper appeared in which an aspect of this problem is discussed, although with a different emphasis \cite{Majumdar09}. In order to connect our results to those in \cite{Majumdar09}, notice that the probability distribution of the purity $P(u)$  is proportional to the volume of the isopurity manifold and therefore 
\begin{equation}
\ln P(u)\simeq N^2 s(u) ,
\end{equation}
where $s$ is the entropy and $N\langle \pi_{AB}\rangle=u$ is the internal energy. See Eq.\ (\ref{eq:1122}).
\section{Acknowledgements}
This work is partly supported by the European Union through the
Integrated Project EuroSQIP.

\end{document}